\documentclass[twocolumn,twocolappendix]{aastex631}

\usepackage{lipsum}
\usepackage{placeins}
\usepackage{amsmath}
\let\tablenum\relax
\usepackage{siunitx}
\usepackage{xfrac}
\usepackage{xcolor}
\usepackage[normalem]{ulem}
\usepackage{orcidlink}  
\usepackage{graphicx}
\usepackage{booktabs}

\shorttitle{\texttt{INDRA} for Characterizing JWST Ice Spectra}
\shortauthors{Prathap et al.}

\begin{document}

\title{Expanding the Ice Inventory of NGC 1333 IRAS 2A with INDRA using JWST Observations: Tracing Organic Refractories and  Beyond}

\correspondingauthor{Liton Majumdar}
\email{liton@niser.ac.in; dr.liton.majumdar@gmail.com}

\author{Prathap Rayalacheruvu\,\orcidlink{0009-0001-6483-7366}}
\affiliation{Exoplanets and Planetary Formation Group, School of Earth and Planetary Sciences, National Institute of Science Education and Research, Jatni 752050, Odisha, India}
\affiliation{Homi Bhabha National Institute, Training School Complex, Anushaktinagar, Mumbai 400094, India}

\author{Liton Majumdar\,\orcidlink{0000-0001-7031-8039}}
\affiliation{Exoplanets and Planetary Formation Group, School of Earth and Planetary Sciences, National Institute of Science Education and Research, Jatni 752050, Odisha, India}
\affiliation{Homi Bhabha National Institute, Training School Complex, Anushaktinagar, Mumbai 400094, India}

\author{W. R. M. Rocha\,\orcidlink{0000-0001-6144-4113}}
\affiliation{Leiden Observatory, Leiden University, PO Box 9513, NL 2300 RA Leiden, The Netherlands}

\author{Michael E. Ressler\,\orcidlink{0000-0001-5644-8830}}
\affiliation{Jet Propulsion Laboratory, California Institute of Technology, 4800 Oak Grove Drive, Pasadena, CA 91109, USA}

\author{Pabitra Ranjan Giri\,\orcidlink{0009-0008-6173-2519}}
\affiliation{Exoplanets and Planetary Formation Group, School of Earth and Planetary Sciences, National Institute of Science Education and Research, Jatni 752050, Odisha, India}
\affiliation{Homi Bhabha National Institute, Training School Complex, Anushaktinagar, Mumbai 400094, India}

\author{S. Maitrey\,\orcidlink{0000-0003-3412-9454}}
\affiliation{Exoplanets and Planetary Formation Group, School of Earth and Planetary Sciences, National Institute of Science Education and Research, Jatni 752050, Odisha, India}
\affiliation{Homi Bhabha National Institute, Training School Complex, Anushaktinagar, Mumbai 400094, India}

\author{Karen Willacy\,\orcidlink{0000-0001-6124-5974}}
\affiliation{Jet Propulsion Laboratory, California Institute of Technology, 4800 Oak Grove Drive, Pasadena, CA 91109, USA}

\author{D. C. Lis\,\orcidlink{0000-0002-0500-4700}}
\affiliation{Jet Propulsion Laboratory, California Institute of Technology, 4800 Oak Grove Drive, Pasadena, CA 91109, USA}

\author{Yuan Chen\,\orcidlink{0000-0002-3395-5634}}
\affiliation{Leiden Observatory, Leiden University, PO Box 9513, NL 2300 RA Leiden, The Netherlands}

\author{P. D. Klaassen\,\orcidlink{0000-0001-9443-0463}}
\affiliation{UK Astronomy Technology Centre, Royal Observatory Edinburgh, Blackford Hill, Edinburgh EH9 3HJ, UK}

\begin{abstract}
In the era of JWST, with its unprecedented sensitivity and spectral resolution, infrared spectral surveys have revealed a rich inventory of ices, including complex organic molecules (COMs), in young stellar objects (YSOs). However, robust methods to decompose and quantify these absorption features particularly across broad spectral ranges, are still under investigation. We present \texttt{INDRA} (\textit{Ice-fitting with NNLS-based Decomposition and Retrieval Algorithm}), a fully Python-based tool that performs continuum and silicate removal, global ice fitting using Weighted Non-Negative Least Squares (NNLS), and estimates column densities and statistical significance. We apply \texttt{INDRA} to NGC 1333 IRAS 2A, a target from the JWST Observations of Young protoStars (JOYS+) program previously studied using local fitting. We derive optical depths via polynomial continuum subtraction and remove silicate absorption using a synthetic model, isolating ice features for global MIRI fitting. Our results are consistent with previous local fits, confirming simple species and COMs, and expand the inventory by identifying additional absorption features from CO$_2$ and NH$_4^+$. We also propose the presence of organic refractories contributing up to 9.6\% in the spectral region of $\sim$5--8\,$\mu$m among the various ice components, whose inclusion significantly improves the global spectral fitting. These broad absorption features, extending across $\sim$5.5--11\,$\mu$m, are likely produced by large, complex molecules containing carbonyl (C=O), hydroxyl (O--H), amine (N--H), and C--H bending modes. Our expanded inventory, now incorporating these organic residues, offers new insights into the chemical evolution of ices in star-forming regions and highlights the importance of global spectral fitting in constraining ice compositions.

\end{abstract}
\section{Introduction} \label{sec:intro}
Molecules form in dense molecular clouds, where low temperatures allow gas-phase species to freeze onto dust grains and also promote chemical reactions on grain surfaces, leading to the formation of icy mantles. These ices are predominantly composed of water (H$_{2}$O), carbon monoxide (CO), methane (CH$_{4}$), and other simple molecules, which later end up in planet-forming disks where future planets and other planetary bodies form \citep{Oberg2021}. Understanding the ice chemistry in young stellar objects (YSOs), an intermediate stage in planet formation, provides a unique window into the evolving conditions from cold molecular clouds to planet-forming disks. The icy mantles not only serve as reservoirs for simple molecules but also play a critical role in the formation of complex organic molecules (COMs). Examining the composition and column densities of ices in YSOs provides valuable insights into the physical and chemical conditions in which these ices form and evolve during star and planet formation \citep{Boogert2015}.
\par

Gas-phase observations have revealed the chemical diversity of both low-mass and high-mass YSOs (e.g., \citealt{Jorgensen2020}). While gas-phase COMs are well characterized through millimeter observations, the identification of solid-phase COMs remained challenging prior to the launch of the James Webb Space Telescope (JWST), due to limited spectral resolution and overlapping features in infrared (IR) absorption spectra. Methanol (CH$_{3}$OH) is the only securely detected COM in interstellar ices, identified via IR absorption spectroscopy using facilities such as the United Kingdom Infrared Telescope (UKIRT), Infrared Space Observatory (ISO), Very Large Telescope (VLT), AKARI, and \textit{Spitzer} (\citealt{Grim1991}; \citealt{Skinner1992}; \citealt{Dartois1999}; \citealt{Gibb2004}; \citealt{Pontoppidan2004}; \citealt{Dartois2003}; \citealt{Thi2006}; \citealt{Chu2020}; \citealt{Boogert2008}; \citealt{Bottinelli2010}; \citealt{Shimonishi2010}). Other COMs, such as ethanol (CH$_{3}$CH$_{2}$OH) and acetaldehyde (CH$_{3}$CHO), have been tentatively identified based on Spitzer and ground-based data, but their detections remain ambiguous due to spectral blending and the absence of distinct spectral features (\citealt{Schutte1999_weak}; \citealt{Oberg2011}; \citealt{Scheltinga2018}).

\par
JWST is revolutionizing our understanding of chemical diversity of star-forming regions by offering unprecedented sensitivity and resolution across the IR spectrum. One of its key instruments, the Mid-Infrared Instrument (MIRI; \citealt{2015PASP..127..584R}, \citealt{2015PASP..127..595W}, \citealt{2023PASP..135d8003W}), operates in four modes, including the Medium-Resolution Spectroscopy (MRS; \citealt{2015PASP..127..646W}, \citealt{2021A&A...656A..57L}, \citealt{2023A&A...675A.111A}). JWST’s MIRI-MRS offers a significant improvement in spectral resolution (R $\sim$ 1300–3700) over the Spitzer Infrared Spectrograph (IRS) (R $\sim$ 60–600), enabling the identification of individual ice absorption bands that were previously blended. This is particularly crucial for studying COMs in the MIRI wavelength range (\citealt{2021A&A...656A..57L}; \citealt{Yang2022}; \citealt{2024A&A...683A.124R}; \citealt{2024A&A...690A.205C}), helping to trace the transport of COMs from the early stages of star formation to later evolutionary phases. Additionally, JWST's full spectral coverage and high sensitivity enable simultaneous measurements of CO and CO$_{2}$ ices, including their isotopologues, facilitating a more comprehensive analysis using carbon isotopic ratios to further constrain evolutionary processes \citep{Brunken2024}. These capabilities provide new opportunities to study the chemical evolution of interstellar ices and their role in the formation of larger COMs during various stages of star and planet formation. In this context, the JWST Guaranteed Time Observation program JOYS (JWST Observations of Young protoStars) aims to characterize the physical and chemical processes occurring in both high-mass and low-mass star-forming regions using near- and mid-infrared spectra of molecules in both the gas and ice phases. Early results from the JOYS program reveal a rich chemical diversity of species in both the gas and ice phases across several sources (\citealt{2024A&A...683A.124R}; \citealt{2024A&A...690A.205C}; \citealt{vanGelder2024A&A}).


\par
Studying the formation and evolution of COMs holds significant importance in star and planet formation studies and remains an active area of research. Studies like \cite{Manigand2020}; \cite{Belloche2020}; \cite{vanGelder2020}; \cite{Jorgensen2020}; \cite{Nazari2021} and \cite{Gieser2021} etc., have elaborated on the formation of COMs. They primarily form on the surfaces of interstellar dust grains during the cold early stages when $T \sim 10$ K, leading to the development of ice mantles. Simple species like CO can undergo hydrogenation and radical recombination in the ice mantle, initiating the formation of COM precursors. The presence of energetic processes such as UV irradiation and cosmic rays further drives the COM chemistry, leading to the formation of even larger COMs. UV photoprocessing and thermal processing both play important roles in the evolution of COM chemistry. For example, the ion CH$_{3}^{+}$, observed in the gas phase, is commonly associated with UV-driven reactions. In the solid phase, species like OCN$^{-}$ can form through acid–base reactions, which are efficiently mediated by both thermal processing and UV photoprocessing of HNCO-containing ices. These chemical pathways, triggered by energetic photons and charged particles, contribute to the buildup of organic refractory residues, including compounds of potential astrobiological significance. Laboratory experiments by \cite{2024MNRAS.534.2305J} showed that irradiation and thermal alteration of simple ices like CH$_{3}$OH, H$_{2}$O and NH$_{3}$ can produce diverse organic molecules which include residual organic compounds containing up to 78 carbon atoms, 188 hydrogen atoms, and 37 oxygen atoms. \cite{2024EPSC...17..800U} carried out similar experiments involving simple ices and found the destruction of pristine frozen compounds and the formation of new species that survived till 300 K. These refractory organics potentially become part of dust grains in planet-forming regions and later agglomerate into comets and asteroids. As a result, some of the organic matter present in the Solar System may have been inherited from the earliest stages of its formation. \cite{2024EPSC...17..596D} have demonstrated how this matter can evolve inside asteroids, parent bodies of meteorites in our solar system using an analytical approach. The residues inherited may undergo further processing depending on the local conditions. For example, studies by \cite{2024A&A...684A.198M} and \cite{2024EPSC...17..367C} have shown how these materials are altered inside asteroids like Ryugu and the Jovian moon Europa respectively. However, in environments with milder processing conditions, such as comets, these materials are more likely to remain unaltered, preserving a record of the early chemistry of the Solar System and offer crucial insights into the origin and evolution of COMs. Studying the chemical inventory of both low-mass and high-mass protostellar environments enhances our understanding of how organic refractories form and survive before becoming incorporated into planetary bodies.

\par

The low-mass Class 0 protostar NGC 1333 IRAS 2A (hereafter IRAS 2A), located in the Perseus molecular cloud ($\sim$299 pc), is a well-known hot corino with a rich inventory of gas-phase COMs; (\citealt{Bottinelli2007}). Gas-phase glycolaldehyde (HCOCH$_{2}$OH), a key prebiotic molecule, has been detected in this source (\citealt{Coutens2015}; \citealt{Taquet2015}; \citealt{desimone2017}). \cite{vanGelder2024A&A} recently confirmed the mid-infrared emission of SO$_{2}$ in IRAS 2A for the first time. Based on the rotational temperature, the spatial extent of SO$_{2}$ emission, and the narrow line widths in ALMA data, they suggested that SO$_{2}$ likely originates from ice sublimation in the central hot core around the protostar, rather than from an accretion shock. Furthermore, IRAS 2A hosts a protobinary system with collimated jets, whose shocks and associated energetic processing are expected to significantly influence ice chemistry (\citealt{Looney2000}; \citealt{Reipurth2002}; \citealt{Sandell1994}; \citealt{Tobin2015}). While gas-phase studies of IRAS 2A have yielded substantial insights, solid-phase studies have been limited due to the low spectral resolution of earlier facilities and the inherently blended nature of ice absorption features. However, observations with JWST have now revealed rich ice absorption signatures with high signal-to-noise (S/N) ratios, enabling more detailed analysis. \cite{2024A&A...683A.124R} studied IRAS 2A in the ice-COM fingerprint region between 6.8 and 8.6~$\mu$m, estimating the column densities of various complex organic and simple ices. Their findings shed light on how these ices may be formed, evolve, and be inherited by icy cometary bodies. They also reported a good correlation between the ice abundances observed in comet 67P and those in IRAS 2A. The detected ices in IRAS 2A include CH$_{3}$CHO, larger organics with multiple carbon atoms such as CH$_{3}$OCH$_{3}$ and C$_{2}$H$_{5}$OH, as well as ions like HCOO$^{-}$ and OCN$^{-}$. \cite{2024A&A...690A.205C} also examined another low-mass protostar, B1-c, and found similar ice components as that of IRAS 2A. Their gas-to-ice comparisons of COMs in both sources revealed that molecules such as CH$_{3}$OCH$_{3}$ and CH$_{3}$OCHO show similar abundance ratios in both phases, while CH$_{3}$CHO and C$_{2}$H$_{5}$OH are more abundant in the ice phase. This suggests that inheritance-driven processes play a significant role, alongside possible gas-phase reprocessing. These findings highlight the complex interplay between ice chemistry, gas-phase reactions, and physical processes in shaping the chemical evolution of COMs in star-forming regions. While these studies have offered valuable insights into the chemistry of IRAS 2A, they were typically limited to specific spectral regions, particularly within the 6.8-8.8 $\mu$m range. This region, however, lies within a broader absorption feature spanning 5-11 $\mu$m, whose origin remains uncertain. Although H$_{2}$O has a bending mode at around 5.8 $\mu$m, it cannot contribute entirely to the observed absorption dip across this interval. In order to account for 5-8 $\mu$m absorption region, \cite{Boogert2008} have analytically constructed a spectrum, defined as C5, by subtracting known features from linear combinations of observed spectra, making it an analytical diagnostic component. Though the nature of the C5 component is not fully explored, they related it to the flat profile of high-temperature H$_{2}$O ice bending mode, the overlap of other negative ions (HCO$_{3}^{-}$, NO$_{3}^{-}$, NO$_{2}^{-}$) or organic refractory residue. We note that the absorption in the 5-8 $\mu$m region is likely due to a combination of all these factors but to what extent remains unclear. Studies of other protostellar sources suggest that processed ices, including refractory organic residues formed by energetic processing, may contribute significantly to the absorption near 6 $\mu$m. For instance, W33 A shows the deepest 6 $\mu$m excess absorption, attributed by \citet{Gibb2002} to a refractory organic component produced by UV processing of icy mantles. Similar components are evident in AFGL 7009S, Mon R2 IRS 3, and NGC 7538 IRS 9, where subtraction of the dominant H$_{2}$O feature leaves residuals well explained by such organic material. The use of organic refractory spectra obtained through experimental measurements can help us understand the extent to which refractories shape the C5 component of \cite{Boogert2008} in the 5-8 $\mu$m absorption region. Furthermore, the absorption bands of H$_{2}$O and CO$_{2}$ long-wards of 10 $\mu$m in the observed spectrum of the source have yet to be thoroughly characterized. Utilizing the full MIRI spectral range ($\sim$5-28 $\mu$m) allows for a more comprehensive characterization of these absorption bands, thereby providing critical insights into the physical and chemical environment of IRAS 2A, including potential signatures of energetic processing. Moreover, this extended coverage is especially important in the COM region, where features can be spectrally blended with signatures from refractory organic residues which are byproducts of energetic processing of COM ices. Access to the full MIRI range enables a clearer distinction between pristine and processed ices, offering a more complete picture of the chemical evolution occurring in the protostar.

\par
\par
This work focuses on quantifying the ice inventory of IRAS 2A across the full MIRI wavelength range, offering crucial insights into the chemical evolution of protostellar ices. To achieve this, we developed INDRA (Ice-fitting with NNLS-based Decomposition and Retrieval Algorithm), a Python-based tool designed to perform spectral decomposition of ice absorption features using laboratory ice spectra. The structure of the paper is as follows: Section \ref{sec:observations} provides an overview of the observations of IRAS 2A. Section \ref{sec:methodology} describes the ice-fitting tool \texttt{INDRA}, detailing the methods used for continuum removal, silicate absorption feature subtraction, and the ice-fitting technique. In Section \ref{sec:results}, we present the global fitting results for the IRAS 2A protostar over the complete MIRI range and compare the derived ice column densities with those obtained by \cite{2024A&A...683A.124R} using a local fitting approach. The discussion is provided in Section \ref{sec:discussions}, and our conclusions are outlined in Section \ref{sec:conclusions}.


\section{Observations}\label{sec:observations}
NGC 1333 IRAS 2A (RA 03$^{h}$28$^{m}$55.57$^{s}$, Dec +31$^{d}$14$^{m}$36.97$^{s}$) was observed using the JWST as part of the guaranteed observation time (GTO) program 1236 (P.I. M. E. Ressler). The observations were carried out with MIRI-MRS. The target was observed using a single pointing in two-point dither mode, with dedicated background observations in the same mode. All three gratings (A, B, C) were employed, providing full wavelength coverage from 4.9 to 28 $\mu$m. The FASTR1 readout mode was utilized, and the integration time for each grating was 111 seconds. The data were processed through the three stages of the JWST calibration pipeline \citep{bushouse_2022_6984366}, using the reference context \texttt{jwst\_0994.pmap} of the Calibration Reference Data System (CRDS; \citealt{2016A&C....16...41G}). The raw \texttt{uncal} data were initially reduced with the \texttt{Detector1Pipeline}, followed by the \texttt{Spec2Pipeline}. In this stage, fringe corrections were applied using the extended source fringe flat, supplemented by a residual fringe correction. The telescope background was subtracted using the dedicated background observation. Subsequently, the \texttt{Spec3Pipeline} was used to produce data cubes for all 12 sub-bands, with both the master background and outlier rejection routines switched off.
\par
The observations revealed continuum emission associated only with the primary component of the IRAS 2A binary system. The spectrum was extracted from the continuum peak at 5.5 $\mu$m, located at RA (J2000) 03$^{h}$28$^{m}$55.57$^{s}$ and Dec (J2000) +31$^{d}$14$^{m}$36.76$^{s}$. The aperture diameter was set to $4 \times 1.22\lambda/D$ to maximize the captured source flux while minimizing noise. The estimated 1$\sigma$ root mean square (rms) noise increases from approximately 0.4~mJy at wavelengths below 15 $\mu$m to a few mJy at 19 $ \mu$m and exceeds 10~mJy for wavelengths longer than 22 $\mu$m.
The spectrum shows a typical profile of an embedded protostar and shows absorption features attributed to various ice molecules. Broad H$_{2}$O absorption bands are visible at 5.5-8 $\mu$m (bending mode) and 10-20 $\mu$m (libration mode). In addition, strong silicate absorption features appear near 9.8 and 18 $\mu$m.
Narrow emission lines present in the spectrum have been masked in this work, as the focus is on ice absorption features. The IRAS 2A spectrum was binned by a factor of four across 8-12 $\mu$m wavelength range to improve the S/N ratio as this region is saturated with silicate emissions.
\FloatBarrier  
\section{Methodology} \label{sec:methodology}
\begin{figure*} [t]
\centering
   \includegraphics[width=2\columnwidth]{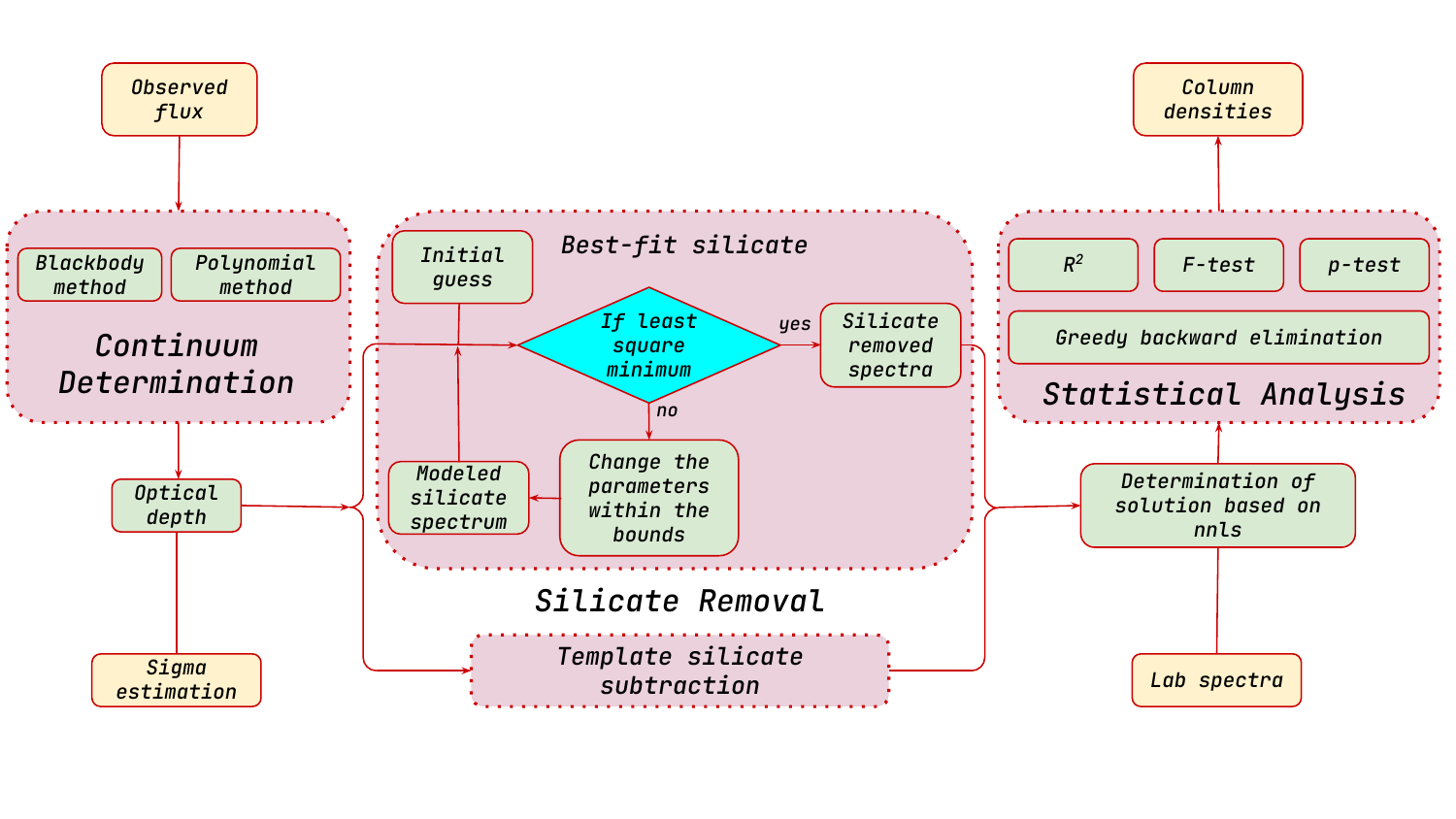}
     \caption{Schematic representation of the methodology for deriving the ice column densities using \texttt{INDRA}. The process includes continuum determination, followed by silicate feature removal and global ice fitting using NNLS. Statistical analysis is performed to assess fit quality and uncertainties after which final column densities are estimated.}
     \label{fig:flow chart}
\end{figure*}

In this section, we provide details about the ice fitting tool \texttt{INDRA}, which is designed for global continuum removal, silicate feature elimination, ice absorption fitting using laboratory data, and estimation of ice column densities. The tool also includes functionality for applying a Savitzky-Golay (Savgol) filter \citep{1964AnaCh..36.1627S, 2005SigPr..85.1429L} to smooth selected laboratory data, along with robust statistical analysis tools to assess the quality of the fit. The various components of the tool are outlined in Figure~\ref{fig:flow chart}.

\subsection{Continuum removal}
In order to constrain the ice absorption features and estimate accurate column densities, it is imperative to determine the continuum and remove it from the observed data before fitting for laboratory ices. However, accurately identifying the continuum SED poses several challenges, particularly in the wavelength range from 5 to 30~$\mu$m, where ice absorption features not only overlap with silicate features but also blend with other ice absorption bands. Additionally, the absorption bands can show a wide range of widths and shapes depending upon the local chemical inventory, further complicating the continuum determination. \cite{2021A&A...654A.158R} have detailed some of the challenges in determining the continuum in this wavelength range. The continuum SED shortwards of 5 $\mu$m  can be better constrained because of the lack of ice absorption features except for some narrow features of CO$_{2}$ and CO ices that appear at around 4.27 and 4.67 $\mu$m respectively. Several methods for constraining the continuum SED at these wavelengths have been documented in the literature.  For example, continuum estimation using Kurucz stellar atmosphere models for wavelengths below 4~$\mu$m \citep{Pontoppidan2004}, and a linear combination of blackbodies to fit the near- and mid-infrared (IR) wavelengths \citep{2002AJ....124.2790I}, can not only estimate the continuum SED but also constrain physical properties such as extinction and effective temperature. 

In this study, we adopt the two continuum estimation methods between 5 and 30~$\mu$m proposed by \citet{2021A&A...654A.158R}. These two methods are:
\par
\noindent (1) A polynomial approach, where a low-order polynomial given by Equation~\ref{eq:polynomial_fitting} is used to estimate the continuum:
\begin{equation}
F_{\text{poly}}(\lambda) = \sum_{k=0}^{n} a_k \lambda^k.
\label{eq:polynomial_fitting}
\end{equation}
Here, $n$ is the order of the polynomial, $\lambda$ is the wavelength, and $a_k$ are the polynomial coefficients that determine the contribution of each power of wavelength to the overall continuum shape.

\par
\noindent (2) A multi-temperature blackbody approach, where the continuum is modeled as a sum of blackbody curves, as described by Equation~\ref{eq:blackbody_fitting}:
\begin{equation}
F_{\text{BB}}(\lambda, T) = \sum_{i=1}^{m} f_i \frac{2hc^2}{\lambda^5} \left( \frac{1}{\exp\left(\frac{hc}{\lambda k_{\text{B}} T_i}\right) - 1} \right).
\label{eq:blackbody_fitting}
\end{equation}
Here, $T_i$ is the temperature of the $i^{\text{th}}$ blackbody, $\lambda$ is the wavelength, $k_{\text{B}}$ is Boltzmann’s constant, $c$ is the speed of light, $h$ is Planck’s constant, and $f_i$ is a scaling factor.

To achieve continuum subtraction, the \texttt{INDRA} tool enables users to interactively select guiding points for polynomial continuum fitting. The tool then identifies the continuum based on these user-defined points and calculates the continuum SED. These guiding points serve as anchors for optical depth calculations. Once selected, the tool fits the continuum using a user-defined polynomial of finite order. It then calculates the optical depth using Equation \ref{eq:optical_depth}, thereby consolidating the absorption features in optical depth scale.
\begin{equation}
\tau(\lambda) = -\ln\left( \frac{F_{\text{obs}}(\lambda)}{F_{\text{cont}}(\lambda)} \right)
\label{eq:optical_depth}
\end{equation}
\subsection{Silicate feature removal}
To isolate ice absorption features in the mid-infrared spectra, we remove the silicate contributions by modeling their synthetic optical depth, followed by baseline correction methods. This is necessary to address the overlapping silicate absorption bands around 9.8 $\mu$m and 18 $\mu$m with key ice features. Previous studies of protostars employed a silicate profile observed towards the galactic centre source (GCS 3) as a template  (\citealt{Boogert1997}, \citealt{Bottinelli2007}). However silicate profiles can come in different shapes and sizes, and using GCS 3 as a template silicate may not completely remove the silicate features. Instead, it can introduce spurious features as demonstrated in figures A.1 and A.2 of \cite{2024A&A...683A.124R}. To overcome this limitation, \texttt{INDRA} tool offers two approaches to removing silicate features: one utilizing a predefined silicate model and another  modeling the synthetic silicate spectra by exploring the silicate composition parameter space.  Both approaches are discussed below.
\subsubsection{Template silicate subtraction}
In this approach, we use a predefined silicate absorption model derived from observational or laboratory data, or educated guess. The known silicate components are scaled to match the observed spectrum, ensuring that their contribution is accurately subtracted. This approach is particularly useful when the silicate composition is well-constrained, allowing for a straightforward removal of the silicate absorption without requiring extensive modeling. However, it requires a precise knowledge of the surface densities and compositions of different silicate components, and their mass fractions which is not always possible. 
\subsubsection{Synthetic silicate model}
In the second approach, we explore the parameter space consisting of different silicate components with varying surface densities and mass fractions. We utilize the optool code \citep{2021ascl.soft04010D} to generate the optical depth spectra of the silicates. The silicate parameters that can be modelled using INDRA are shown in the Table \ref{tab:grain_params}.
\begin{table}[ht]
\centering
\caption{Silicate grain parameters}
\begin{tabular}{ll}
\hline
\hline
\textbf{Parameter} &  \textbf{Typical Values} \\
\hline
Core composition    & Pyroxene, Carbon etc. \\
Core mass fraction  & Pyroxene (0.82), Carbon (0.18) \\
Surface density     & $\sim$10$^{5}$ g\,cm$^{-2}$ \\
Mantle composition  & H$_{2}$O \\
Mantle fraction     &  0--0.5 \\
Grain geometry      & DHS, CDE \\
Porosity            & 0.8 \\
\hline
\end{tabular}
\label{tab:grain_params}
\vspace{2mm}
\parbox{0.95\linewidth}{
\small
\textbf{Notes.} DHS: Distribution of Hollow Spheres; CDE: Continuous Distribution of Ellipsoids. These grain geometries are used to model irregular silicate grain shapes.
}
\end{table}

To determine the synthetic silicate model, we simultaneously fit both the 9.8 $\mu$m and 18 $\mu$m bands. Once the synthetic silicate optical depths of different components are computed, a baseline correction is applied to each spectrum to remove any continuum variations or underlying broad features that could interfere with the fitting process. Following the baseline correction, the spectra are recombined using scaling factors applied to each silicate species to best match the observed data, and in particular to match the 9.8 $\mu$m and 18 $\mu$m absorption  features. In this process, we optimize the scaling factors, the mass fractions of silicates and their surface densities using a non-linear least squares fitting procedure as shown in the Figure \ref{fig:flow chart} and thereby determine the best fit for the silicate absorption profiles. 

We note that the placement of the continuum is closely tied to the silicate absorption, particularly at wavelengths longer than 7.4~$\mu$m. In this region, the continuum shape is primarily dictated by the depth and extent of the silicate profile - higher continuum placement corresponds to stronger inferred silicate absorption. Consequently, using appropriate silicate models allows for the effective removal of their contribution, helping to isolate the underlying ice absorption features. At shorter wavelengths ($\lesssim$7.4~$\mu$m), the observed spectrum contains relatively featureless regions (e.g., 4.8--5.6~$\mu$m), which provide more reliable anchors for continuum placement. Nevertheless, the continuum shape over the silicate region should still be considered to ensure consistency across the entire spectrum. We note that significant uncertainties remain in both the continuum placement and the silicate modeling. A more rigorous treatment of these uncertainties would require detailed radiative transfer modeling, which is beyond the scope of the present comparative study. As demonstrated in previous works (e.g., \citealt{2024A&A...683A.124R,2025A&A...693A.288R}), such detailed modeling is not essential for ice studies like this. In practice, a variety of approaches have been used to estimate and remove the silicate absorption. In some cases, a polynomial continuum is first fitted to the spectrum, followed by the application of a simple silicate absorption model \citep{2024A&A...683A.124R} or a multi-spline function \citep{2025A&A...693A.288R} to isolate the ice features. More sophisticated modeling is needed to simultaneously account for both the continuum and the silicate features in a physically consistent manner. However, our primary objective in this work is not to model the continuum and silicate features in detail, but rather to assess the robustness of ice column density estimates using a global continuum approach and to expand the chemical inventory of IRAS 2A leveraging the full MIRI spectra of the source.

\subsection{Laboratory ice data}\label{sec:labdata}
The laboratory ice data used in \texttt{INDRA} consists of 772 ice spectra, provided in the format of wavenumber versus absorbance. The data are compiled from  publicly available ice databases such as the Leiden Ice Database\footnote{\href{https://icedb.strw.leidenuniv.nl}{https://icedb.strw.leidenuniv.nl}}, the NASA Cosmic Ice Laboratory \footnote{\href{https://science.gsfc.nasa.gov/691/cosmicice/spectra.html} {https://science.gsfc.nasa.gov/691/cosmicice/spectra.html}}, UNIVAP \footnote{\href{https://www1.univap.br/gaa/nkabs-database/data.htm}{https://www1.univap.br/gaa/nkabs-database/data.htm}} and the Optical Constant Database \footnote{\href{https://ocdb.smce.nasa.gov/}{https://ocdb.smce.nasa.gov/}}  that have comprehensive spectroscopic measurements of ices relevant to astrophysical environments. Some key ices include CO, H$_{2}$O, CH$_{4}$, NH$_{3}$, and organic molecules which have been observed in various astrophysical environments including  low mass star forming regions. The spectra are measured in the laboratory under various physical conditions  representing those of star-forming regions, providing us a wide range of data useful for ice fitting of YSOs. Before using the laboratory data in \texttt{INDRA}, we conducted a survey of the ices which are potentially contributing to the observed optical depths. This is because the spectra of same ice components are recorded at different temperatures. Though the band profiles are temperature-dependent, yet most absorption features remain largely unaffected posing a risk of overfitting if all the components are included. To mitigate this, we selected 76 components for the final fitting, the details of which are found in  Appendix \ref{appendix:labdata}. These spectra are essential for accurately fitting the observed ice absorption features. To ensure proper fitting, the spectra are preprocessed, which involves applying baseline corrections to select ice components to enhance the clarity of absorption features or to isolate minor component features that might otherwise be obscured by stronger ones. For this work, we correct the baselines of specific ice components, particularly those involving organic molecules in water mixtures where H$_2$O features dominate (see Appendix~\ref{appendix:baselines}). Once compiled, the data can be used in \texttt{INDRA} fitting, where we further normalize the laboratory spectra to facilitate comparison with observational data.

\subsection{Ice fitting using weighted NNLS}
This study employs a weighted Non-Negative Least Squares (weighted NNLS) algorithm 
to fit laboratory ice absorption spectra to the observed spectra of the YSO. The weighted-NNLS method ensures that all fitted coefficients remain non-negative, which is physically meaningful for the absorption spectra and ensures that the fitted components remain above the sigma noise estimate. The laboratory ice absorption spectra consist of several ices each with a distinct absorption profile. The observed spectrum can be modeled as a linear combination of these profiles scaled by appropriate non-negative coefficients. The weighted-NNLS method is applied to determine these scaling factors and thereby determining the contribution of each such ice to the absorption spectra. The following is a description of the algorithm.\\

Let the observed spectrum be represented as a vector $\mathbf{y} \in \mathbb{R}^n$ with each element $y_i$ corresponding to the observed flux at wavelength $\lambda_i$. Similarly let $\mathbf{A} \in \mathbb{R}^{n \times m}$ represent the matrix of laboratory ice absorption spectra where each column $\mathbf{A}_j$ is the absorption profile of a specific ice component. The objective of the fitting process is to find the non-negative coefficient vector $\mathbf{x} \in \mathbb{R}^m$ where each element $x_j$ represents the scaling factor of the corresponding ice component such that the residual between the observed and modeled spectrum is minimum.

The weighted-NNLS problem can be formulated as:

\begin{equation}
\min_{\mathbf{x}} \sum_{i=1}^{n} w_i^2 \left( y_i - \sum_{j=1}^{m} A_{ij} x_j \right)^2
\label{eq:nnls_criteria}
\end{equation}

Here, $w_i$ is the weight given by $\frac{1}{\sigma_i}$, where $\sigma_i$ is the observational uncertainty at wavelength $\lambda_i$. The weighting ensures that data points with larger uncertainties have a smaller contribution to the total residual.

The NNLS algorithm solves for the minimization criteria given by Equation \ref{eq:nnls_criteria} subject to the constraint that $x_j \geq 0$ for all $j$. This ensures that the fit does not include any unphysical negative contributions from any of the ice components.

Convergence is reached when the residual cannot be reduced further while maintaining the non-negativity constraint on the coefficients. By using the weighted-NNLS method, we ensure that the laboratory ice absorption profiles are optimally scaled to match the observed spectrum, providing robust estimates of the ice column densities in the observed source.

\subsection{Estimation of statistical significance}
Accurate noise estimation is crucial in spectroscopic analysis, particularly when identifying weak absorption features in the optical depth spectrum and ascertaining the accuracy of the spectral fitting. A common approach, which has also been used by \cite{2024A&A...683A.124R}, is to assume a fixed 10\% noise level across the spectrum. While this assumption may be reasonable for high-S/N spectra with uniform noise at all wavelengths, it does not account for variations in baseline fluctuations, instrumental response or wavelength-dependent noise. Consequently, such a fixed uncertainty can lead to either an over-estimation or under-estimation of the actual noise, potentially affecting the reliability of spectral fits. To obtain a more accurate estimate of the noise, we employ a polynomial-based approach similar to that of \cite{2024A&A...690A.205C}. Instead of assuming a fixed percentage, the noise level is estimated at multiple spectral regions devoid of strong absorption features, and an average value is taken across these regions. Appropriate polynomials are used to model and subtract the baseline fluctuations, ensuring that the noise estimation reflects the true uncertainties in the data. The noise level at each region is calculated using:  

\begin{equation} \label{eq:noiselevel}
\sigma = \sqrt{\frac{\sum_{i=1}^{N} (y_i - \bar{y})^2}{N}}
\end{equation}  

where \( y_i \) represents the polynomial-subtracted optical depth at each spectral channel within the selected region, and \( \bar{y} \) is the mean value of \( y_i \). This method provides a robust uncertainty estimate, crucial for assessing the significance of weak absorption features and the reliability of the fits. By averaging over multiple regions, it captures the global noise characteristics across the wavelength range rather than relying on a single estimate. The computed noise level is then incorporated into our statistical analysis to ensure a reliable interpretation of the spectral fits.  

\subsubsection{Greedy backward elimination as a check against overfitting}
As a check against overfitting, we use the greedy backward elimination approach based on a chi-square statistical evaluation using the Equation \ref{eq:chi}:
\begin{equation} \label{eq:chi}
\chi^2 = \sum_i \frac{(O_i - E_i)^2}{\sigma_i^2},
\end{equation}
with $O_i$ representing the observed data points, $E_i$ the model values for those data points, $\sigma_i$ the estimated uncertainty given by Equation \ref{eq:noiselevel}. This method iteratively removes each component from the model while preserving the rest and evaluates the \(\chi^2\) value to assess the goodness of fit. At each step, we systematically remove one component at a time  and fit the rest of the components using weighted-NNLS. The component whose removal results in the smallest increase in \(\chi^2\) is identified as the least significant and is tentatively eliminated. To ensure whether a specific component meaningfully contributes to the fit, we compute the p-value. This test is based on the null hypothesis that the component in question does not contribute significantly to the model, meaning any change in the fit quality upon removing the component is purely due to random fluctuations rather than a true effect. To test this, we compare the fit quality before and after the component is removed, using the difference in \(\chi^2\) values (\(\Delta \chi^2 = \chi^2_{\text{before}} - \chi^2_{\text{after}}\)). Under the null hypothesis, the test statistic follows a chi-square distribution with degree of freedom, $\nu$, equal to the number of omitted components. The p-value is then computed from the cumulative distribution function (CDF) of the \(\chi^2\) distribution:

\begin{equation}
p = 1 - F_{\chi^2}(\Delta \chi^2, \nu)
\end{equation}

where the function \(F_{\chi^2}(\Delta \chi^2, \nu)\) represents the cumulative probability of observing a chi-square value up to \(\Delta \chi^2\) under the null hypothesis.  A high p-value (\(p > 0.05\)) suggests that the difference in chi-square is small and likely due to random fluctuations rather than a meaningful improvement in fit. In this case, the removed component is deemed statistically insignificant, justifying its elimination. Conversely, if the p-value falls below a predefined threshold (\(p \leq 0.05\)), the change in chi-square is large enough to indicate that the removed component significantly contributes to the model, and it should be retained.  By evaluating the statistical significance of component removal using p-values, we ensure that only the most relevant components are retained in the fit, thereby preventing overfitting while maintaining an optimal representation of the observed data.

\subsubsection{R$^{2}$ statistic to assess the goodness-of-fit}
The goodness-of-fit is assessed using the coefficient of determination (\( R^2 \)), which quantifies how well the model explains the variance in the observed data as each component is being added. It is given by:

\begin{equation}
R^2 = 1 - \frac{\sum_{i=1}^{n} (y_i - \hat{y}_i)^2}{\sum_{i=1}^{n} (y_i - \bar{y})^2}
\end{equation}

where \( y_i \) are the observed values, \( \hat{y}_i \) are the model-predicted values, and \( \bar{y} \) is the mean of the observed data. An \( R^2 \) value close to 1 indicates that the model explains most of the variance in the data, while a value near 0 suggests poor explanatory power. While calculating the R$^{2}$ statistic as each component is added, we consider only its contributing region, defined as the wavelength range where the component's contribution exceeds the sigma level calculated for the source.
\subsubsection{F statistic to evaluate the relevance of minor components}
We note that while GBE helps identify components that are important for the overall model, it does not provide a quantitative measure of statistical significance, especially for minor components. To address this, we performed an F-test, which evaluates whether adding a specific component significantly improves the model fit or not and to what extent. The F-value for each component is computed using Equation \ref{eq:ftest}.

\begin{equation} \label{eq:ftest}
F = \frac{\left( \frac{\text{RSS}_{\text{test}} - \text{RSS}_{\text{full}}}{\text{df}_{\text{test}} - \text{df}_{\text{full}}} \right)}{\left( \frac{\text{RSS}_{\text{full}}}{N - \text{df}_{\text{full}}} \right)}
\end{equation}

where $\text{RSS}_{\text{full}}$ is the residual sum of squares (RSS) for the full model (including the tested component), $\text{RSS}_{\text{test}}$ is the RSS for the reduced model (excluding the tested component), $\text{df}_{\text{full}}$ and $\text{df}_{\text{test}}$ are the degrees of freedom of the full and reduced models, respectively, and $N$ is the total number of data points. The F-values are computed as the ratio of the variance explained by each component to the residual variance and indicate the relative contribution of each component to the overall fit. In this sense, it offers an additional check along with GBE which uses \(\chi^2\) statistic. Note that we use only the relevant wavelength window of each component specific to its contribution to the fitting while evaluating F-value.  A higher value indicates that the tested component significantly improves the fit by reducing the residual variance in that wavelength window. Empirical thresholds for spectral fitting applications suggest the following classification of F-values:
\begin{itemize}
    \item Strong contribution: $F > 10$
    \item Moderate contribution: $2 < F < 10$
    \item Weak contribution: $F < 2$
\end{itemize}

We note that these ranges are not strictly defined, and their interpretation depends on the sample size, degrees of freedom, etc.\\

The statistical significance of each component is further evaluated using the corresponding p-value, which represents the probability of obtaining an F-statistic as extreme as observed under the null hypothesis. This time the p-value is computed as:

\begin{equation}
p = P(F > F_{\text{obs}})
\end{equation}

where $F_{\text{obs}}$ is the observed F-statistic. A lower p-value (typically $p < 0.05$) suggests that the inclusion of the component significantly improves the model.

\subsubsection{Estimation of confidence intervals}
To quantify uncertainties in the best-fit coefficients for the ice absorption modeling and to estimate confidence intervals in the column densities, we generate perturbed coefficients by sampling values linearly spaced around each best-fit coefficient. Specifically, for each coefficient \( C_i \), perturbed values were sampled uniformly between \( C_i - \epsilon C_i \) and \( C_i + \epsilon C_i \), where \( \epsilon \) is a fractional error factor defining the perturbation range.

For each perturbed coefficient set, a synthetic spectrum \( y_{\mathrm{model}} \) was generated, and its fit to the observed optical depth spectrum \( y_{\mathrm{obs}} \) is evaluated using the normalized chi-square statistic given by:

\begin{equation}
\chi^2_{\mathrm{norm}} = \frac{\sum \left( \frac{y_{\mathrm{obs}} - y_{\mathrm{model}}}{\sigma_y} \right)^2}{\sum y_{\mathrm{model}}^2},
\label{eq:chi_squared_norm}
\end{equation}

where \( \sigma_y \) is the observational uncertainty.

Confidence intervals are then determined for each coefficient by identifying the range of \( C_i \) values corresponding to an acceptable increase in \( \chi^2_{\mathrm{norm}} \). The lower and upper bounds in the coefficients are then used to calculate the minimum and maximum optical depths, setting limits on the column densities for each ice, computed as described in Section~\ref{sec:column_density_estimation}.

The correlations between different ice components vary with wavelength due to the presence of complex mixtures, which can induce shifts in absorption band positions and alter band shapes. As a result, uncertainties and parameter correlations are not uniform across the spectral range, and a global analysis may not effectively capture these localized effects. Therefore, we perform the analysis in a sliding-window, local manner across the spectrum to account for wavelength-dependent uncertainties. This localized approach enables robust estimation of coefficient significance, their correlations, and the associated error bars in the derived column densities.

For each coefficient, the overall lower bound is taken as the minimum of all lower bounds obtained from individual sliding windows, while the overall upper bound is taken as the maximum of all upper bounds from these windows. This method ensures a comprehensive estimation of the confidence intervals.

\subsection{Column density estimation}\label{sec:column_density_estimation}
Once the best fit solution is found, we calculate the column density of each ice using Equation \ref{eq:colden}, 

\begin{equation} \label{eq:colden}
N_{\text{ice}} = \frac{1}{A} \int_{\nu_1}^{\nu_2} \tau_{\text{lab}}(\nu) d\nu
\end{equation}

where \( A \) represents the vibrational mode band strength of the molecule. The band strengths vary with the chemical environment, making their accurate determination crucial for reliable column density estimates. While calculating the column densities, we incorporate corrected band strength values from the literature. The appendix Table \ref{appendix:bandstrengths} provides the list of all the band strengths used in this work. It is to be noted that derivation of band strengths is inherently dependent on ice density, leading to typical uncertainties of approximately 15\% for pure ices and 30\% for mixed ices (\citealt{Rachid2022}; \citealt{slav2023}).


\section{Results}
\label{sec:results}
In this section we present the fitting results for IRAS 2A across the entire MIRI wavelength range (5–28 $\mu$m). This includes the continuum removal, silicate subtraction, global ice fitting, statistical analysis and the final estimation of ice column densities. Additionally, we compare our results with previous study that employed a local fitting approach. 
\subsection{Continnum determination in IRAS 2A} \label{continuum-determination-IRA2A}
\begin{figure}[h]
    \includegraphics[width=0.5\textwidth]{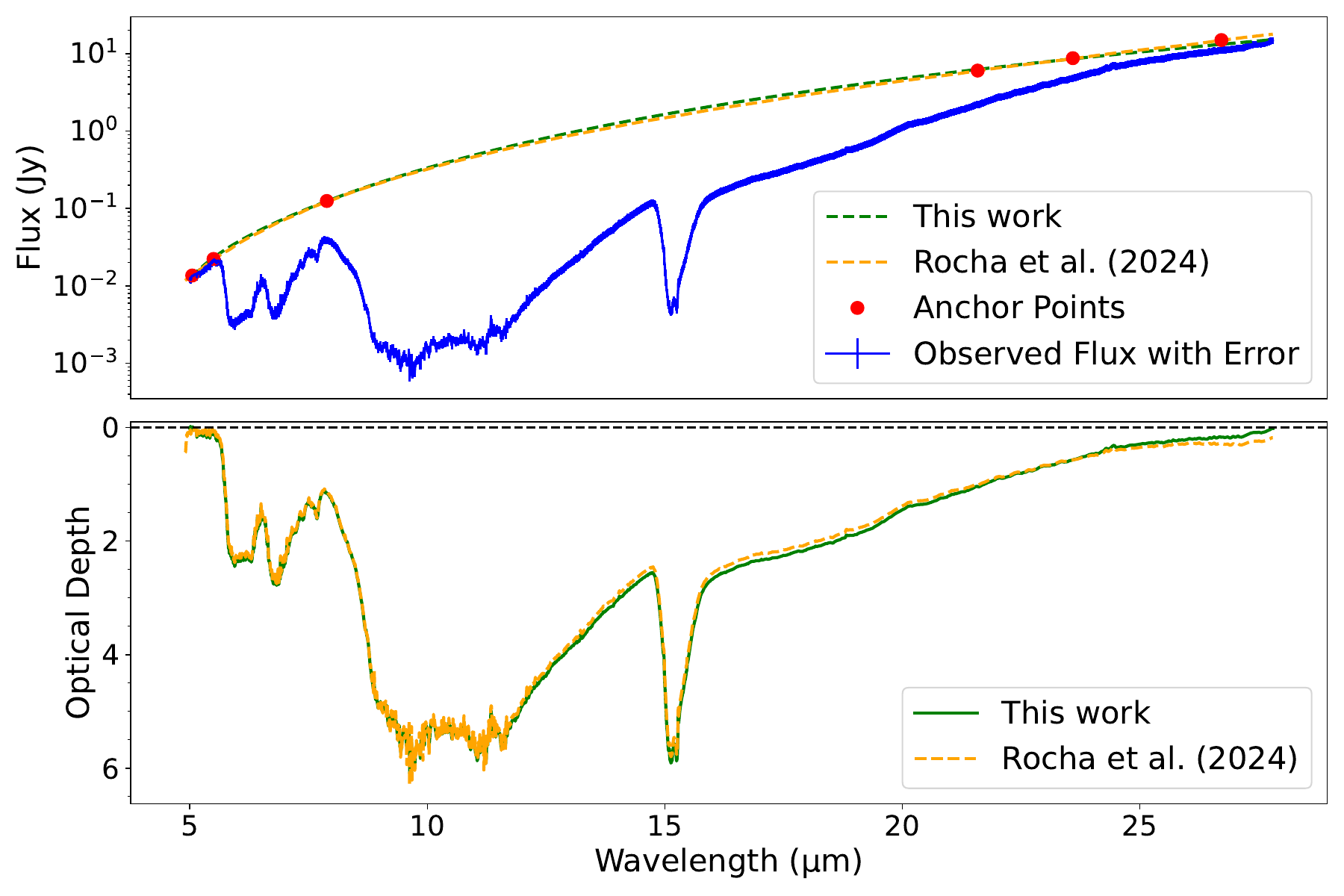}
    \caption{Continuum determination in the source IRAS 2A. The top panel displays the observed flux (blue) on a logarithmic scale, along with the continuum (green) traced using a guided polynomial fit. The anchor points used for continuum fitting are marked as red dots. The bottom panel presents the corresponding estimated optical depths (green). For comparison, data from \citealt{2024A&A...683A.124R} is also plotted (orange).}
    \label{fig:continnum-iras2a}
\end{figure}

The continuum fit and the corresponding optical depths of the source IRAS 2A are presented in  Figure \ref{fig:continnum-iras2a}. The continuum and the optical depths calculated by \cite{2024A&A...683A.124R} are also presented for a comparison. 
The SED of IRAS 2A is characteristic of embedded protostars, both high-mass and low-mass, exhibiting an increasing slope toward longer wavelengths (\citealt{Gibb2004}, \citealt{Boogert2008}). The observed SED can be modeled using a blackbody approach, assuming contributions from warm dust at wavelengths shorter than 20 $\mu$m and cold envelope material at longer wavelengths. However, in this work, a third-order guided polynomial is used to determine the continuum.  
This choice is motivated by several factors: (1) the lack of reliable constraints on the temperature components required for blackbody modeling, (2) the flexibility of the polynomial approach in tracing complex absorption features across MIRI, and (3) to maintain consistency with the continuum shape adopted in \citet{2024A&A...683A.124R}, enabling a direct comparison of the resulting ice column densities and evaluation of how different fitting methods influence the derived ice column densities. We use anchor points to trace the continuum and the same are shown in Figure~\ref{fig:continnum-iras2a}. Please note that some points were placed slightly above the observed data to better approximate the continuum used in the earlier study.  However, we emphasize that significant uncertainties are inherently associated with the choice of the continuum, as silicate absorption can extend to longer wavelengths making it challenging to trace it accurately. A detailed treatment of this continuum ambiguity is beyond the scope of the present work. Once the continuum is traced, we convert the spectra into optical depth units using Equation \ref{eq:optical_depth}.  
\begin{figure}[h]
    \includegraphics[width=0.5\textwidth]{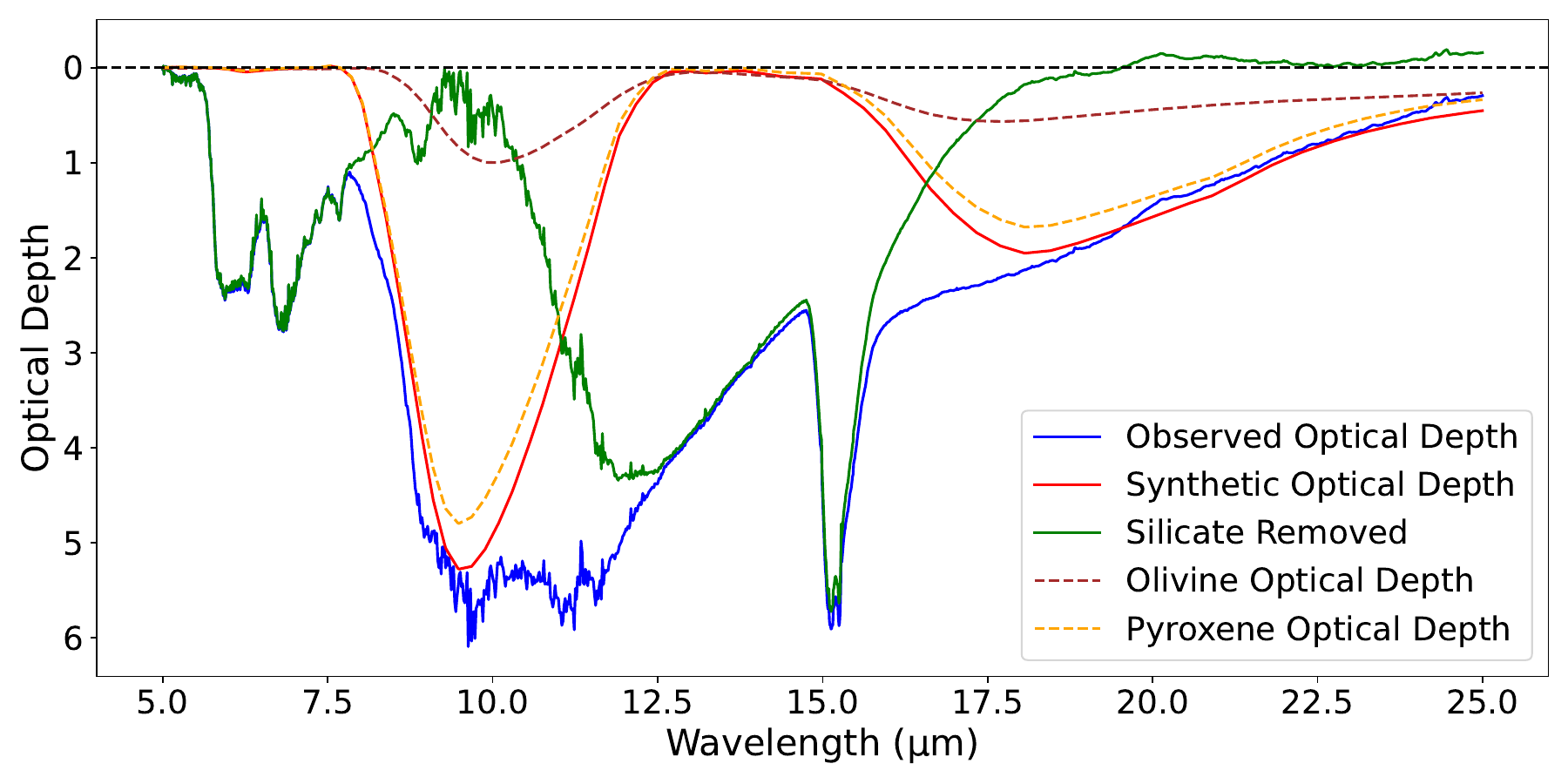}
    \caption{Silicate removal in the source IRAS 2A. The observed optical depth spectrum is shown in blue, while the modeled silicate absorption is represented by the solid red line. The individual silicate components contributing to the overall absorption are olivine (dotted brown) and pyroxene (dotted orange). The final silicate-subtracted spectrum is shown in green.}
    \label{fig:silicate_removed}
\end{figure}
\FloatBarrier  
\begin{figure*}[t]
    \includegraphics[width=\textwidth]{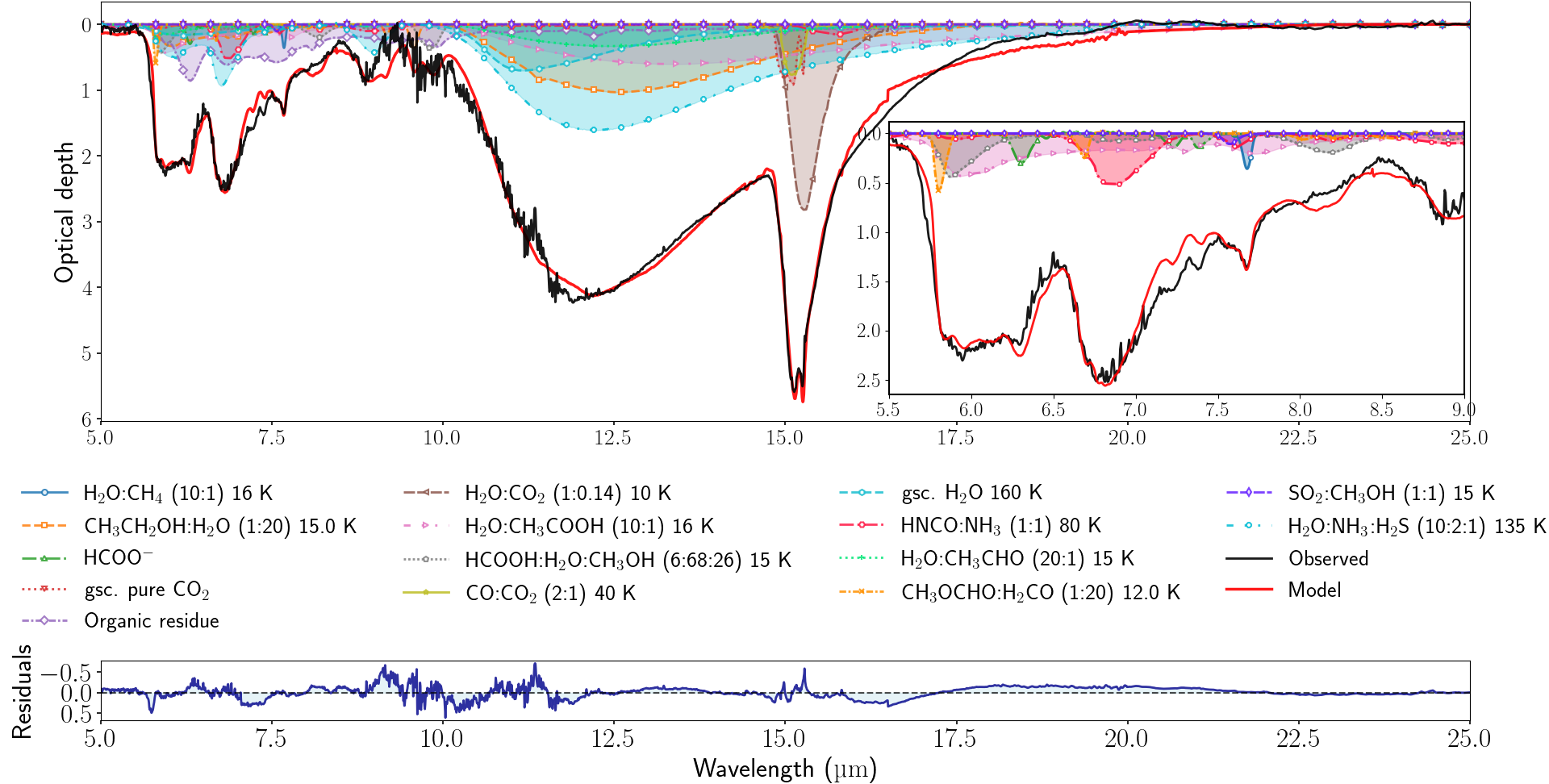}
    \caption{Global ice-fitting results across the entire MIRI spectrum (5–25 $\mu$m). The top panel shows the observed optical depth spectrum (solid black) alongside the modeled optical depth spectrum (solid red), which represents the contribution from 15 best-fit ice components. Each contributing component is displayed using different colors, markers, and line styles and are mentioned in the middle panel as a legend. The zoomed-in version of the contribution from select components (weak components) are shown in the inset plot. The bottom panel presents the residuals corresponding to the fit. \textit{Note} - Organic residue refers to the residue spectra obtained by irradiating H$_{2}$O:NH$_{3}$:CH$_{3}$OH:CO:CO$_{2}$ ice that shows broad features correspond to various functional groups, including carbonyl (-C=O), carboxylate (-COO$^{-}$), amine (-NH$_{2}$), and -CN containing species along with hexamethylenetetramine (HMT); polyoxymethylene (POM).  Abbreviations: gsc. = grain shape corrected.}
    \label{fig:globalfit-iras2a}
\end{figure*}
\subsection{Silicate feature removal in IRAS 2A} \label{silicate-removal-IRAS2A}
The spectrum of IRAS 2A in optical depth units is shown in the bottom panel of  Figure \ref{fig:continnum-iras2a}. Broad silicate absorption features appear near 9.8 $\mu$m, where they partially overlap with organic ice bands, and near 18 $\mu$m, where they overlap with H$_{2}$O and CO$_{2}$ absorption features. In particular, organic ice absorption features embedded within the 9.8 $\mu$m region become more distinct once the silicate features and local continuum are removed \citep{2024A&A...683A.124R}. Additionally, the true shape of the CO$_{2}$ band at $\sim$15 $\mu$m emerges more clearly after accounting for the overlapping silicate feature near 18 $\mu$m. The silicate removed spectra of IRAS 2A is shown in  Figure \ref{fig:silicate_removed}.  For this work, we adopt a template silicate subtraction method, leveraging the well-constrained silicate model from \cite{2024A&A...683A.124R} to ensure consistency and accuracy in isolating the ice absorption features. 
Briefly, the template silicate model consists of two components: amorphous pyroxene (Mg$_{0.5}$Fe$_{0.5}$SiO$_{3}$) and olivine (MgFeSiO$_{4}$). Each of these component is assumed to be mixed with carbon, a common constituent of interstellar grains. The volume fractions of silicate and carbon are set to 82\% and 18\% respectively. This approach is in line with various protostellar and diffuse interstellar medium models, where carbon fractions can vary between 15 and 30\% (\citealt{Woitke2016, Pontoppidan2005, Weingartner2001}). The grain size distribution is modeled using a power-law size distribution with an exponent of 3.5, and grain sizes ranging from 0.1 to 1 $\mu$m, consistent with interstellar grain models. To account for the irregular geometry of the dust grains, we adopt a distribution of hollow spheres, a method that provides more realistic fits to the silicate absorption features than simple spherical grain models.
\subsection{\texttt{INDRA} global fitting results}
The MIRI spectrum of IRAS 2A shows a high S/N and distinct absorption features spanning 5–28 $\mu$m. At shorter wavelengths $\lesssim$ (10 $\mu$m), the spectrum is primarily shaped by absorption from H$_{2}$O, organic molecules and organic refractory materials, and silicates. At longer wavelengths ($\gtrsim$ 10 $\mu$m), strong absorptions from H$_{2}$O, CO$_{2}$, and silicates dominate the spectrum. By subtracting the silicate contributions, the remaining absorption features between 5 and 20 $\mu$m can be primarily attributed to ices. The fitting results for the entire MIRI wavelength range of the source IRAS 2A is shown in  Figure \ref{fig:globalfit-iras2a}. Our spectral fitting analysis reveals the presence of multiple ice components including simple species such as H$_{2}$O, CO$_{2}$, CH$_{4}$ as well as COMs like CH$_{3}$OH, CH$_{3}$CH$_{2}$OH, etc. \cite{2024A&A...683A.124R} carried out local fitting in the COMs finger-print region (6.8 - 8.6 $\mu$m) of the source and estimated the column densities of COMs present in that region along with H$_{2}$O. In this work, we analyze the entire MIRI spectrum and decompose it using a laboratory ice database containing 76 ices, out of which 15 contribute to the global minimum solution. Each of these 15 components shows a peak optical depth that exceeds the noise level in the observed spectrum, indicating appreciable contributions. Their statistical significance are evaluated in Section \ref{sec:stats}.
\subsubsection{Absorption bands of simple ices H$_{2}$O, CO$_{2}$, NH$_{3}$, SO$_{2}$}
\begin{figure*}[t]    \includegraphics[width=\textwidth]{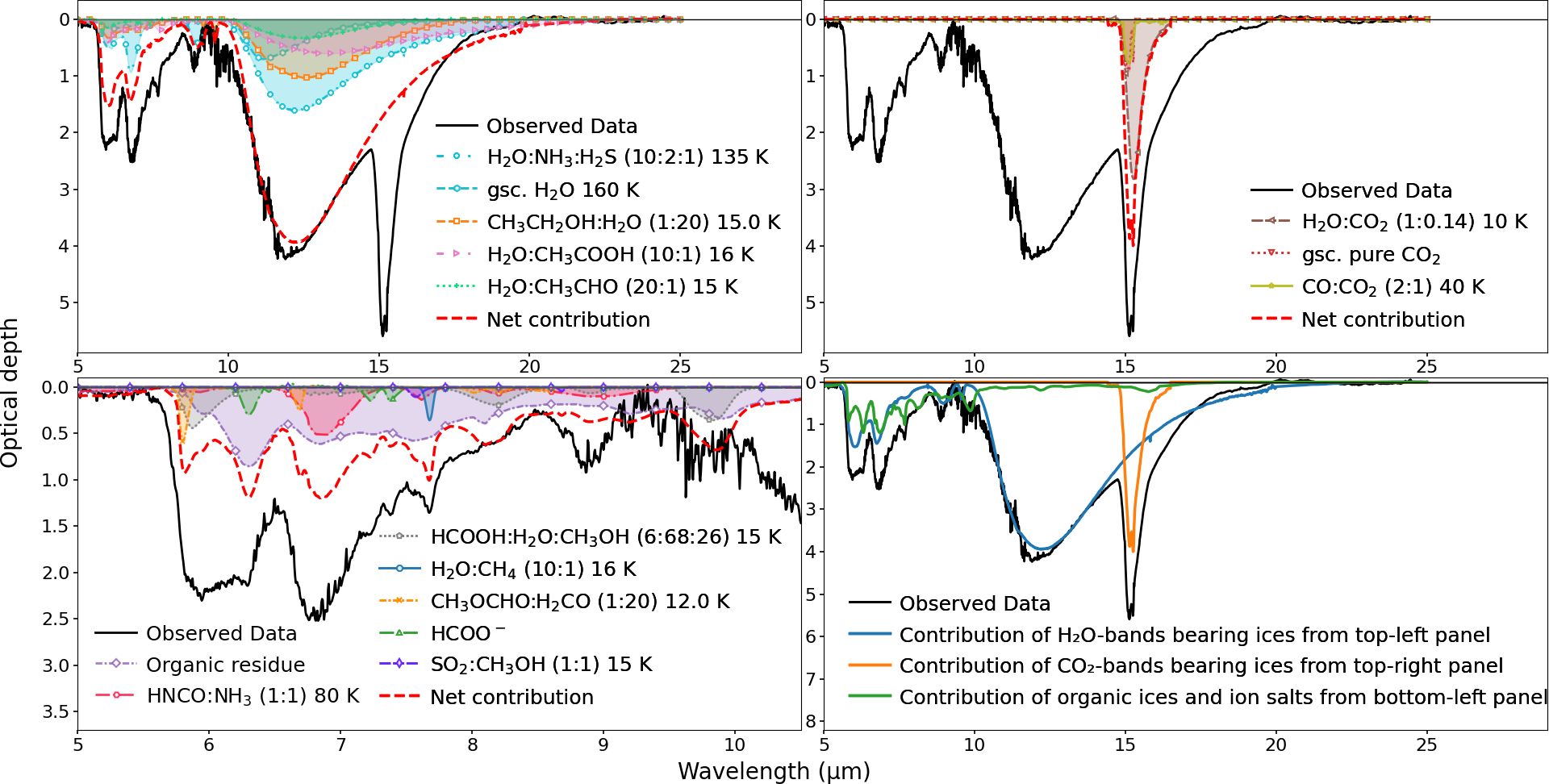}
    \caption{Best-fit decomposition of the MIRI spectrum of IRAS 2A, highlighting the contributions from different ice components. The observed optical depth spectrum is shown in solid black across all panels. Each contributing component is displayed using different colors, markers, and line styles. \textit{Top-left panel:} Contributions from ice components bearing dominant H$_{2}$O bands. These components have absorption features spanning from 5-20 $\mu$m. Their cumulative contribution to the observed spectrum is shown in dashed red. \textit{Top-right panel:} Contributions from ices bearing CO$_{2}$ bands. These components have absorption features spanning from 14-17 $\mu$m. Their cumulative contribution to the observed spectrum is shown in dashed red. \textit{Bottom-left panel:} Contributions from ice components that include organic ice mixtures, ions and organic residue. These components have significant absorption features spanning from 5-11 $\mu$m. Their combined contribution to the overall spectrum is shown in dashed red. \textit{Bottom-right panel:} A summary panel showing the individual contributions (shown in solid colors) to the observed spectrum from the three categories: H$_{2}$O bands bearing ices, CO$_{2}$ band bearing ices, and organic ice mixtures along with ions and organic residue.}
    \label{fig:components}
\end{figure*}

The MIRI spectra of IRAS 2A show broad absorption features of H$_{2}$O due to the bending mode at around 6 $\mu$m and the libration mode at around 13 $\mu$m. The 6 $\mu$m region is particularly complex due to significant overlap with absorption features from COMs and other simple molecules.  Similarly, the 13 $\mu$m region coincides with strong absorption from silicates and CO$_{2}$, resulting in blended features that require a combination of pure and mixed ice components for spectral decomposition. The top right panel in  Figure \ref{fig:components} shows the H$_{2}$O and CO$_{2}$ related best-fit components. All these components are dominant in nature and contribute much to the shape of the overall spectra. H$_{2}$O ice appears in both pure and mixed forms, contributing to the bending mode at $\sim$6 $\mu$m and the libration mode at $\sim$13 $\mu$m. We apply grain shape correction to the 160 K pure H$_{2}$O ice using Mie theory and assuming small grains where it shows a blue wing excess in the 13 $\mu$m libration band.
Similarly, the MIRI spectrum of IRAS 2A has a distinct absorption at 15 $\mu$m due to CO$_{2}$ . CO$_{2}$ ice appears both in its pure form and as part of mixed ice components. The following components are used to fit the spectrum: pure CO$_{2}$ ice, 40 K ice mixture CO: CO$_{2}$ in 2:1 ratio, 10 K ice mixture H$_{2}$O: CO$_{2}$ in 1:0.14 ratio. Similar ice components were used by \cite{Pontoppidan2008} and \cite{2025A&A...693A.288R}. The observed double peak in the 15 $\mu$m band is likely a result of ice distillation or segregation due to thermal processing and is well fit by pure CO$_{2}$ ice. This component shows a prominent blue wing, contributing to the shortward side of 15 $\mu$m band, whereas the  10 K H$_{2}$O:CO$_{2}$ ice mixture contributes to the red wing of this band. Both \cite{Pontoppidan2008} and \cite{2025A&A...693A.288R} found CO$_{2}$ ice in H$_{2}$O ice-rich environments, which is also the case in the present work. Additionally, contributions from CO ice dominated regions play a role.
\par

Sulfur dioxide (SO$_{2}$) shows strong absorption features in the mid-infrared (MIR) region, particularly near 7.7 and 8.5\,$\mu$m, which correspond to its asymmetric ($\nu_3$) and symmetric ($\nu_1$) stretching modes, respectively. We isolated both these bands of SO$_{2}$ from a mixture SO$_{2}$:CH$_{3}$OH as shown in the Figure \ref{fig:SO2}. Ammonia (NH$_{3}$) is also found in the fits. The absorption band at 8.9 $\mu$m in the observed optical depth spectrum can be related to this ice. Two ice components as shown in Table \ref{tab:simple_ices} carry this band in the fits with the H$_{2}$O:H$_{2}$S based mixture showing significant contribution to the 8.9 $\mu$m band. Notably, this region also overlaps with some minor bands from CH$_{3}$CH$_{2}$OH, CH$_{3}$OH, CH$_{3}$CHO, and SO$_{2}$, which can lead to blending and complex shape to the optical depth spectrum. The deviation between the model and the observations in this region may be due to uncertainties in the organic refractory baseline, the placement of the global continuum, the silicate subtraction, or a combination of these factors.
\begin{table}[h!]
\centering
\caption{Ice mixtures containing H$_{2}$O, CO$_{2}$, and SO$_{2}$}
\label{tab:simple_ices}
\resizebox{\columnwidth}{!}{%
\begin{tabular}{@{}lcc@{}}
\toprule
Ice Mixture & Ratio & T (K) \\
\midrule
H$_{2}$O components \\
H$_{2}$O:NH$_{3}$:H$_{2}$S & 10:2:1 & 135 \\
CH$_{3}$CH$_{2}$OH:H$_{2}$O & 1:20 & 15 \\
H$_{2}$O:CH$_{3}$COOH & 10:1 & 16 \\
H$_{2}$O:CH$_{3}$CHO & 20:1 & 15 \\
Pure gsc. H$_{2}$O & --- & 160 \\
\midrule
CO$_{2}$ components \\
H$_{2}$O:CO$_{2}$ & 1:0.14 & 10 \\
CO:CO$_{2}$ & 2:1 & 40 \\
Pure gsc. CO$_{2}$ & --- & --- \\
\midrule
SO$_{2}$ component \\
SO$_{2}$:CH$_{3}$OH & 1:1 & 10 \\
\midrule
NH$_{3}$ component \\
H$_{2}$O:NH$_{3}$:H$_{2}$S & 10:2:1 & 135 \\
NH$_{3}$:HNCO$^{\dagger}$ & 1:1 & 80 \\
\bottomrule
\end{tabular}%
}
\vspace{1mm}

\smallskip
\small \textit{Note:} \texttt{gsc.} denotes a grain shape–corrected component; $\dagger$ denotes the spectrum of the residual component. For references, see Appendix~\ref{appendix:labdata}. 
\end{table}
\subsubsection{Absorption bands of NH$_{4}^{+}$, OCN$^{-}$ and HCOO$^{-}$ ion salts}
Table \ref{tab:salt_mixtures} lists all the ice components that show the spectral features of ions. In our global fitting results, we found the Ammonium ion (NH$_{4}^{+}$) in three distinct ice components: within the NH$_{3}$:HNCO ice residue that is responsible for OCN$^{-}$ ion, in the H$_{2}$O:NH$_{3}$:CH$_{3}$OH:CO:CO$_{2}$ ice residue that is responsible for the organic residue shown in Figure \ref{fig:organic_residue} and the H$_{2}$O:NH$_{3}$:H$_{2}$S (10:2:1) ice mixture at 135 K. The 6.7 $\mu$m band of these ice components can be assigned to the $\nu_4$ mode of NH$_4^+$ \citep{wagner_vibrational_1950}. The broad absorption dip centered around 6.6~$\mu$m in the observed optical depth spectrum of IRAS 2A is well explained by these three components, and the model provides an excellent agreement with the data in this region.
\begin{table}[h!]
\centering
\caption{Ice mixtures containing ion salts}
\label{tab:salt_mixtures}
\resizebox{\columnwidth}{!}{%
\begin{tabular}{@{}lccc@{}}
\toprule
Ice Mixture & Ratio & T (K) & Species \\
\midrule
H$_{2}$O:NH$_{3}$:H$_{2}$S & 10:2:1 & 135 & H$_{2}$O,\newline NH$_{4}^+$, NH$_{3}$ \\
NH$_{3}$:HNCO$^{\dagger}$ & 1:1 & 80 & OCN$^{-}$,\newline NH$_{4}^+$, NH$_{3}$ \\
H$_{2}$O:NH$_{3}$:CH$_{3}$OH:CO:CO$_{2}^{\dagger}$ & --- & --- & NH$_{4}^+$ \\
H$_{2}$O:NH$_{3}$:HCOOH$^{\dagger}$ & 100:2.6:2 & 14 & HCOO$^{-}$ \\
\bottomrule
\end{tabular}%
}
\vspace{1mm}

\smallskip
\small \textit{Note:} $\dagger$ denotes spectra of the residual components. (For references, see Appendix~\ref{appendix:labdata}.)
\end{table}

Of all the components, Cynate ion (OCN$^{-}$) contributes to several bands in the 5-10 $\mu$m region and to the $\sim$ 15 $\mu$m band. The spectrum of this ion salt was obtained by \cite{Novozamsky2001} using an 80 K ice mixture of NH$_{3}$:HNCO in 1:1 ratio. The spectral evolution of the mixture when it is gradually being warmed, reveals features at approximately 6.67, 7.72, 8.26, and 15.87 $\mu$m, which are not associated with the original species. As the temperature increases, these bands grow in intensity while the characteristic bands of HNCO and NH$_{3}$ diminish, indicating proton transfer between HNCO and NH$_{3}$. The band at 6.67 $\mu$m corresponds to the vibrational mode of NH$_{4}^{+}$, whereas the features at 7.72, 8.26, and 15.87 $\mu$m are assigned to OCN$^{-}$. 
\par
Another ion identified in our fitting is Formate ion (HCOO$^{-}$), which contributes to the bands at 6.32, 7.24, and 7.40 $\mu$m. This result is consistent with the findings of \citet{2024A&A...683A.124R}. The spectrum is obtained using the hyper-quenching technique on NH$_{4}$COOH salts obatined via acid base reactions involving the 14 K ice mixture  of H$_{2}$O:NH$_{3}$:HCOOH in a ratio 100:2.6:2 \citep{Galvez2010}. To analyze the HCOO$^{-}$,  we isolated its ice bands using a local baseline subtraction method. HCOO$^{-}$ exhibits a strong band at 6.32 $\mu$m, which blends with the intense H$_{2}$O absorption at 6 $\mu$m, contributing to the excess in this region. Additionally, its 7.2 $\mu$m band overlaps with contributions from HCOOH. Therefore, the 7.4 $\mu$m band is used to estimate HCOO$^{-}$ column densities. 
\FloatBarrier
\subsubsection{Absorption bands of organic molecules}\label{subsec:organic molecules}
The Table \ref{tab:com_mixtures} summarizes the organic ice components appearing in the fitting. Components where the H$_{2}$O bands are retained as mentioned in Table \ref{tab:com_mixtures} are shown in the top left panel of the Figure \ref{fig:components} and the mixtures where H$_{2}$O bands are removed (refer to the  Appendix \ref{appendix:baselines}) are shown in the bottom left panel of the figure to aid visual inspection. All the organic molecules with secure and tentative detections as reported in \cite{2024A&A...683A.124R} are found in this work.
\begin{table}[h!]
\centering
\caption{Organic ice mixtures appearing in the fitting}
\label{tab:com_mixtures}
\resizebox{\columnwidth}{!}{%
\begin{tabular}{@{}lccc@{}}
\toprule
Ice Mixture & Ratio & T (K) & H$_{2}$O Bands \\
\midrule
H$_{2}$O:CH$_{4}$         & 10:1     & 16 & Removed \\
HCOOH:H$_{2}$O:CH$_{3}$OH & 6:68:26  & 15 & Removed \\
CH$_{3}$OCHO:H$_{2}$CO    & 1:20     & 12 & --- \\
CH$_{3}$CH$_{2}$OH:H$_{2}$O & 1:20   & 15 & Retained \\
H$_{2}$O:CH$_{3}$COOH     & 10:1     & 16 & Retained \\
H$_{2}$O:CH$_{3}$CHO      & 20:1     & 15 & Retained \\
\bottomrule
\end{tabular}%
}
\vspace{1mm}

\smallskip
\small \textit{Note:} See Section~\ref{sec:band_removal} and Appendix~\ref{appendix:baselines} for more on H$_{2}$O band removal. (References in Appendix~\ref{appendix:labdata})
\end{table}

 \par
In our analysis, the observed absorption excess at 7.7 $\mu$m is attributed to Methane (CH$_{4}$), which is present in the fits as an H$_{2}$O:CH$_{4}$ ice mixture. CH$_{4}$ has a band at $\sim$ 7.7 $\mu$m as shown in Figure \ref{fig:H2O_CH4}. Due to the interaction with H$_{2}$O, CH$_{4}$ shows band broadening, leading to a better match with observations when this mixture is considered alongside the OCN$^{-}$, SO$_{2}$ and CH$_{3}$COOH ice bands in the region. An absorption excess in the blue wing of the band is primarily due to OCN$^{-}$, and SO$_{2}$ whereas CH$_{3}$COOH contributes to the red wing of the band as shown in the inset plot in the Figure \ref{fig:globalfit-iras2a}.
\par
In our fits, Formic acid (HCOOH) appears as a HCOOH:H$_{2}$O:CH$_{3}$OH ice mixture which contributes to several absorption features across the MIRI range, including distinct bands at 5.83 $\mu$m ($\nu_S$(C=O)), 6.06 $\mu$m ($\nu_S$(C=O)), 7.21 $\mu$m ($\nu_B$(OH) \& $\nu_B$(CH)), 8.26 $\mu$m ($\nu_S$(C–O)), 9.32 $\mu$m ($\nu_B$(CH)), 10.75 $\mu$m ($\nu_B$(OH)), and 14.18 $\mu$m ($\nu_B$(OCO)). Additionally, HCOOH's interaction with H$_{2}$O results in feature at 5.88 $\mu$m. Blending effects arise due to the presence of other species in the mixture. Methanol (CH$_{3}$OH) shows absorption bands at 6.85, 8.85, and 9.75 $\mu$m, overlapping with HCOOH features and contributing to the complexity of the spectrum. The band at 9.74 $\mu$m is due to the C-O stretch and fits well within the observed spectrum. This band falls in a region where the absorption band is saturated and a factor multiplication of 2 is applied to the column density estimations which is also aligned with \cite{2024A&A...683A.124R}. The deviation between the model and the observed optical depth spectrum around 8.26 $\mu$m region may be due to the uncertainties in the refractory material baseline, the placement of the global continuum or any combination of these factors.  
\par
Methylformate (CH$_{3}$OCHO) appears as a mixture with H$_{2}$CO in the fits. Its pure form shows vibrational bands at 5.8 $\mu$m (C=O stretch), 8.26 $\mu$m (C–O stretch), 8.58 $\mu$m (CH$_3$ rocking), 10.98 $\mu$m (O–CH$_{3}$ stretch), and 13.02 $\mu$m (OCO deformation) \citep{Scheltinga2021}. However, in the CH$_{3}$OCHO:H$_{2}$CO mixture, matrix interactions lead to band shifts and altered widths. The O–CH$_{3}$ stretching mode is blue-shifted, and the OCO deformation mode also shifts in position. The CH$_{3}$ rocking mode at 8.58 $\mu$m overlaps with the CH$_{2}$ wagging mode of H$_{2}$CO at 8.49 $\mu$m, producing a blended, broader feature in this region. The C=O stretch at 5.8 $\mu$m overlaps with the strong H$_{2}$CO band, making it unreliable for column density estimation. Instead, the relatively stable C–O stretching mode at 8.26 $\mu$m is used to determine the column density of CH$_{3}$OCHO. The H$_{2}$CO column density is estimated using its feature near 8.02 $\mu$m.
\par
Ethanol (CH$_{3}$CH$_{2}$OH) is identified in our fits as a H$_{2}$O mixture. The weaker vibrational modes at 7.24 $\mu$m (CH$_{3}$ symmetric deformation), 7.4–7.7 $\mu$m (OH deformation), and 7.85 $\mu$m (CH$_{2}$ torsion) fall in relatively clean spectral regions. However, its stronger bands at 9.17 $\mu$m (CH$_{3}$ rocking), 9.51 $\mu$m (CO stretch), and 11.36 $\mu$m (CC stretch) are blended with dominant ice features of H$_{2}$O, as well as the broad silicate feature at $\sim$9.7 $\mu$m. The 7.24 $\mu$m band of ethanol-water mixture is relatively easy to detect as H$_{2}$O induces a peak shift, making it a promising candidate for identification \citep{Scheltinga2018}. 
\par
Acetic acid (CH$_{3}$COOH) shows absorption bands at 7.3 $\mu$m and 7.7 $\mu$m. In our fits, it appears as a mixture with H$_{2}$O, resulting in a broadened profile at 7.7 $\mu$m. This band overlaps to certain extent with the some bands of OCN$^{-}$, SO$_{2}$, CH$_{4}$, CH$_{3}$OCHO, CH$_{3}$CH$_{2}$OH and HCOOH whereas 7.3 $\mu$m falls rather in a clean region.  However, the intensity of the 7.7 $\mu$m band is twice that of the 7.3 $\mu$m band and is used to determine its column density.
\par
Acetaldehyde (CH$_{3}$CHO) is detected in our fits as part of an H$_{2}$O mixture in the ratio 1:10. It shows four significant absorption features in the 5.5–12.5 $\mu$m range, with the most prominent being the CO stretching mode at 5.8 $\mu$m. However, this band overlaps with other interstellar molecules like H$_{2}$CO, HCOOH, and NH$_{2}$CHO, making its identification challenging. Similarly, the 6.995 and 8.909 $\mu$m bands coincide with CH$_{3}$OH features, further complicating detection. The CH$_3$ symmetric deformation + CH wagging mode at 7.427 $\mu$m, which has minimal overlap with common ice components, is the most reliable for identifying it and the same is used to calculate the column density.
\begin{figure}[t]
    \includegraphics[width=0.5\textwidth]{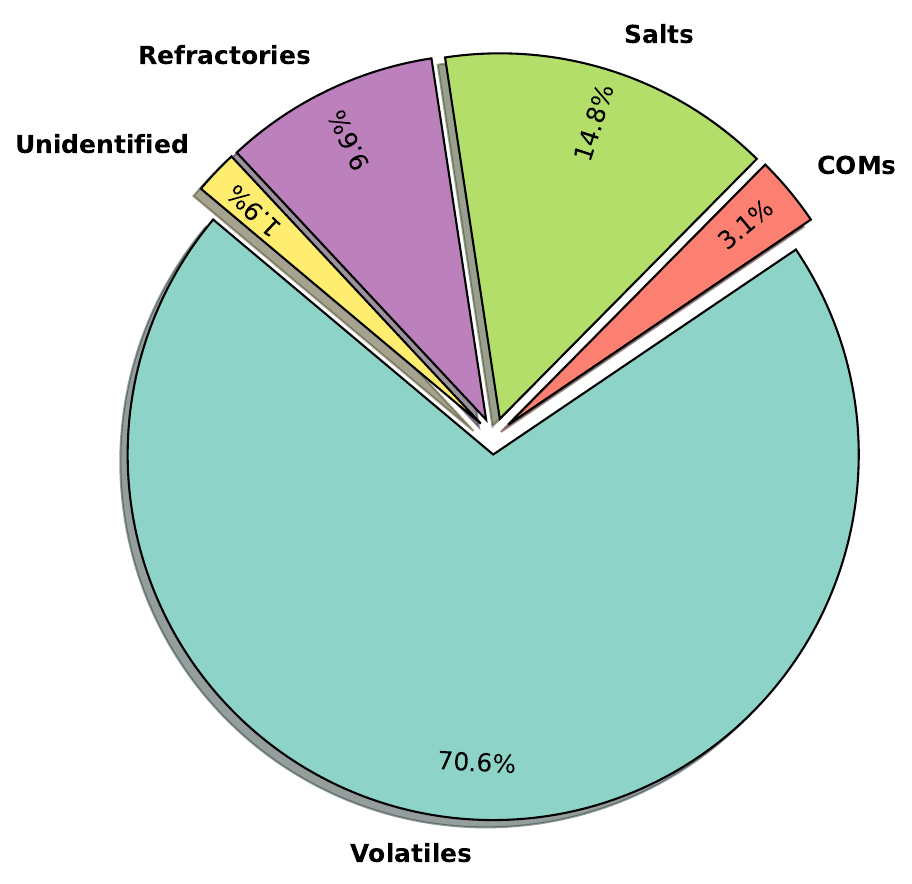}
    \caption{Percentage contribution to the spectral area of different ice groups in the 5–8 $\mu$m range. The model provides a good fit across the region as shown in the Figure \ref{fig:globalfit-iras2a}, except for the 7.1–7.5 $\mu$m interval. Appropriate polynomial baselines are used to separate the different ice groups present in the same ice mixture.}
    \label{fig:area_pie}
\end{figure}
\begin{figure*}[t]
    \includegraphics[width=\textwidth]{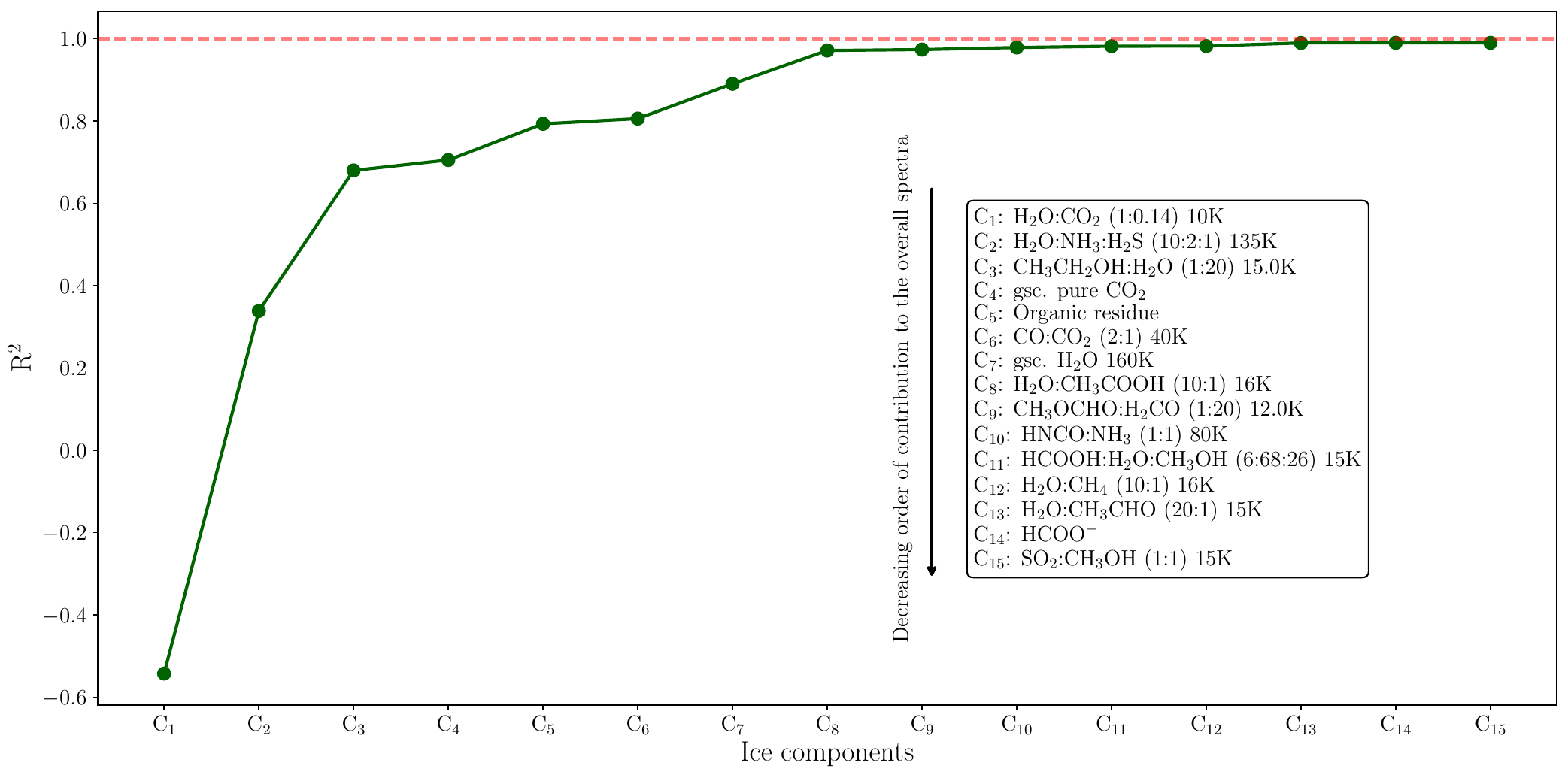}
    \caption{R$^{2}$ value as each component is added to the model. The R$^{2}$ for individual components is shown in green dots. The cumulative R$^{2}$, the normalized sum of cumulative R$^{2}$, is also plotted in blue dots to illustrate the relative change in R$^{2}$. Components are added in descending order of their contribution, which is defined as the maximum optical depth of the respective component.}
    \label{fig:rsquare}
\end{figure*}
\subsubsection{Organic refractories}
The bottom left panel in  Figure \ref{fig:components} shows the significant presence of organic refractory materials, which contribute to the excess absorption dip in the 5–10 $\mu$m region of the observed optical depth, alongside the H$_{2}$O bending mode. These materials are refractories even at room temperature, having high molecular mass up to 200 amu, and contain several molecules, radicals, and other fragments. The organic refractories show broad absorption features spanning from $\sim$5.5 $\mu$m to $\sim$11 $\mu$m in the MIRI range, indicating the presence of various functional groups including carbonyl (C=O), hydroxyl (O-H), amine (N-H), and C-H bending modes. Notably, this spectral region also coincides with absorption bands of volatile organics such as COMs, including aldehydes, esters, carboxylic acids, and amides, leading to significant spectral blending. The overlapping absorption features contribute to the overall depth of the observed bands making it challenging to isolate individual molecular contributions. This necessitates the determination of the local continuum in order to accurately isolate the absorption features of volatiles which is shown in \cite{2024A&A...683A.124R}.
\par
The spectra of organic refractories are obtained by \cite{Caro2003} who  irradiated 12 K mixture  of H$_{2}$O:NH$_{3}$:CH$_{3}$OH:CO:CO$_{2}$ in the ratio 2:1:1:1:1 using an hard UV dose of 0.25 photon molec$^{-1}$. The residue spectrum is shown in Figure \ref{fig:organic_residue}. The strongest peak due to the -COO$^{-}$ antisymmetric stretch of carboxylic acid salts [(R-COO$^{-}$)(NH$_{4}^{+}$)] is observed at around 6.3 $\mu$m. Hexamethylenetetramine (HMT) is also identified through the 8.10 ($\nu_{22}$) and 9.93 ($\nu_{21}$) $\mu$m stretch of CN. The peak at 9.21 $\mu$m characteristic of  glycolic acid is due to ammonium glycolate. It also accounts for the 20\% of the feature present at 6.30 $\mu$m due to the COO$^{-}$ stretching mode. The rest is due to other acid salts like   ammonium glycerate (HOCH$_{2}$CH(OH)COO$^{-}$)(NH$_{4}^{+}$) and ammonium oxamate (NH$_{2}$COCOO$^{-}$)(NH$_{4}^{+}$). The minor absorption features at 5.74 $\mu$m and 5.95 $\mu$m are attributed to the C=O stretching mode of esters (R–C(=O)–O–R$^{'}$) and primary amides (R–C(=O)–NH$_{2}$), respectively. A weak band due to NH$_{2}$ deformation in primary amides arises between 6.06–6.17 $\mu$m confirms the presence of primary amides.
\par
The model, which includes refractory material spectra, accurately reproduces the observed absorption features in the 5–10\,$\mu$m region, particularly between 5 and 7.12\, $\mu$m, with the exception of some localized mismatches between 7.1–7.5\, $\mu$m and at 8.26 and 8.9\, $\mu$m. We suspect the mismatch in the 7.1–7.5\, $\mu$m region may arise from unidentified or missing components. Similarly, the deviations observed at 8.26 and 8.9\,$\mu$m may be due to uncertainties in the baseline of refractory material spectra. To better understand the relative contributions of refractory materials and various ice types to the 5–8\,$\mu$m optical depth spectrum, we estimate the percentage area covered by each ice group, as illustrated in Figure~\ref{fig:area_pie}. Volatile ices dominate the absorption area (70.6\%), followed by salts (14.8\%), refractory organics (9.6\%), COMs (3.1\%), and unidentified components constitute (1.9\%). These findings highlight the importance of including organic refractory materials in the model to accurately reproduce the observed spectrum. The implications of refractory materials in astrophysical environments, including their formation efficiencies, are discussed further in Section~\ref{sec:discussions}.

\subsection{Statistical interpretation} \label{sec:stats}
\begin{figure*}[t]
    \includegraphics[width=\textwidth]{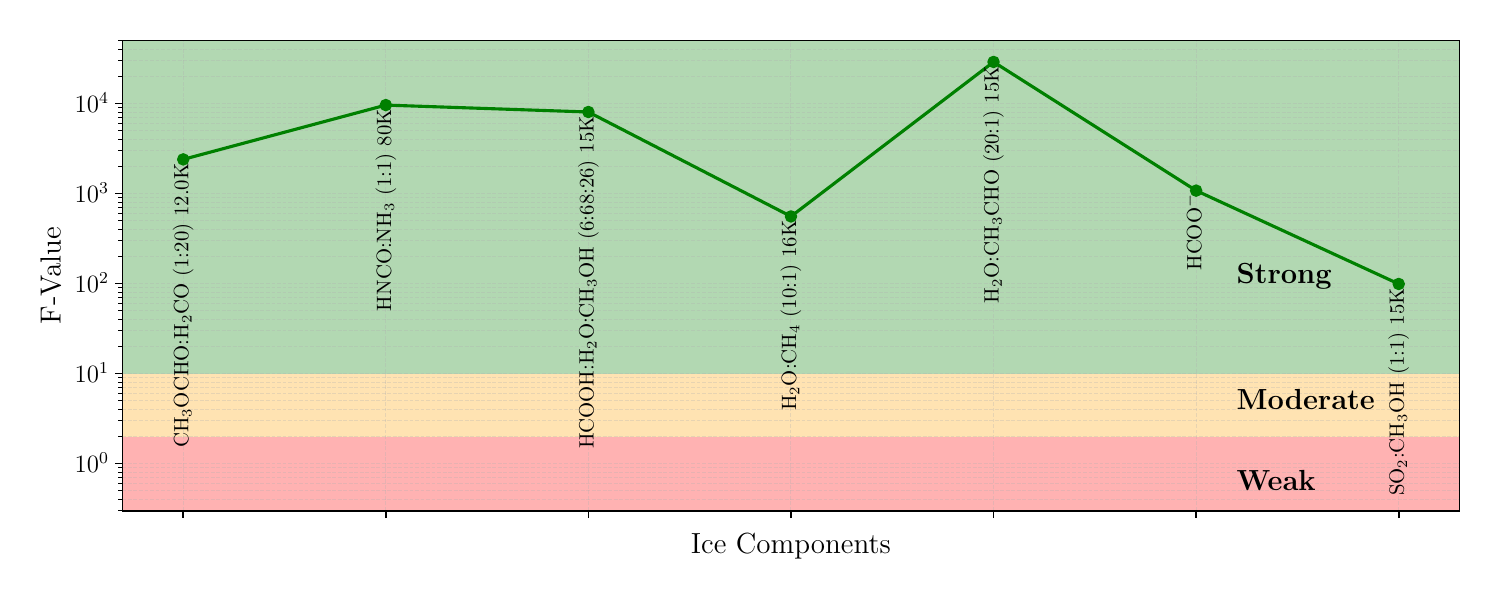}
    \caption{F-values for each minor ice component as it is sequentially added to the model, illustrating the relative change in variance. A higher F-value signifies a greater contribution of that component to the overall fit. Components are categorized as strong (in green region), moderate (in yellow region), or weak (in red region) based on the relative change in variance.}
    \label{fig:ftest}
\end{figure*}
Figure \ref{fig:rsquare} presents the R$^{2}$ analysis of the spectral fitting process where components are added sequentially in descending order of their contribution, defined by the maximum optical depth of each component in the MIRI range. As shown in Figure \ref{fig:rsquare}, the first few components, mainly H$_2$O and CO$_2$ mixtures together with organic residues, play a dominant role in improving the fit and lead to a rapid increase in the R$^{2}$ value. These species are expected to be major constituents of interstellar and protostellar ices due to their observed abundance in astrophysical environments. The model accuracy improves significantly till the first eight components whose contribution to the optical depth spectrum exceeds at least 0.6 in optical depth units. Beyond this point, adding relatively minor components (contributing $\lesssim$ 0.6 optical depth units to the spectrum) yields only marginal improvements, indicating that the overall spectral profile is largely governed by a few dominant species.
\par
To assess the significance of these minor components, we performed Greedy Backward Elimination, F-tests, and p-tests. Among the 15 components included in the best-fit model, 8 are dominant, while the remaining ones contribute mostly to localized features. As evident from Figure~\ref{fig:rsquare}, the R$^2$ value stabilizes after the eighth component, suggesting weak contributions from additional components, COM ices. To further evaluate the impact of adding components, p-values calculated based on the \(\chi^2\) distribution remain consistently low ($\lesssim$ 0.05), indicating that these minor components are locally relevant. While computing statistical values, we consider only the `contributing region,' defined as the wavelength range where the component's contribution exceeds the noise level (sigma). We have also performed an F-test  to measure the extent to which each component is contributing and  the results are shown in  Figure \ref{fig:ftest}. All minor ice components have a significant influence on the spectral fitting and are essential to the fitting procedure. Among these, CH$_3$CHO contributes the most to the overall optical depth spectrum due to its association with H$_2$O bands, whereas the SO$_2$ mixture contributes the least, adding only $\sim$0.1 optical depth units in narrow local regions. Therefore, we included all of them in the final solution, as they were identified as relevant by the Greedy Backward Elimination method. Each component contributes to the optical depth by more than one sigma, and their inclusion leads to a noticeable change in the model’s relative variance, as evident from Figure~\ref{fig:ftest}.

\begin{table*}[t]
\caption{\label{tab:ice_cd} Ice column densities obtained towards IRAS 2A using global fitting estimations using \texttt{INDRA} and local fitting results from \cite{2024A&A...683A.124R}. These values are compared to literature values for other objects.}
\renewcommand{\arraystretch}{1.5}
\centering 
\begin{tabular}{lccccccc}
\hline\hline

Specie &  \multicolumn{2}{c}{$N_{\rm{ice}}$ ($10^{17}$ cm$^{-2}$)} & \multicolumn{2}{c}{$X_{\rm{H_2O}}$ (\%)} & \multicolumn{3}{c}{Literature (\% H$_2$O)}\\
\cline{2-8}
 & Global & Local & Global & Local & LYSOs & MYSOs & Comet 67P/C-G$^{(m)}$\\
\hline
H$_2$O (13~$\mu$m) & 324.00$_{205.00}^{504.00}$&300$\pm$12 & 100 & 100 & 100 & 100 & 100\\

HCOOH & 3.97$_{2.15}^{5.34}$&3.0$_{1.7}^{5.3}$ & 1.23 & 1.0 &$<0.5-$4$^{(c)}$ & $<0.5-$6$^{(d)}$ & 0.013$\pm$0.008\\

OCN$^-$ & 4.90$_{2.68}^{6.50}$& 3.7$_{3.3}^{6.6}$ & 1.51 & 1.2  & $<0.1-$1.1$^{(h)}$ & 0.04$-$4.7$^{(i)}$ & ...\\

H$_2$CO & 15.10$_{8.28}^{19.90}$&12.4$_{6.6}^{19.7}$ & 4.66 & 4.1 & $\sim$6$^{(g)}$ & $\sim$2$-$7$^{(b)}$ & 0.32$\pm$0.1\\

HCOO$^-$ (7.4~$\mu$m) & 1.58$_{2.22}^{8.27}$ &1.4$_{0.4}^{2.4}$ & 0.49 & 0.15 & $\sim$0.4$^{(g)}$ & $<$0.3$-$2.3$^{(b)}$ & ...\\

CH$_3$OH & 13.88$_{7.52}^{18.66\ddagger}$ & 15, 23$^{\dagger}$ & 4.28 & 5.0, 7.6  &$<1-$25$^{(d)}$ & $<3-$31$^{(d)}$ & 0.21$\pm$0.06\\

CH$_3$CH$_2$OH & 5.70$_{3.65}^{8.89}$&3.7$_{0.5}^{4.5}$ & 1.76 & 1.2 & ... & $<$1.9$^{(e)}$  & 0.039$\pm$0.023\\

CH$_3$CHO & 1.94$_{1.13}^{3.15}$&2.2$_{1.4}^{2.8}$ & 0.60 & 0.7 & ... & $<$2.3$^{(e)}$ & 0.047$\pm$0.017\\

CH$_4$ & 4.55$_{2.43}^{6.30}$&4.9$_{3.2}^{7.5}$ & 1.40 & 1.6 & $<$3$^{(a)}$ & 1$-$11$^{(b)}$ & 0.340$\pm$0.07\\

CH$_3$COOH & 1.05$_{0.65}^{1.65}$&0.9$_{0.6}^{1.3}$ & 0.32 & 0.3 &  ... & ... & 0.0034$\pm$0.002\\

CH$_3$OCHO & 0.34$_{0.18}^{0.44}$& 0.2$_{0.1}^{0.4}$ & 0.10 & 0.1 & $<$2.3$^{(f)}$ & ... & 0.0034$\pm$0.002\\

SO$_2$ & 0.63$_{0.30}^{0.98}$ & 0.6$_{0.4}^{1.9}$ & 0.19 & 0.2 &  0.08$-$0.76$^{(a)}$ & $<0.9-$1.4$^{(b)}$ & 0.127$\pm$0.100\\

\hline
 \multicolumn{7}{c}{\hspace{2cm}New species found in this work}\\
\hline
\hline
 \multicolumn{7}{c}{\hspace{2cm}Simple ice species}\\
\hline
NH$_{4}^{+}$ & 37.30$_{23.10}^{55.20}$$^{\star}$& ... & 11.51 & ... & $\sim$ 2.2$-$15.9$^{(g)(p)}$& $\sim$ 4.9$-$28$^{(g)(p)}$ & ...\\

CO$_{2}$ (15~$\mu$m) & 43.50$_{35.40}^{93.60}$ & ... & 13.43 & ... & $\sim$11.09$-$65.12$^{(o)}$ & ... & 4.7$\pm$1.4\\

NH$_{3}$ & 16.20$_{10.00}^{24.00}$& ... & 5.00 & ... & $\sim$ 3$-$8$^{(g)}$ & ... & 0.67$\pm$0.20\\
\hline

NH$_{4}^{+\maltese}$ & 0.75 & \textbf{...} & 0.23 &  ... & $\sim$ 2.2$-$15.9$^{(g)(p)}$& $\sim$ 4.9$-$28$^{(g)(p)}$ & ... \\

HMT$^{\maltese}$ & 3.15 & ... &0.9 & ...& ... & ... & Negligible$^{(n)}$ \\

Carboxylic acid salt$^{\maltese}$ & 1.66 & ... &0.5 & ...& ... & ... & ... \\

POM$^{\maltese}$ & 0.39 & ... &  0.1 & ...& ... & ... & Negligible$^{(n)}$ \\

Amide$^{\maltese}$ & 0.21 & ... & 0.06 & ...& ... & ... & ... \\

Ester$^{\maltese}$ & 0.17 & ... & 0.05 & ...& ... & ... & ... \\


\hline
\end{tabular}
\tablecomments{$^a$\citet{Oberg2008}, $^b$\citet{Gibb2004}, $^c$\citet{Oberg2011}, $^d$\citet{Schutte1999_weak}, $^e$\citet{Scheltinga2018}, $^f$\citet{Scheltinga2021}, $^g$\citet{Boogert2008}, $^h$\citet{vanBroekhuizen2005}, $^i$\citet{Boogert2022}, $^j$\citet{Rachid2021}, $^k$\citet{Rachid2022}, $^l$\citet{slav2023}, $^m$\citet{Rubin2019},
$^n$\citet{Hanni2022}, $^o$\citet{Pontoppidan2008}, $^p$\citet{Slavicinska2025}, $^{\ddagger}$ values are reported with a factor of 2, $^{\dagger}$ values are reported with a factor of 2 or 3, $^{\maltese}$ features belonging to organic refractory salt are shown separately as there are $\lesssim$ 20 \% uncertainties in the estimates due to local baseline placements \citep{Caro2003}}, $^{\star}$ contributions from H$_{2}$O:NH$_{3}$:H$_{2}$S mixture, OCN$^{-}$ ion salts.
\end{table*}
\subsubsection{Confidence intervals}
In Appendix~\ref{appendix:stats}, we present the confidence intervals for the model components. Figures~\ref{fig:corner_h2o_1}--\ref{fig:corner_h2o_3} show the correlations between components containing H$_2$O bands across three different spectral windows in the 5--20~$\mu$m range. The confidence contours indicate that all components contribute significantly to the fit. Most components show notable mutual correlations, reflecting the spectral overlap inherent to H$_2$O-rich mixtures. In particular, the H$_2$O:CH$_3$COOH and pure H$_2$O components display strong degeneracy, as evidenced by their elongated elliptical confidence contours, suggesting either spectral overlap or similar contributions within these windows. The H$_2$O:CH$_3$CHO component shows a large degree of uncertainty, as reflected in its broad and diffuse contour shapes.

Similarly, the CO$_2$-bearing components shown in Figures~\ref{fig:corner_co2_1}--\ref{fig:corner_co2_3} show strong mutual correlations, again indicating significant spectral overlap. In particular, the H$_2$O:CO$_2$ mixture displays highly elongated confidence contours, pointing to substantial uncertainty in its contribution. The CO:CO$_2$ mixture and pure CO$_2$ components also show degeneracy, suggesting considerable spectral overlap, as illustrated in Figure~\ref{fig:components}. Additionally, the similar widths of their contours imply comparable contributions within the examined windows.

The correlation plots for components without H$_2$O and CO$_2$ bands are shown in Figures~\ref{fig:corner_minor_1}--\ref{fig:corner_minor_3}. These components, which include species such as HCOOH, CH$_3$OCHO, SO$_2$, HCOO$^{-}$, and complex organic residues, generally show weaker mutual correlations compared to the H$_2$O- and CO$_2$-bearing ices, suggesting more confined spectral contributions. Their confidence contours are typically compact and less elongated, indicating lower degeneracy and more distinct contributions in this spectral region. The SO$_2$ mixture, however, shows considerable uncertainty within its region of influence.

\subsection{Column densities and comparison with local fitting results}
The ice column densities of different species obtained through global fitting are presented in Table~\ref{tab:ice_cd}. For comparison, we also include column densities derived from local fitting by \citet{2024A&A...683A.124R}, shown in Figure~\ref{fig:cdcompare}. Additionally, abundances relative to H$_2$O observed toward various sources are presented in Table \ref{tab:ice_cd} and are compared in Figures~\ref{fig:comet} and~\ref{fig:lyso_myso}. The column densities derived from global fitting fall within the ranges estimated by local fitting, indicating that both approaches yield consistent results. As shown in Figure~\ref{fig:cdcompare}, most ice species agree within a factor of two between the two methods, with the exception of HCOO$^{-}$, whose global fitting estimate is 2.6 times higher than the local fitting result. 

This discrepancy in HCOO$^{-}$ can be interpreted in two ways. On one hand, the global fit includes its strong absorption band at 6.32\,$\mu$m, which may support the higher estimate. On the other hand, the value could be overestimated due to fitting uncertainties in the 7.1–7.5\,$\mu$m region, where missing components may lead the model to assign extra absorption to HCOO$^{-}$. A similar situation is observed for CH$_3$CH$_2$OH, which also shows a slightly higher estimate, though still within a factor of two.

\section{Discussions}\label{sec:discussions}
\subsection{Global vs local fitting}
\begin{figure*}[t]
    \includegraphics[width=\textwidth]{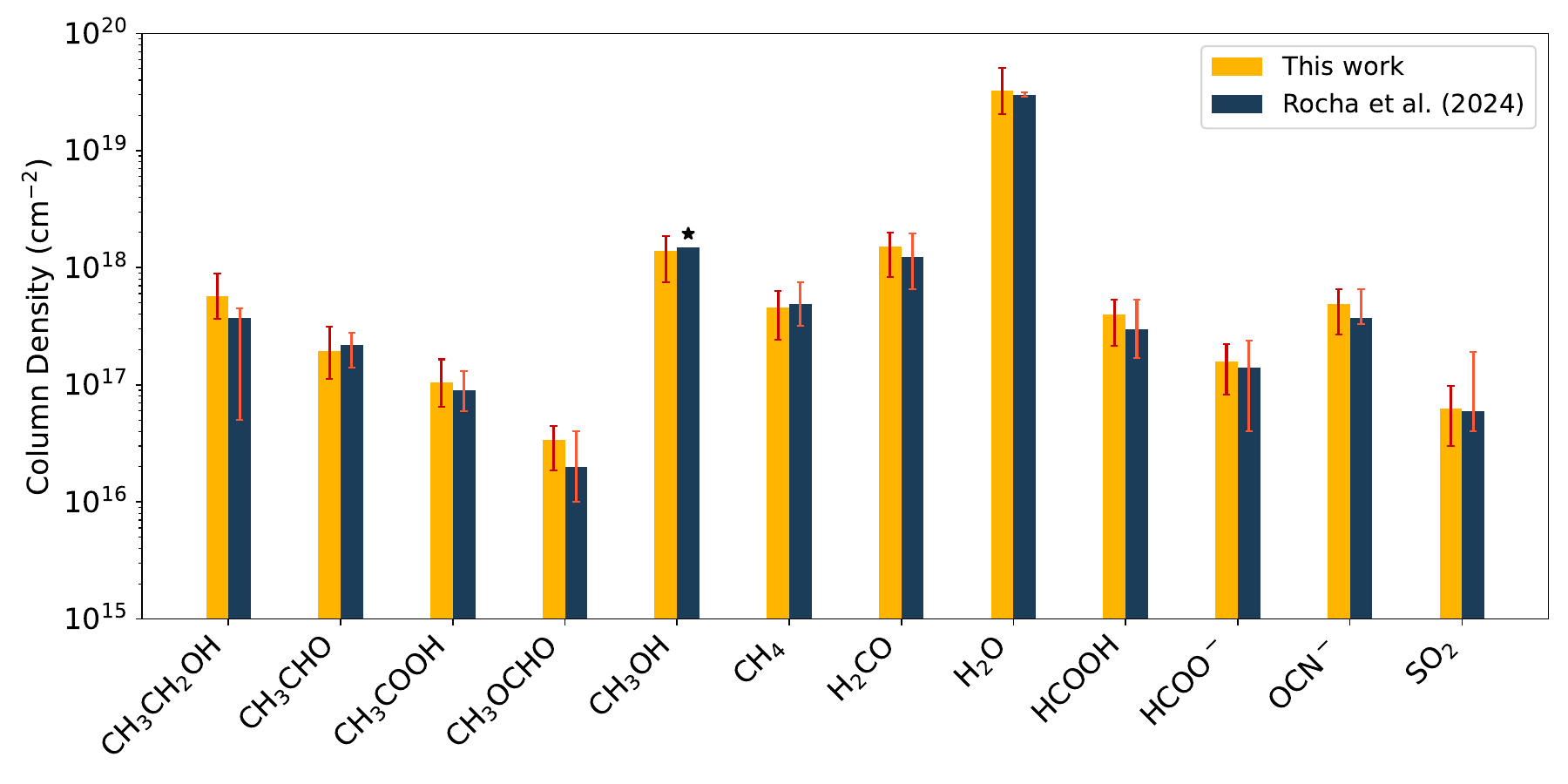}
    \caption{Comparison of ice column densities derived in this work (gold), using global fitting, with those from \cite{2024A&A...683A.124R} (dark blue) who used local fitting. The x-axis represents different ice species, while the y-axis shows the column densities in cm$^{2}$. Error bars, shown in red, indicate the uncertainties associated with each measurement. For CH$_{3}$OH, a factor of 2 or 3 is used in the two fitting methods denoted by star. This agreement highlights the robustness of our global fitting approach using \texttt {INDRA} in capturing the ice inventory consistently with previous studies.}
    \label{fig:cdcompare}
\end{figure*}

From  Figure \ref{fig:cdcompare}, it is evident that global and local fitting yield comparable column densities. In local fitting, the baseline is determined within a narrow spectral window requiring a precise knowledge of the placement of guiding points to anchor the continuum. This method is particularly sensitive to user-defined inputs and can result in the exclusion or misrepresentation of certain components. For example, \citet{2024A&A...683A.124R} have shown in their Appendix J, how the components such as CH$_{3}$CHO and CH$_{3}$COOH may be excluded from the fit solely due to the choice of guiding point placement. Since ice spectra contain overlapping absorption bands, determining the continuum placement can be challenging. However, global fitting relies on a single baseline across the entire spectral range and simultaneously considers multiple features of the same ice at different wavelengths. This approach ensures a more self-consistent decomposition of spectral components, reducing uncertainties associated with local continuum selection. For example, in the IRAS 2A optical depth spectrum, the absorption band due to OCN$^{-}$ at 15.87 $\mu$m overlaps with CO$_{2}$ and H$_{2}$O, making it almost unrecognizable. In such cases, one might tend to attribute the entire absorption to CO$_{2}$ and H$_{2}$O ices, potentially overlooking the contribution from OCN$^{-}$. However, global fitting accounts for this issue by considering multiple spectral regions where OCN$^{-}$ shows distinct features, ensuring that its contribution is properly constrained. Also, if the absorption features of different ices are not severely blended with dominant ones, regardless of whether the baseline is determined locally or globally, the overall column densities remain largely unaffected, as they are primarily dictated by the unique absorption strengths and intrinsic band profiles of ices which is the case with CH$_{4}$, SO$_{2}$ and CH$_{3}$OH as shown in the Figure \ref{fig:cdcompare}. 
Another drawback of the local fitting approach is that it can disrupt physically meaningful correlations between absorption bands of the same species, potentially leading to inconsistent or unreliable estimates across those bands. For example, HCOO$^{-}$ shows an intense absorption band at 6.32 $\mu$m, which is nearly twice as strong as its weaker bands at 7.2 and 7.4 $\mu$m. Previous study that employed local fitting within the 6.8–8.6 $\mu$m region was limited to the weaker bands for column density estimation. As a result, these estimates may have been less reliable due to the lower band strengths. In contrast, the global fitting approach adopted in this work leverages the full MIRI range, allowing the inclusion of the stronger 6.32 $\mu$m feature. This leads to a more robust constraint on the HCOO$^{-}$ abundance, yielding column densities that are nearly twice those derived from local fitting method. However, we caution that this higher value may also be influenced by the spectral gap in the model and observed spectrum due to the absence of certain components in the 7.1–7.5\,$\mu$m region, which could lead the model to overfit HCOO$^{-}$ in the region.
\subsection{Treatment of H$_{2}$O, CO$_{2}$ and CH$_{3}$OH bands in some of the ice mixtures} \label{sec:band_removal}
In our spectral fitting analysis, the presence of H$_{2}$O and CH$_{3}$OH bands in the ice mixtures alongside the features of COMs carries important physical implications. In general, the inclusion of H$_{2}$O and CH$_{3}$OH absorption features in ice mixtures provides compelling evidence that these species are embedded in H$_{2}$O- and CH$_{3}$OH-rich matrices, a condition supported by laboratory experiments. Therefore, our general approach has been to retain these bands in ice mixtures wherever possible. However, exceptions were made in cases where the inclusion of dominant H$_{2}$O or CH$_{3}$OH bands introduced inconsistencies or degeneracy in the spectral fit, primarily due to overlapping contributions from different ice mixtures or uncertain baselines. In three of the COM mixtures, as listed in Table~\ref{tab:com_mixtures}, we retain the H$_{2}$O bands to reflect the physical co-existence of H$_{2}$O and COMs in the ice matrix. The specific cases where dominant band removal was necessary are described below.

For the HCOOH:H$_{2}$O:CH$_{3}$OH (6:68:26) 15 K mixture, the spectrum near the 6 $\mu$m H$_{2}$O bending mode displayed a negative baseline as shown in the lower left panel of the Figure~\ref{fig:H2O_HCOOH}. After the baseline correction, the 6 $\mu$m band showed a significant improvement in its relative peak strength. Including the H$_{2}$O contribution from this mixture introduced inconsistencies in the modeled spectrum and adversely affected the stability of the fit. Given that H$_{2}$O is already well-accounted for by other mixtures in the model, we removed the H$_{2}$O bands from this component to ensure stability in the fitting.

For the H$_{2}$O:CH$_{4}$ (10:1) 16 K mixture, the observed feature near 7.7~$\mu$m in the IRAS 2A optical depth spectrum could be reliably reproduced only when the dominant H$_{2}$O bands from this mixture were excluded. Retaining these bands led to overfitting of water features that were already adequately modeled by other components, which degraded the quality of the overall fit in the region. The feature at 7.7~$\mu$m is shaped by the combined contributions of CH$_{4}$ (7.67~$\mu$m), SO$_{2}$ (7.60~$\mu$m), and the OCN$^{-}$ ion (7.62~$\mu$m). The improved fit quality around this region can be observed in Figure~\ref{fig:globalfit-iras2a}, and the corresponding CH$_{4}$ column density estimation compared to the local fitting is shown in Figure~\ref{fig:cdcompare}.

For the SO$_{2}$:CH$_{3}$OH (1:1) 10 K mixture, isolation of the SO$_{2}$ feature was necessary due to the presence of strong CH$_{3}$OH absorption band at 9.74~$\mu$m. In the optical depth spectrum of IRAS 2A, this band is already well addressed by CH$_{3}$OH band from HCOOH:H$_{2}$O:CH$_{3}$OH mixture. As a result, including the additional CH$_{3}$OH contribution from the SO$_{2}$ mixture introduced redundancy and led to the rejection of the component. To address this, we removed the CH$_{3}$OH band from this mixture, isolating the SO$_{2}$ signatures at 7.60 and 8.68 $\mu$m with the former band being relatively intense one. The column densities of SO$_{2}$ and CH$_{3}$OH are comparable with the local fitting case and the same is shown in the Figure \ref{fig:cdcompare}.

Grain shape corrections were applied exclusively to some pure H$_{2}$O and CO$_{2}$ ices, as these are dominant ices and they can interact with grains which can affect the band shapes. In contrast, such corrections were not applied to weaker ice components, which are not susceptible to grain shape effects (\citealt{Brunken2025, Ehrenfreund1996, Dartois2006, Dartois2022}). For mixtures containing dominant bands (e.g., H$_{2}$O-rich mixtures), applying grain shape corrections is not straightforward, as the effects depend on factors such as ice layer thickness, density, and the internal structure of the mixture. We also note that further experimental and modeling studies are needed to better understand the grain shape effects on band shapes.
\subsection{Formation pathway of the detected ices}
Using global fitting, we identified multiple ice components contributing to the observed optical depth of IRAS 2A. These include simple molecules like H$_{2}$O, CH$_{4}$, and NH$_{3}$, ions such as OCN$^{-}$ and NH$_{4}^{+}$,  as well as COMs such as CH$_{3}$OH, CH$_{3}$OCHO, and HCOOH. Additionally, we noted the possible presence of organic refractory salts. In this section we discuss the formation pathways of these ices.
\par
Simple molecules such as H$_{2}$O, CH$_{4}$, NH$_{3}$, and SO$_{2}$ form through a combination of gas-phase and grain-surface reactions in interstellar environments. Water ice primarily forms on dust grains through the hydrogenation of atomic oxygen \citep{Molpeceres2018}, while in warmer regions, gas-phase reactions involving oxygen and molecular hydrogen also contribute \citep{Glassgold2009}. Methane is produced via the stepwise hydrogenation of atomic carbon on grain surfaces, a process that occurs efficiently at low temperatures in dense molecular clouds (\citealt{Qasim2020, Lamberts2022}). Similarly, ammonia forms through successive hydrogenation of atomic nitrogen on icy dust grains (\citealt{Hiraoka1995, Hidaka2011, Fedoseev2014}), whereas in gas phase through a series of ion-molecular reactions (\citealt{Herbst1973, Scott1997, Fedoseev2014}). Sulfur dioxide, on the other hand, can originate from both gas-phase reactions and surface chemistry. These simple molecules act as fundamental building blocks for more complex ices, undergoing further processing as star and planet formation proceeds.
\par
The IRAS 2A environment harbors several ionic species. Laboratory experiments have demonstrated that such ions can form under purely non-energetic conditions, primarily through acid–base reactions within the ice. However, the contribution of energetic processing cannot be excluded completely. Among these ions, NH$_{4}^{+}$ forms in interstellar ices primarily  through acid-base reactions involving NH$_{3}$ and strong acids such as HNCO (grain-surface product). A proton transfer can readily occur in the presence of the strongly nucleophilic NH$_{3}$ with even a modest energy input, such as a slight temperature increase or UV radiation \citep{1993MNRAS.261...83H}. HCOO$^{-}$ ion is typically formed through acid-base reactions in interstellar ices. Additionally, it can also be produced through surface chemistry involving CO and OH radicals on interstellar grains or via energetic processing (e.g., UV irradiation or cosmic-ray-induced reactions) of ice mixtures containing H$_{2}$O, CO, and H$_{2}$CO molecules. OCN$^{-}$ forms efficiently under interstellar conditions through acid-base reactions \citep{Gerakines2025} or processes involving  photolysis or irradiation of CO and NH$_{3}$ containing ices \citep{2001ApJ...550.1140H}. It has been observed in interstellar ices \citep{Grim1987, Schutte2003, van_broekhuizen_quantitative_2004} and is the primary carrier of the interstellar XCN band. 
CH$_{3}$CHO and CH$_{3}$CH$_{2}$OH are chemically linked in interstellar ices. Laboratory studies have shown that hydrogen bombardment of CH$_{3}$CHO ice can produce CH$_{3}$CH$_{2}$OH via sequential hydrogenation through the CH$_{3}$CH(OH) radical intermediate, and that this pathway can proceed with 22 \% efficiency at temperatures as low as 10~K through quantum tunneling \citep{Bisschop2007, Chuang2020}. However, this reaction has a relatively high activation barrier \citep{Fedoseev2022, Sivaramakrishnan2010}, and recent quantum chemical calculations suggest that CH$_{3}$CHO is comparatively resistant to direct hydrogenation to ethanol in icy environments \citep{Molpeceres2025}. Chemical modeling instead points to a dominant top-down surface pathway in which CH$_{3}$CH$_{2}$OH is converted to CH$_{3}$CHO via hydrogen abstraction \citep{Jin2025}, likely enhanced by the higher ice abundance of ethanol. The relative importance of these pathways, particularly under the warmer and more energetic conditions typical of protostellar environments, remains uncertain. Further experimental and modeling studies are needed to quantify their efficiencies.  CH$_{3}$COOH in the interstellar medium may form through gas-phase processes involving species such as CH$_{3}$CO$^{+}$, H$_{2}$O, CH$_{3}$OH$_{2}^{+}$, and HCOOH \citep{huntress1979, ehrenfreund2000}, though these pathways alone do not fully account for observed abundances \citep{mehringer1997, remijan2002}. Laboratory and theoretical studies suggest that grain-surface chemistry involving radicals like CH$_{3}$, HOCO, CH$_{3}$CO, and OD in irradiated ices is a more efficient and likely dominant formation route \citep{bennett2007, bergantini2018, kleimeier2020, ahmad2020}. CH$_{3}$OCHO forms primarily through surface reactions involving CH$_{3}$O and HCO radicals. These radicals originate from CO hydrogenation on icy dust grains, a process that occurs efficiently at low temperatures ($\sim$10–20 K) (e.g., \cite{Watanabe2002}; \cite{Fuchs2009}). During this sequence, CO is first hydrogenated to form H$_{2}$CO, which can further react with hydrogen to produce CH$_{3}$OH. Meanwhile, the CH$_{3}$O and HCO radicals, which are intermediates in this process, can recombine to form CH$_{3}$OCHO (e.g., \citealt{Chuang2016}; \citealt{Garrod2022}; \citealt{Chen2023}).
\par
The formation of COMs within icy mantles on dust grains is primarily driven by the condensation of volatile species such as H$_{2}$O, CO, CO$_{2}$, and NH$_{3}$ in dense molecular clouds (\citealt{caselli1999, Boogert2015}). Upon exposure to UV radiation from YSOs or external cosmic sources, these ices undergo photodissociation, leading to the generation of reactive radicals. The recombination of these species results in the synthesis of more COMs, including HMT, ammonium salts of carboxylic acids, amides, esters, and polyoxymethylene (POM)-related species (\citealt{Gibb2001, garrod2006, chuang2017}). H$_{2}$O ice plays a catalytic role in facilitating these reactions by promoting radical diffusion and stabilizing intermediates. Laboratory experiments confirm that UV photoprocessing of ice mixtures, such as H$_{2}$O:CH$_{3}$OH:NH$_{3}$:CO:CO$_{2}$ at temperatures ($\sim$12 K), significantly enhances the formation of refractory organic residues \citep{Caro2003}. The transformation of these organics into nonvolatile refractory materials occurs through further energetic processing. Silicate grains, as suggested by \citet{2024GeCoA.387...98H}, can incorporate these organics within their interlayer spaces, preserving them as interstellar material evolves into larger interplanetary objects like asteroids and comets.
\subsection{Contributions from polar environment}
\begin{figure*}[t]
    \includegraphics[width=\textwidth]{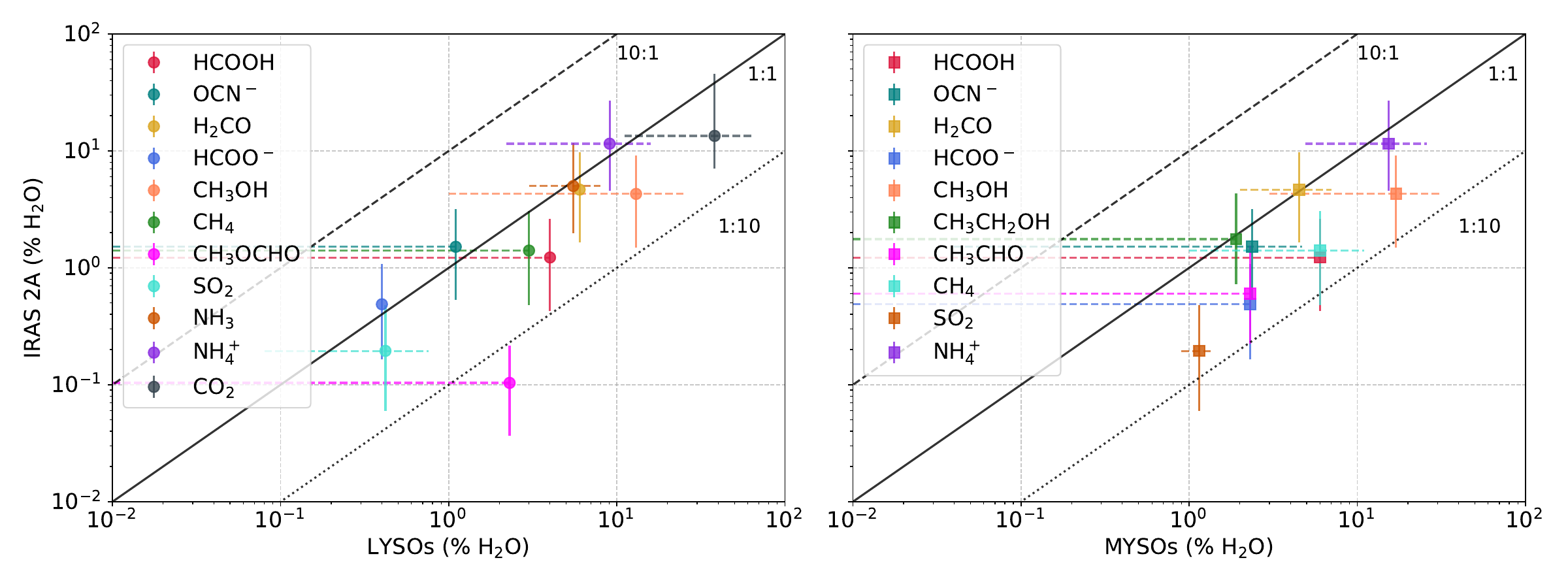}
    \caption{Comparison of ice abundances as a percentage of H$_{2}$O of IRAS 2A with those of LYSOs (left panel) and MYSOs (right panel). The solid line represents 1:1 reference, while the dotted and dashed lines indicate abundances that are a reduced and increased by a factor of 10 respectively. Table~\ref{tab:ice_cd} lists the values used for the comparison of LYSOs and MYSOs with IRAS 2A.)}
    \label{fig:lyso_myso}
\end{figure*}
We observed that most of the organics are associated with water, suggesting a polar environment, which is also consistent with previous work by \cite{2024A&A...683A.124R}. The presence of COMs such as CH$_{3}$CHO, CH$_{3}$CH$_{2}$OH, and CH$_{3}$COOH within a H$_{2}$O-rich ice matrix indicates that water plays a dominant role in shaping the chemical environment of IRAS 2A. Water, being highly polar, strongly influences the spectral band shapes of these molecules, leading to broader features and peak shifts compared to their absorption in apolar environments. Additionally, laboratory experiments have demonstrated that COMs in polar ice environments exhibit different spectral characteristics compared to those in apolar matrices, further supporting the H$_{2}$O-dominated nature of the ice. Observations from the JWST-Ice Age program \citep{McClure2023} suggest that CH$_{3}$OH, the most abundant COM, commonly coexists with H$_{2}$O in cold prestellar clouds, which is also the case for IRAS 2A. This coexistence may point to the influence of polar interactions in shaping ice chemistry, although recent modeling by \cite{JimenezSerra2025} indicates that methanol formation is predominantly driven by CO hydrogenation, independent of the ice's polar or apolar nature.  Furthermore, the experiments by \cite{Caro2003} suggests the role of H$_{2}$O as a catalyst in the formation of large organic molecules at cryogenic temperatures which ultimately leads to the formation of COMs and refractory materials. The significance of a polar environment is also supported by the fact that the observed 7.2 $\mu$m feature in the ISO spectrum toward W 33A was better reproduced by \cite{Bisschop2007} when HCOOH is present in polar environment rather than in CO or CO$_2$ dominated regions. We have also observed the HCOOH appearing as a tertiary mixture that includes H$_{2}$O. This highlights the importance of complex ice-phase interactions through surface reactions involving H, OH, and HCO radicals, primarily in polar environments rich in H$_{2}$O and CH$_{3}$OH. Altogether, the available evidence points toward a predominantly polar chemical environment in IRAS 2A, although further chemical modeling and laboratory experiments are needed to fully confirm this interpretation.
\subsection{Astrophysical importance of the detected ices}
\begin{figure}[h]
    \includegraphics[width=0.5\textwidth]{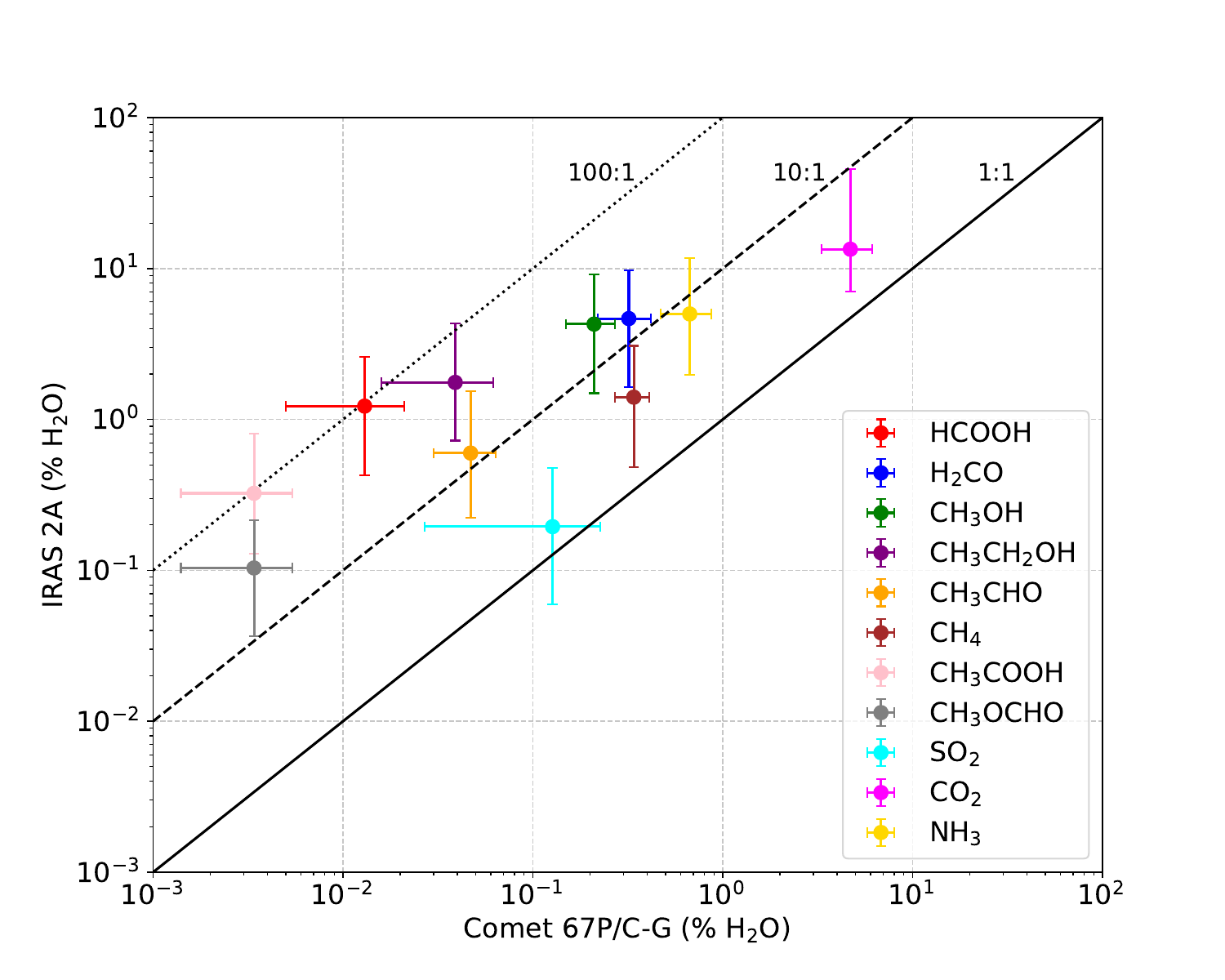}
    \caption{Comparison of ice abundances as a percentage of H$_{2}$O of IRAS 2A with those of comet 67P/C-G from \cite{Rubin2019}. The solid line represents 1:1 reference, while the dashed and dotted lines indicate cometary abundances that are reduced by factors of 10 and 100, respectively.}
    \label{fig:comet}
\end{figure}

The detection of various ices in IRAS 2A provides crucial insights into the chemical composition and evolution of planetary objects. All the identified ices which include H$_{2}$O, CH$_{4}$, NH$_{3}$, CO, CO$_{2}$, and COMs such as CH$_{3}$CHO, CH$_{3}$CH$_{2}$OH, and CH$_{3}$COOH, have previously been observed in different astrophysical environments. Figures~\ref{fig:lyso_myso} and~\ref{fig:comet} show the presence of various ice species and comparison of the ice estimates from this work with those found in LYSOs and MYSOs, as well as in comet 67P. Note the good agreement between the estimated ice column densities and the observations toward LYSOs in the Figure~\ref{fig:lyso_myso}. Most estimates lie within a factor of 10 except CH$_{3}$OCHO which appears to be an outlier. However, it is important to note that this apparent discrepancy is based on a single observation reported by \citet{Scheltinga2021}. This study sets an upper limit of approximately 2.3 \% relative to H$_{2}$O for CH$_{3}$OCHO in the low-mass protostar HH 46. Unlike the gas phase, where CH$_{3}$OCHO has been detected toward numerous sources including molecular clouds, direct ice-phase identifications remain limited due to strong spectral blending with other species.  So, given that the available ice-phase constraint is based on a single upper-limit measurement and that spectral blending complicates its identification, more comprehensive ice observations are needed before drawing firm conclusions about its abundance relative to H$_{2}$O in low-mass protostellar environments.  The left panel of the Figure \ref{fig:lyso_myso} indicates that IRAS 2A's ice composition is broadly comparable to both LYSOs and MYSOs. The left panel of the Figure~\ref{fig:lyso_myso} suggests that IRAS 2A is chemically more similar to LYSOs, supporting its classification as a low-mass protostar in a relatively early evolutionary stage. However, further observational and modeling studies are needed to evaluate the chemical similarity. Figure~\ref{fig:comet} presents a similar comparison with observations from comet 67P \citep{Rubin2019}. If the ice inventory from earlier phases is inherited by cometary bodies without significant alteration, one would expect a one-to-one correlation, as indicated by the solid line in Figure~\ref{fig:comet}. However, the fractional abundances of most species relative to water are systematically depleted in comet 67P compared to IRAS 2A. This suggests that ices inherited from earlier stages undergo substantial processing before being incorporated into cometary nuclei. Notably, the COMs and their precursors, such as HCOOH and H$_{2}$CO, exhibit the strongest depletion, by factors $\gtrsim$10 in comet 67P. In contrast, simpler volatiles like NH$_{3}$, SO$_{2}$, CH$_{4}$, and CO$_{2}$ fall within a range of $\sim$ 1–10. This pattern likely reflects the thermal, radiative processing of ices and outgassing of materials during the protostellar-to-disk transition and after, where fragile COMs are easily destroyed or converted, while simpler volatiles can efficiently be re-condensed at low temperatures. These findings support the view that chemical inheritance from the protostellar phase is neither complete nor direct, and that significant chemical evolution occurs between the early envelope stage and the comet formation.
\par
The detection of ions like OCN$^{-}$, NH$_{4}^{+}$, and HCOO$^{-}$ in IRAS 2A indicates an active chemical evolution. These ions are primarily formed through proton-transfer and acid–base reactions within the ice, although the contribution of energetic processing cannot be excluded. The presence of HCOOH in this ice mixture highlights a chemically rich environment where hydrogen bonding interactions and acid-base equilibria play a crucial role in stabilizing ions and shaping the observed spectral signatures. Additionally, the detection of various COMs and their strong spectral association with H$_{2}$O suggest that water ice is a key driver in their formation, either by acting as a chemical catalyst or through hydrogenation pathways that transform simpler molecules into more complex species. The polar nature of the ice environment influences COM formation by stabilizing reactive intermediates, enabling radical accumulation and recombination under energetic processing, and potentially modifying spectral band shapes, as seen in JWST observations. Further, the detection of CH$_{3}$COOH, HCOOH and CH$_{3}$OH highlights their potential role in prebiotic chemistry. This supports the idea that IRAS 2A harbors a chemically dynamic and evolutionarily significant ice reservoir, with potential implications for the prebiotic chemistry inherited by nascent planetary systems. 
 \par
The inheritance of the refractory organics by carbonaceous chondrites (CR and CM types) further strengthens the connection between interstellar ice chemistry and the complex organic inventory in planetary systems \citep{2024EPSC...17..596D}. Additionally, HMT is known to desorb around 40 K, and its possible presence in protostellar sources, such as IRAS 2A, could be a useful tracer of cold regions where thermal desorption is beginning to release COMs into the gas phase \citep{2024MNRAS.534.2305J}. The detection of HMT in select carbonaceous chondrites (e.g., Murchison, Murray, and Tagish Lake) highlights its potential role in prebiotic chemistry, as its degradation can yield formaldehyde and ammonia which are key ingredients for amino acid formation \citep{Yasuhiro2020}. However, high-resolution mass spectroscopic data from the Rosetta mission at comet 67P did not confirm the presence of polymers like POMs or HMT, suggesting that thermal processing may have altered these materials in cometary environments \citep{Hanni2022}. The presence or absence of HMT in extraterrestrial materials remains a key question in astrobiology, as its detection could provide significant insight into the molecular pathways that link interstellar ice chemistry with the organic inventory of early planetary bodies. If HMT is widely present in primitive extraterrestrial materials, it could represent a crucial prebiotic reservoir. Furthermore, its possible presence in protostellar sources like IRAS 2A may serve as a valuable probe of early-stage star-forming regions, shedding light on the survival and evolution of prebiotic molecules in cold astrophysical environments.

\section{Conclusion} \label{sec:conclusions}
Based on the spectral fitting of IRAS 2A using a global continuum across the entire JWST MIRI range, along with a comparison of the derived column densities with local fitting values and those observed toward YSOs and comet 67P, the following conclusions can be drawn.
\begin{enumerate}
    \item The agreement between our derived ice column densities using global fitting and those from previous studies \citep{2024A&A...683A.124R} which used local fitting confirms that global fitting can be performed with accuracy comparable to local fitting, demonstrating the robustness of our approach.
    \item In complex ice spectra, absorption features of different species often overlap significantly with dominant ice features, making the identification of weaker components challenging, as seen in the case of OCN$^{-}$ feature blended with H$_{2}$O and CO$_{2}$ bands. Local fitting, which focuses on isolated spectral regions, may assign the blended features to more dominant species, leading to under-estimation of minor components or over-estimation of major components. In contrast, global fitting leverages multiple spectral regions where each species shows distinct features, ensuring a more accurate and constrained determination of ice composition.
    \item The presence of broad absorption features from $\sim$5.5\,$\mu$m to $\sim$11\,$\mu$m suggests contributions from various functional groups, including carbonyl (C=O), hydroxyl (O--H), amine (N--H), and C--H bending modes, indicating the presence of organic refractory materials. These refractory organics are crucial for understanding irradiation-driven chemical evolution in star- and planet-forming regions, and are likely incorporated into comets, asteroids, and C-type chondrites.

    \item The global fitting approach not only improves the accuracy of column density estimates by leveraging the full spectral range and all ice absorption bands but also reveals a rich chemical inventory. This includes simple volatiles such as H$_2$O, ionic species like NH$_4^{+}$, COMs and their precursors, as well as refractory organic residues, which together highlight the chemical complexity of interstellar ices. These findings provide insights into the diverse physical and chemical processes at play in star-forming environments.

    \item The spectral association of COMs with H$_2$O suggests that IRAS~2A hosts a predominantly polar environment. Furthermore, H$_2$O may actively influence COM chemistry, including the formation of organic refractory materials, by acting as a catalyst.

    \item The good correlation of the estimated ice column densities of IRAS 2A with those of both LYSOs and MYSOs within a factor of 10 suggests a chemically linked evolutionary pathway, with IRAS 2A likely representing an early or intermediate stage of a LYSO.
    \item The systematic depletion of COMs and to some extent, simple volatiles in the cometary ices of 67P relative to H$_2$O suggests that ices inherited from earlier protostellar phases undergo significant thermal and radiative processing, outgassing, and chemical transformation. This implies that the inheritance of pristine ices is incomplete and that ice compositions are actively reprocessed both prior to and after incorporation into cometary bodies.

\end{enumerate}

\section*{Acknowledgments}
P.R. and L.M. express their sincere thanks to Ewine F. van Dishoeck for detailed discussions related to the development of INDRA and for sharing her valuable feedback, which significantly improved the results and discussions presented in the paper. P.R. and L.M. especially thank Katie Slavicinska for sharing NH$_{4}$SH ice spectroscopy data, which has been used in the INDRA global fitting. L.M. also thanks the JOYS+ team members for their enthusiastic comments, which have helped improve the capabilities of INDRA over time. L.M. further thanks Murthy Gudipati for insightful discussions on ice spectroscopy. L.M. acknowledges financial support from DAE, Government of India, which enabled this work. This research was carried out in part at the Jet Propulsion Laboratory, California Institute of Technology, under a contract with the National Aeronautics and Space Administration (NASA). Also, special thanks to Astrochemistry group in Leiden which is supported by the Netherlands Research School for Astronomy (NOVA), by funding from the European Research Council (ERC) under the European Union’s Horizon 2020 research and innovation programme (grant agreement No. 101019751 MOLDISK), by the Dutch Research Council (NWO) grants TOP-1 614.001.751 and 618.000.001, and by the Danish National Research Foundation through the Center of Excellence “InterCat” (Grant agreement no.: DNRF150). We would like to thank the anonymous referee for constructive comments that helped improve the manuscript.
\clearpage
\appendix
\onecolumngrid
\section{List of laboratory data}\label{appendix:labdata}
Table~\ref{tab_list} presents the laboratory ice components used in the ice-fitting analysis of IRAS~2A in the 5--28~$\mu$m region. These data were compiled from various databases, as indicated in the table and discussed in Section~\ref{sec:labdata}. In total, 76 ice spectra are included, consisting of both pure ices and mixtures with defined proportions. For species measured at multiple temperatures, we provide a single entry indicating the corresponding temperature range, rather than listing each spectrum separately. To improve the spectral fits and maintain physical consistency, we used grain-shape-corrected laboratory data for pure H$_2$O and CO$_2$ ices, obtained from \citet{2024A&A...683A.124R} and \citet{Pontoppidan2008}, respectively. Further discussion of grain-shape effects and correction methods can be found in \citet{Ehrenfreund1996, Dartois2006, Dartois2022, Brunken2025}.

\begin{table*}[h]
\caption{\label{tab_list} Laboratory data used in this work.}
\renewcommand{\arraystretch}{1.0}
\centering 
\begin{tabular}{lcccccc}
    \hline
    Component & Absorption Features$^{\ddagger}$ & Ratio & T (K) & Database & References \\
    \hline
    pure H$_2$O $^\maltese$ & H$_2$O & – & 15–160 & LIDA & \cite{Oberg2007} \\
    pure H$_2$O UV irr. & H$_2$O & – & 10-10.2 & LIDA & \cite{gerakines1996ultraviolet} \\
    H$_2$O:CH$_4$ & CH$_4$ & 10:1 & 15–16 & UNIVAP & \cite{2017MNRAS.464..754R} \\
    H$_2$O:CH$_4$ & CH$_4$ & 10:0.6 & 16 & UNIVAP & \cite{2017MNRAS.464..754R}\\
    NH$_3$:CH$_3$OH & NH$_3$, CH$_3$OH & 1:1 & 15 & UNIVAP & \cite{Rocha2020}  \\
    H$_2$O:CH$_3$CHO & H$_2$O, CH$_3$CHO & 20:1 & 15–120 & LIDA & \cite{Scheltinga2018} \\
    pure CH$_3$CHO & CH$_3$CHO & – & 15–120 & LIDA & \cite{Scheltinga2018} \\
    pure HCOOCH$_3$ & HCOOCH$_3$ & – & 15-120 & LIDA & \cite{Scheltinga2021} \\
    CH$_3$CH$_2$OH:H$_2$O & CH$_3$CH$_2$OH, H$_2$O & 1:20 & 15–160 & LIDA & \cite{Scheltinga2018} \\
    CH$_3$CN & CH$_3$CN & – & 15–150 & LIDA & \cite{Rachid2022} \\
    H$_2$O:CO$_2$ & CO$_2$ & 1:0.14 & 10 & LIDA & \cite{Ehrenfreund1997} \\
    CO:CO$_2$ & CO$_2$ & 2:1 & 15-130 & LIDA & \cite{Schutte1999_weak} \\
    pure CO$_2$$^\maltese$ & CO$_2$ & – & – &  & \cite{Pontoppidan2008}  \\
    H$_2$O:NH$_3$:HCOOH & HCOO$^{-}$ & 100:2.6:2 & 14 & LIDA & \cite{Galvez2010} \\
    H$_2$O:CH$_3$COOH & H$_2$O, CH$_3$COOH & 10:1 & 16 & NASA & NASA Cosmic Ice Laboratory$^{\dagger}$ \\
    H$_2$O:CH$_3$OH:NH$_3$:CO:CO$_2$ 
      & Org. res.$^{\star}$
      & 2:1:1:1:1 & 12 & - & \cite{Caro2003}  \\
    HCOOH:H$_2$O:CH$_3$OH & HCOOH, CH$_3$OH & 6:68:26 & 15 & LIDA & \cite{Bisschop2007}\\
    HNCO:NH$_3$ & OCN$^-$ & 1:1 & 80 & LIDA & \cite{Novozamsky2001} \\
    pure CH$_4$ & CH$_4$ & – & 12.0 & OCdb & \cite{Hudgins1993} \\
    CH$_3$COCH$_3$:H$_2$O & CH$_3$COCH$_3$ & 1:20 & 15 & LIDA & \cite{Rachid2020} \\
    CH$_3$COCH$_3$:H$_2$CO & CH$_3$COCH$_3$, H$_2$CO & 1:20 & 15–150 & LIDA & \cite{Scheltinga2021}\\
    pure SO$_2$ & SO$_2$ & – & 10.0 & LIDA & \cite{Boogert1997} \\
    SO$_2$:CH$_3$OH & SO$_2$ & 1:1 & 10 & LIDA & \cite{Boogert1997} \\
    H$_{2}$O:NH$_{3}$:H$_{2}$S & H$_{2}$O, NH$_{4}^+$, NH$_{3}$  & 10:2:1 & 135 & LIDA & \cite{Slavicinska2025} \\
    \hline
\end{tabular}
\label{tab:ice_mixtures}
\tablecomments{ $^{\dagger}$\href{https://science.gsfc.nasa.gov/691/cosmicice/spectra/refspec/Acids/CH3COOH/ACETIC-W.txt}{NASA Cosmic Ice Laboratory reference spectrum for CH$_3$COOH}; $^{\star}$ Organic residue consists of species with diverse functional groups including carbonyl (-C=O), carboxylate (-COO$^{-}$), amine (-NH$_{2}$), and -CN containing species. They also consist of carboxylic acid salts, hexamethylenetetramine (HMT), polyoxymethylene (POM); $^{\ddagger}$ Only the absorption features of the ice components for the listed species are included in the database; $^\maltese$ Grain shape corrected ices.\\ Abbreviations: irr. = irradiated ice. }
\end{table*}
\clearpage
\section{A note on the band strengths of ices used in this work}
\label{appendix:bandstrengths}
The band strengths of molecular ices vary significantly depending on whether they exist in a pure or mixed state. This is due to the intermolecular interactions that alter the vibrational properties of molecules. In pure ices, absorption features arise from the vibrational modes of individual molecules which in fact depend on molecular dipole moment and crystalline structure etc. However, when molecules are present as mixtures, interactions with other species can lead to shifts in band positions, changes in intensity and modifications of band profiles. Consequently, there can either be enhancement or suppression of vibrational modes depending on the nature and strength of that interaction. The latter depends on the physical conditions like temperature as well. For example, \cite{2007A&A...476..995B} found that in H$_{2}$O:CO ice mixtures, band intensities vary systematically with concentration, with H$_{2}$O features weakening and the free OH stretch strengthening as CO increases. \cite{Bouilloud2015} have derived the correction factors for the band strengths of CO and CO$_{2}$ ices based on the works of \cite{Gerakins1995}.  Such variations in band strengths of ices can impact the ice column density estimations. And, hence it is required to correct the band strengths of ices in order to prevent under or over estimation of abundances. 
\begin{table*}[h]
\caption{\label{ice_bs} List of vibrational transitions and band strengths of molecules considered in this paper.}
\centering 
\begin{tabular}{ccclccccc}
\hline\hline
 Chemical formula & Name & $\lambda \; [\mu \mathrm{m}]$ & $\nu \; \mathrm{[cm^{-1}]}$ & Identification & $\mathcal{A} \; \mathrm{[cm \; molec^{-1}]}$ & References\\
\hline
H$_2$O & Water & 13.20    & 760 & libration & $\mathrm{3.2 \times 10^{-17}}$ & [1][14]\\
CH$_4$    & Methane       & 7.67    & 1303 & CH$_4$ def.   & $\mathrm{8.4 \times 10^{-18}}$ & [1][2][14]\\
SO$_2$   & Sulfur dioxide    & 7.60    & 1320 & SO$_2$ stretch.     & $\mathrm{3.4 \times 10^{-17}}$ & [2]\\
H$_2$CO   & Formaldehyde    & 8.04    & 1244 & CH$_2$ rock.     & $\mathrm{1.0 \times 10^{-18}}$ & [1]\\
CH$_3$OH   & Methanol    & 9.74    & 1026 & C$-$O stretch.     & $\mathrm{1.8 \times 10^{-17}}$ & [1]\\
 & & & & & $\mathrm{1.56 \times 10^{-17}}$ & [15] \\
HCOOH   & Formic acid     & 8.22    & 1216 & C$-$O stretch.     & $\mathrm{2.9 \times 10^{-17}}$ & [1]\\
CH$_3$CHO   & Acetaldehyde    & 7.41  & 1349 & CH$_3$ s-def./CH wag.     & $\mathrm{4.1 \times 10^{-18}}$ & [3][17]\\
CH$_3$CH$_2$OH      & Ethanol     & 7.23    & 1383 & CH$_3$ s-def.     & $\mathrm{2.4 \times 10^{-18}}$ & [4]\\
CH$_3$OCHO & Methyl formate  & 8.25    & 1211 & C$-$O stretch.   & $\mathrm{2.52 \times 10^{-17}}$ & [5]\\
   &   &     &  &    & $\mathrm{2.28 \times 10^{-17}}$ & [5]\\
CH$_3$COOH & Acetic acid  & 7.82    & 1278 & OH bend.      & $\mathrm{4.57 \times 10^{-17}}$ & [6] \\
HCOO$^-$ (B1) & Formate ion  & 7.23    & 1383 & C$-$H def.   & $\mathrm{8.0 \times 10^{-18}}$ & [16]\\
HCOO$^-$ (B2) & Formate ion   & 7.38    & 1355 & C$-$O stretch.  & $\mathrm{1.7 \times 10^{-17}}$ & [16]\\
OCN$^-$ & Cyanate ion   & 7.62    & 1312 & Comb. (2$\nu_2$)  & $\mathrm{7.45 \times 10^{-18}}$ & [6]\\
NH$_{4}^{+}$ & Ammonium ion & 6.85 & 1428 & $\nu_4$ (asym. bend.) & $\mathrm{4.4 \times 10^{-17}}$ &  [11]\\
NH$_{3}$ & Ammonia & 9.34 & 1070 & Umbrella mode & $\mathrm{2.1 \times 10^{-17}}$ & [1]\\
CO$_{2}$ & Carbon dioxide & 15.27 & 654 & & 1.45$\times \mathrm{1.6 \times 10^{-17, a}}$ & [12]\\
HMT & Hexamethylenetetramine & 8.1 & 1234 & CN stretch. & $\mathrm{2.6 \times 10^{-18}}$  &  [13] \\
HMT & Hexamethylenetetramine & 9.3 & 1007 & CN stretch. & $\mathrm{5 \times 10^{-18}}$  &  [13] \\
CAsalt & Carboxylic acid salt & 6.3 & 1586 &  & $\mathrm{6 \times 10^{-17}}$  &  [13] \\
Amide & Amide & 5.9 & 1680 &  & $\mathrm{3.3 \times 10^{-17}}$  &  [13] \\
Ester & Ester & 5.74 & 1742 &  & $\mathrm{2 \times 10^{-17}}$  &  [13] \\
POM & polyoxymethylene & 9.1 & 1098 &  & $\mathrm{9.7 \times 10^{-18}}$  &  [13] \\
\hline
\end{tabular}
\tablecomments{\footnotesize
Abbreviation of vibrational modes: def. = deformation; stretch. = stretching; s-def. = symmetric stretching; rock. = rocking; sym. = symmetric; asym. = asymmetric;  Comb. = combination; wag. = wagging; bend. = bending; CAsalt = carboxylic acid salt; HMT = hexamethylenetetramine; POM = polyoxymethylene.}
[a] Correction factor derived from \citealt{Bouilloud2015, Brunken2024}, 
[1] \citet{Bouilloud2015}, [2] \citet{Boogert1997}, [3] \citet{Hudson2020}, [4] \citet{Boudin1998}, [5] \citet{Scheltinga2021}, [6] \citet{2024A&A...683A.124R}, [7] \citet{Rachid2020}, [8] \citet{Rachid2021}, [9] \citet{slav2023}, [10] \citet{Hudson2005},
[11] \citet{Schutte2003}, [12] \citet{Gerakins1995}, [13] \citet{Caro2003}, [14] \citet{Hudgins1993}, [15] \citet{Luna2018},
[16] \citet{Schutte1999_weak}, [17] \citet{Scheltinga2018} 
\end{table*}
\clearpage
\section{Baseline corrections }\label{appendix:baselines}
In spectroscopic analysis, local baseline correction is essential for accurately extracting individual absorption features, especially when multiple components contribute to a broad spectral profile. Many ices produce overlapping absorption bands, leading to blended features that obscure individual contributions. In some cases, as in the case of organic molecules, the presence of strong, broad absorption bands from dominant species, say H$_{2}$O, can mask weaker features, making their fitting unreliable. By applying a local baseline correction, we can effectively remove underlying continuum variations and overlapping absorptions, ensuring that the extracted spectral features accurately represent the ices of interest. Polynomial baselines are effective in tracing the local continuum and this approach is widely used in many studies \citep{2024A&A...683A.124R, 2024A&A...690A.205C, Caro2003}. This approach helps in determining ice column densities, and enhances the reliability of spectral fitting.
\begin{figure*}[h]
    \includegraphics[width=\textwidth]{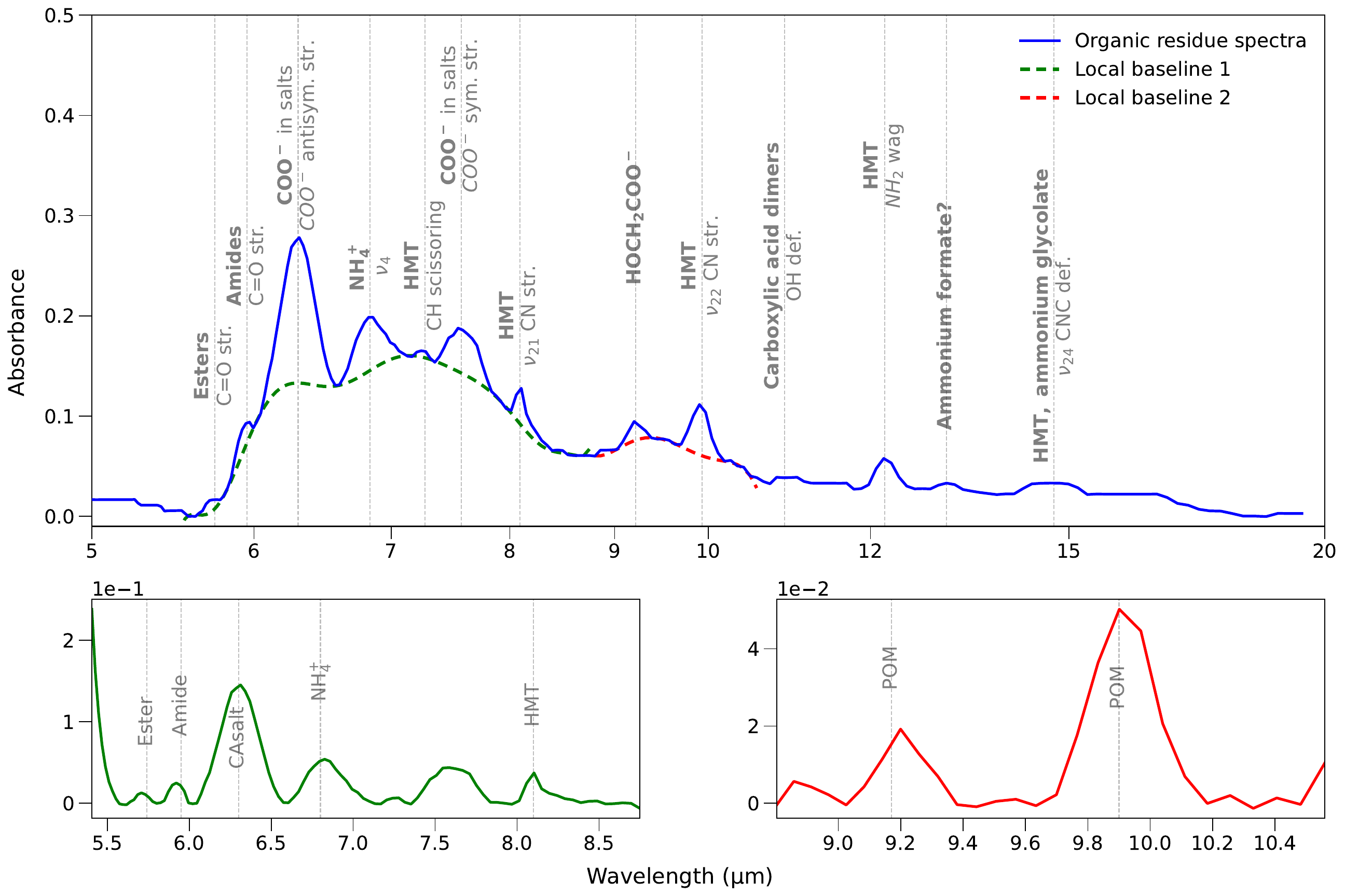}
    \caption{IR spectrum of the residue (blue) from the irradiated ice mixture H$_{2}$O:NH$_{3}$:CH$_{3}$OH:CO:CO$_{2}$ (2:1:1:1:1) at 12~K, adapted from \cite{Caro2003}. The top panel shows the spectrum in the 5-20~$\mu$m region, along with local polynomial baselines (dashed lines) used to isolate absorption features. These features are also indicated by labels in the text and grey dotted lines. They correspond to functional groups such as carbonyl (-C=O), carboxylate (-COO$^{-}$), amine (-NH$_{2}$), and -CN containing species. Because these features are blended within a broad profile, local baselines were required for accurate column density calculations. The bottom panel shows the baseline-corrected features (solid lines). Column densities were estimated using Equation~\ref{eq:colden} and band strengths listed in Appendix~\ref{appendix:bandstrengths}. Uncertainties due to baseline selection are $\lesssim 20\%$. See \cite{Caro2003} for details.\\
    \footnotesize
    Abbreviations: CAsalt = carboxylic acid salt; HMT = hexamethylenetetramine; POM = polyoxymethylene.}
    \label{fig:organic_residue}
\end{figure*}
\begin{figure*}[h]
    \includegraphics[width=\textwidth]{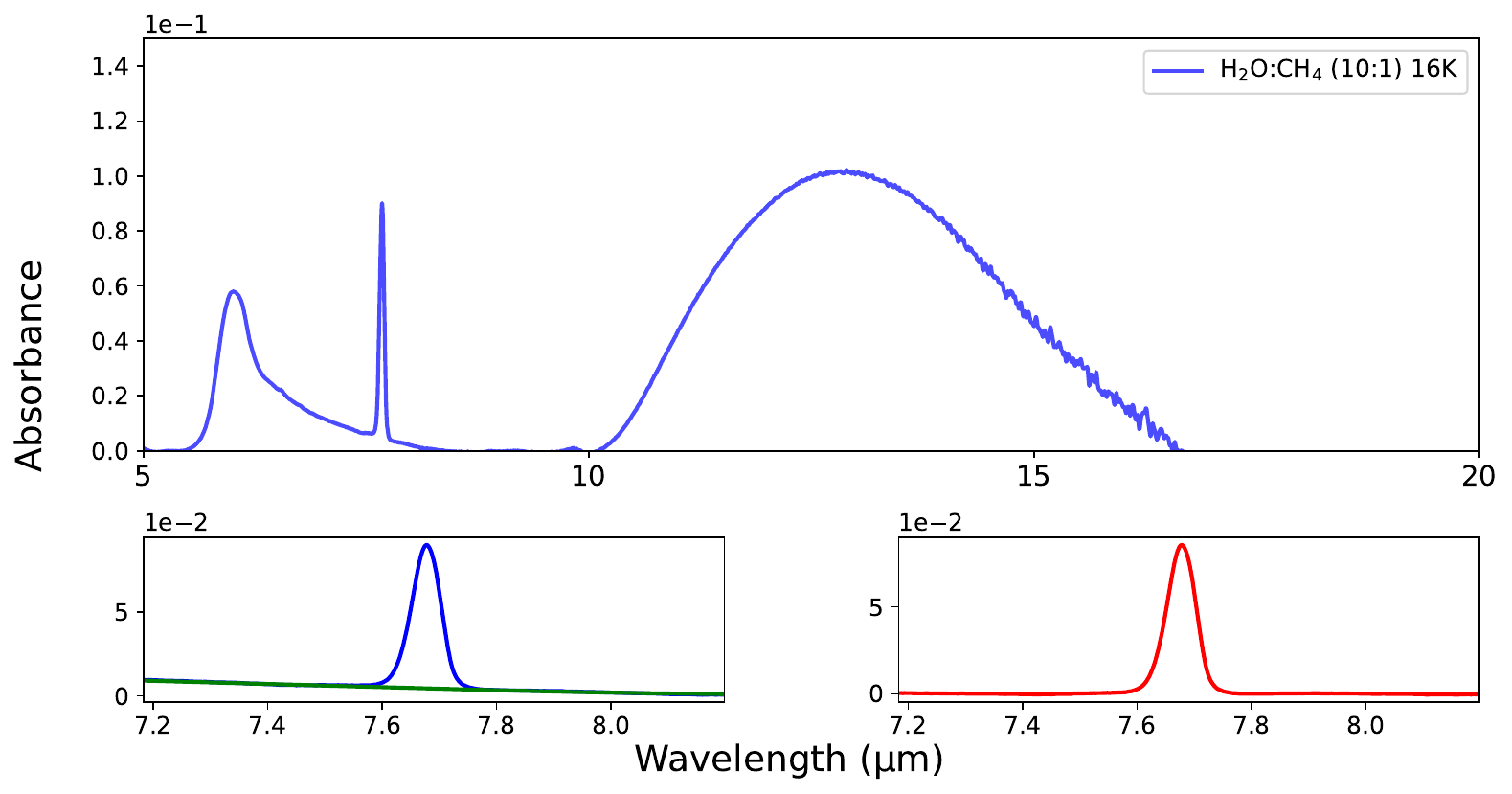}
    \caption{The absorption profile of H$_{2}$O : CH$_{4}$ (10:1) ice mixture \citep{2017MNRAS.464..754R} is shown in blue. The bottom-left shows the CH$_{4}$ bending mode at 7.6 $\mu$m in blue and the corresponding local polynomial baseline in green. The resultant isolated band of CH$_{4}$ is shown in red.}
    \label{fig:H2O_CH4}
\end{figure*}
\begin{figure*}[h]
    \includegraphics[width=\textwidth]{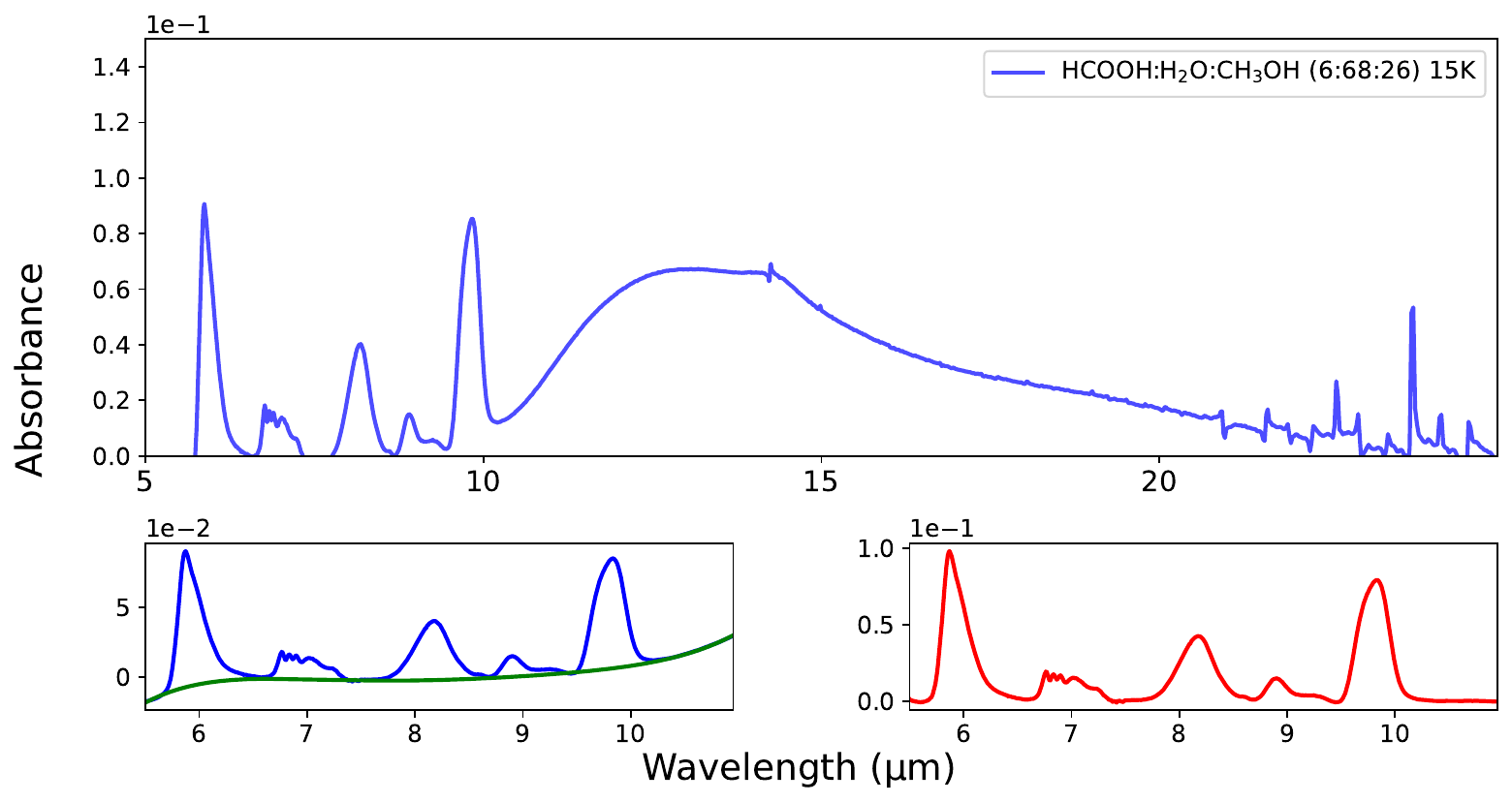}
    \caption{The absorption profile of 15 K ice mixture HCOOH:H$_{2}$O:CH$_{3}$OH (6:68:26) \citep{Bisschop2007} is shown in blue in the top panel. Features other than H$_{2}$O are shown in the bottom-left panel in blue and the corresponding polynomial baseline in green. The H$_{2}$O subtracted spectra is shown in red.}
    \label{fig:H2O_HCOOH}
\end{figure*}
\begin{figure*}[h]
    \includegraphics[width=\textwidth]{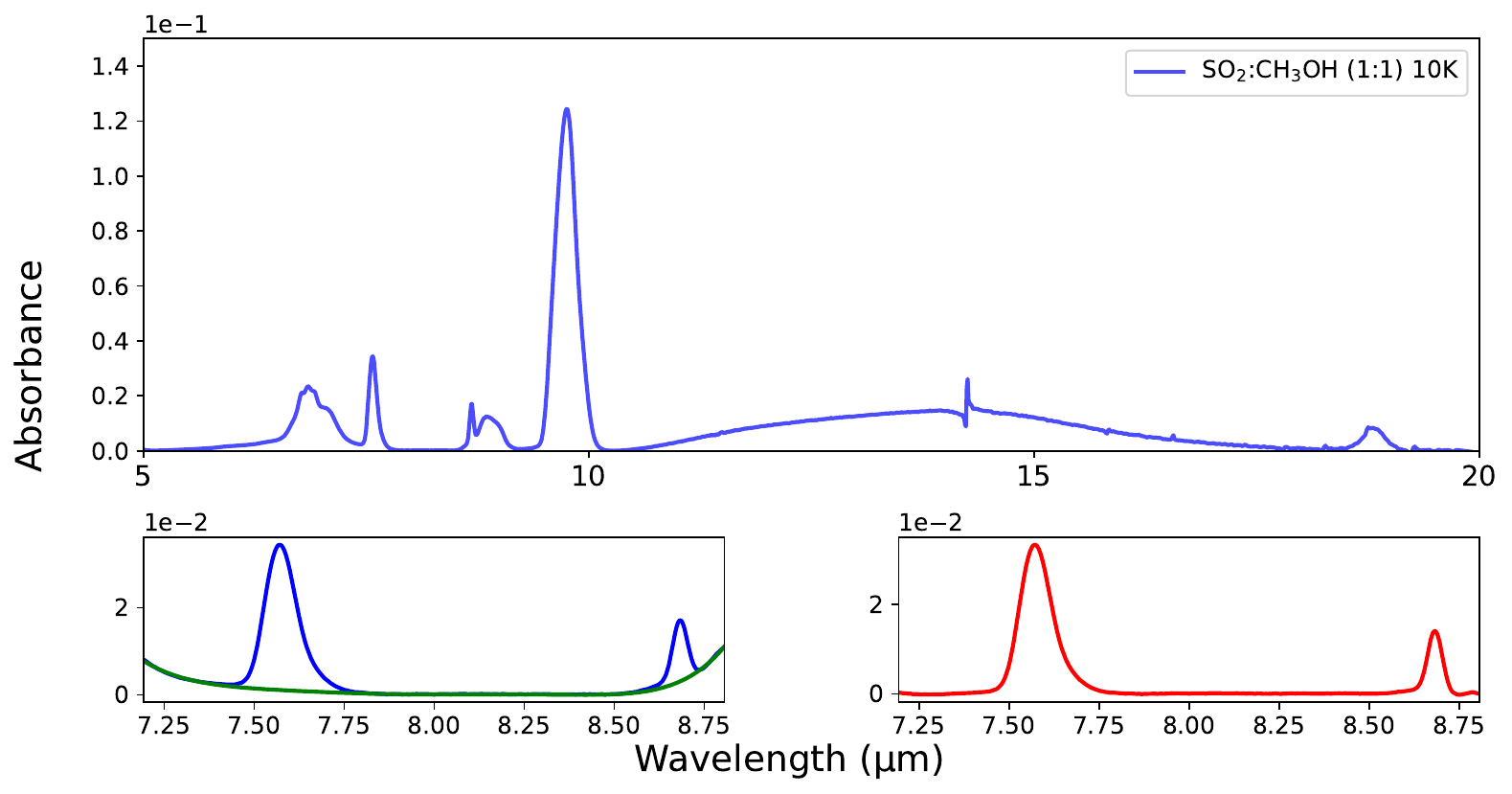}
    \caption{The absorption profile of SO$_{2}$:CH$_{3}$OH (1:1) 10 K ice mixture is shown in blue. The bottom-left panel shows the SO$_{2}$ features at 7.57 and 8.68 $\mu$m in blue and the corresponding local polynomial baseline in green. The resultant baseline-corrected profiles are shown in red in bottom-right panel.}
    \label{fig:SO2}
\end{figure*}
\clearpage
\section{Statistical results} \label{appendix:stats}
\onecolumngrid
\begin{figure*}[htbp]
    \includegraphics[width=\textwidth]{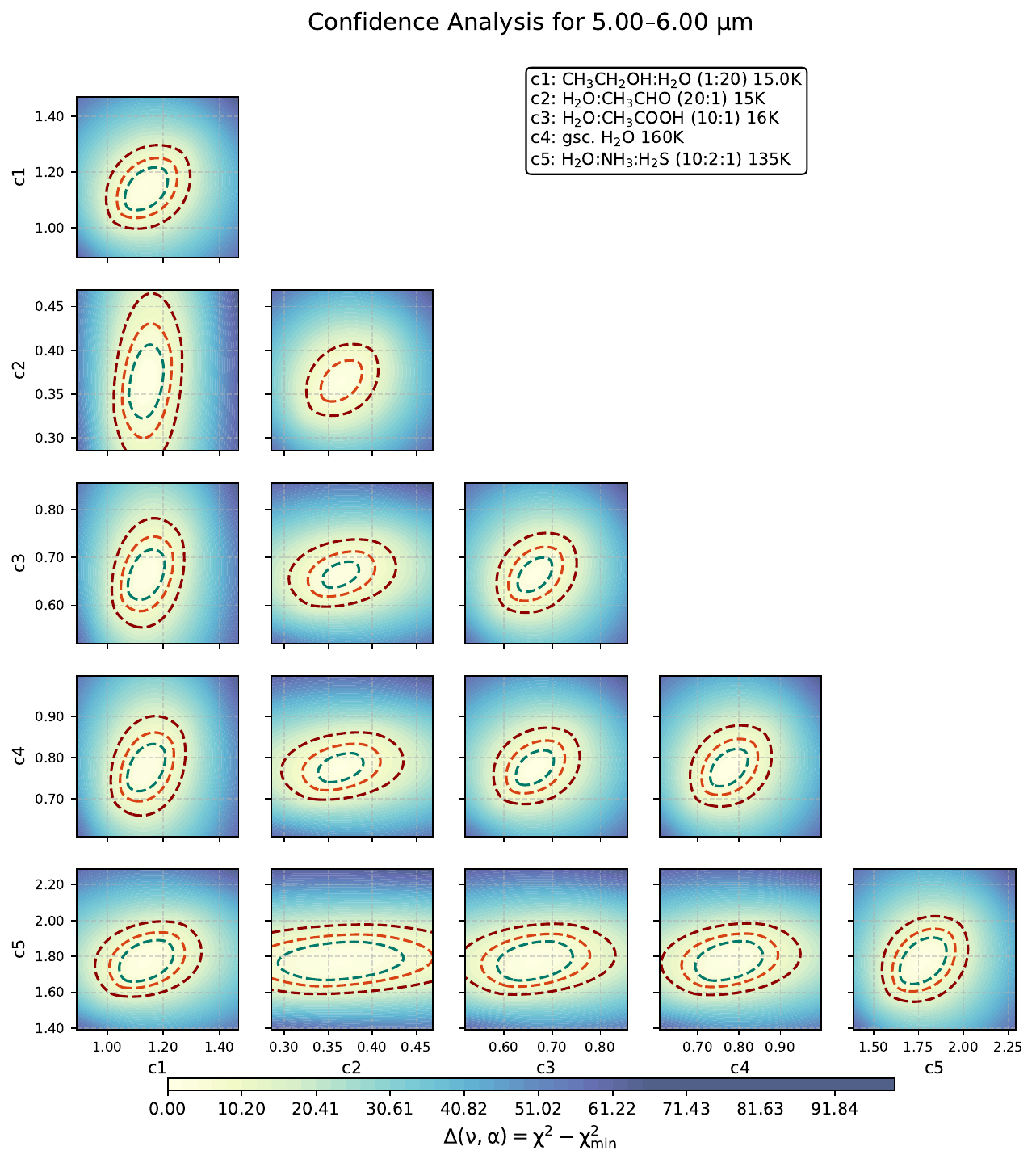}
    \caption{Confidence interval analysis of the components having H$_{2}$O bands based on $\chi^{2}$ contributing in 5-6 $\mu$m region. The blue, red and brown dotted lines indicate the 1, 2 and 3 $\sigma$ levels. The coefficients of the remaining components are kept constant while performing this analysis.}
    \label{fig:corner_h2o_1}
\end{figure*}
\begin{figure*}[h!]
    \includegraphics[width=\textwidth]{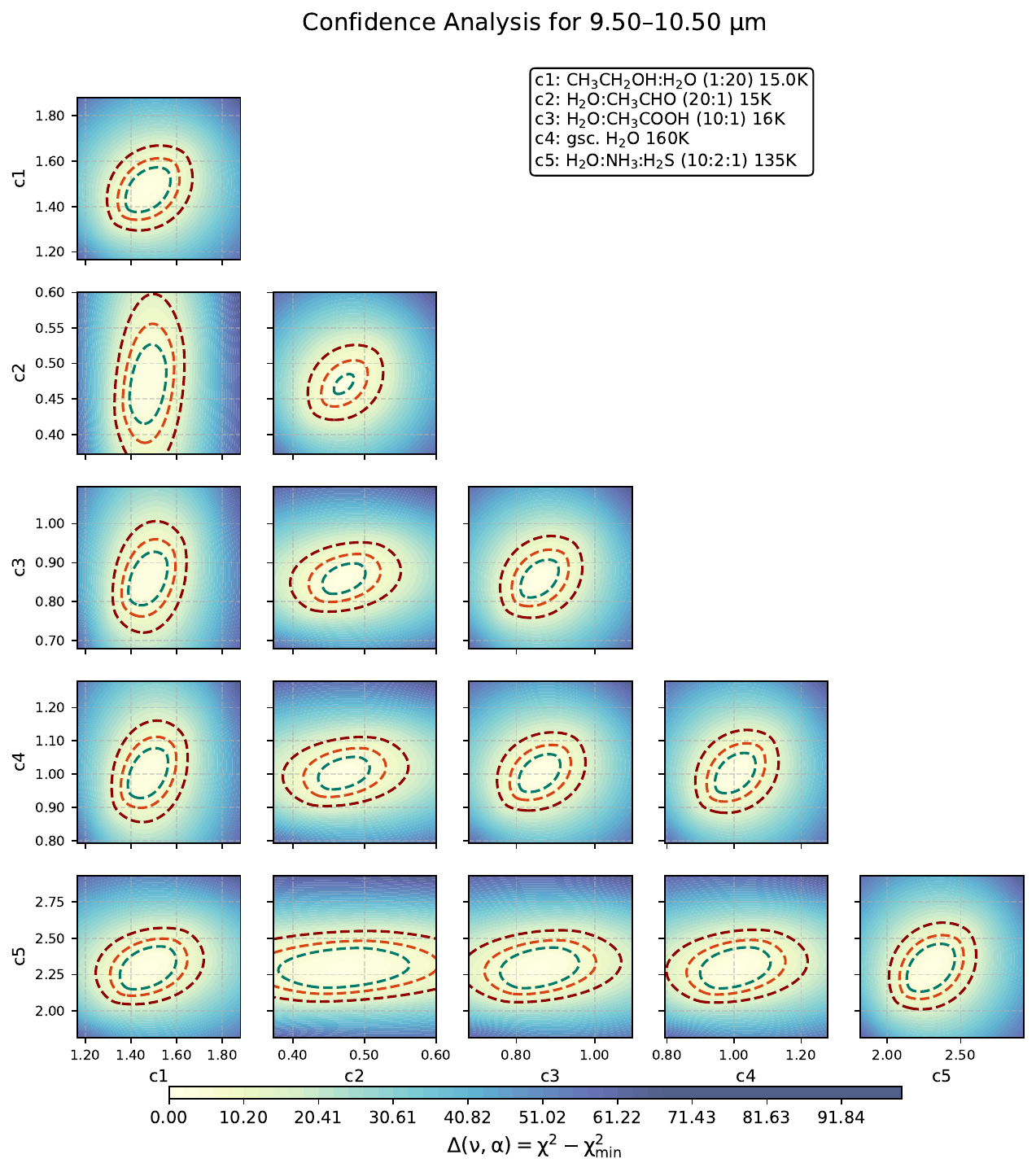}
    \caption{Confidence interval analysis of the components having H$_{2}$O bands based on $\chi^{2}$ that are contributing in 9.5-10.5  $\mu$m region. The blue, red and brown dotted lines indicate the 1, 2 and 3 $\sigma$ levels. The coefficients of the remaining components are kept constant while performing this analysis.}
    \label{fig:corner_h2o_2}
\end{figure*}
\begin{figure*}[h!]
    \includegraphics[width=\textwidth]{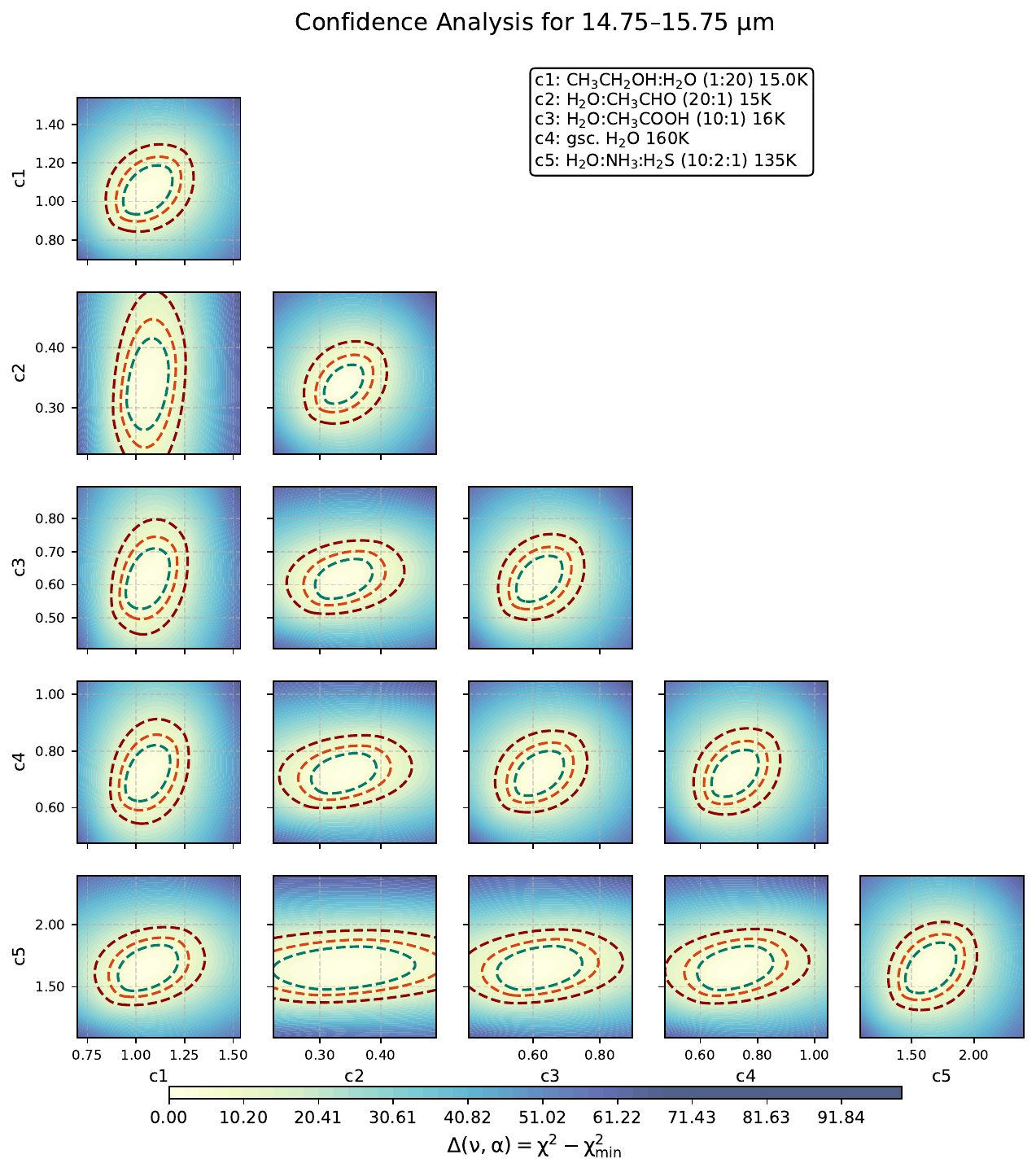}
    \caption{Confidence interval analysis of the components having H$_{2}$O bands based on $\chi^{2}$ in 14.75-15.75 $\mu$m region. The blue, red and brown dotted lines indicate the 1, 2 and 3 $\sigma$ levels. The coefficients of the remaining components are kept constant while performing this analysis.}
    \label{fig:corner_h2o_3}
\end{figure*}
\begin{figure*}[h!]
    \includegraphics[width=\textwidth]{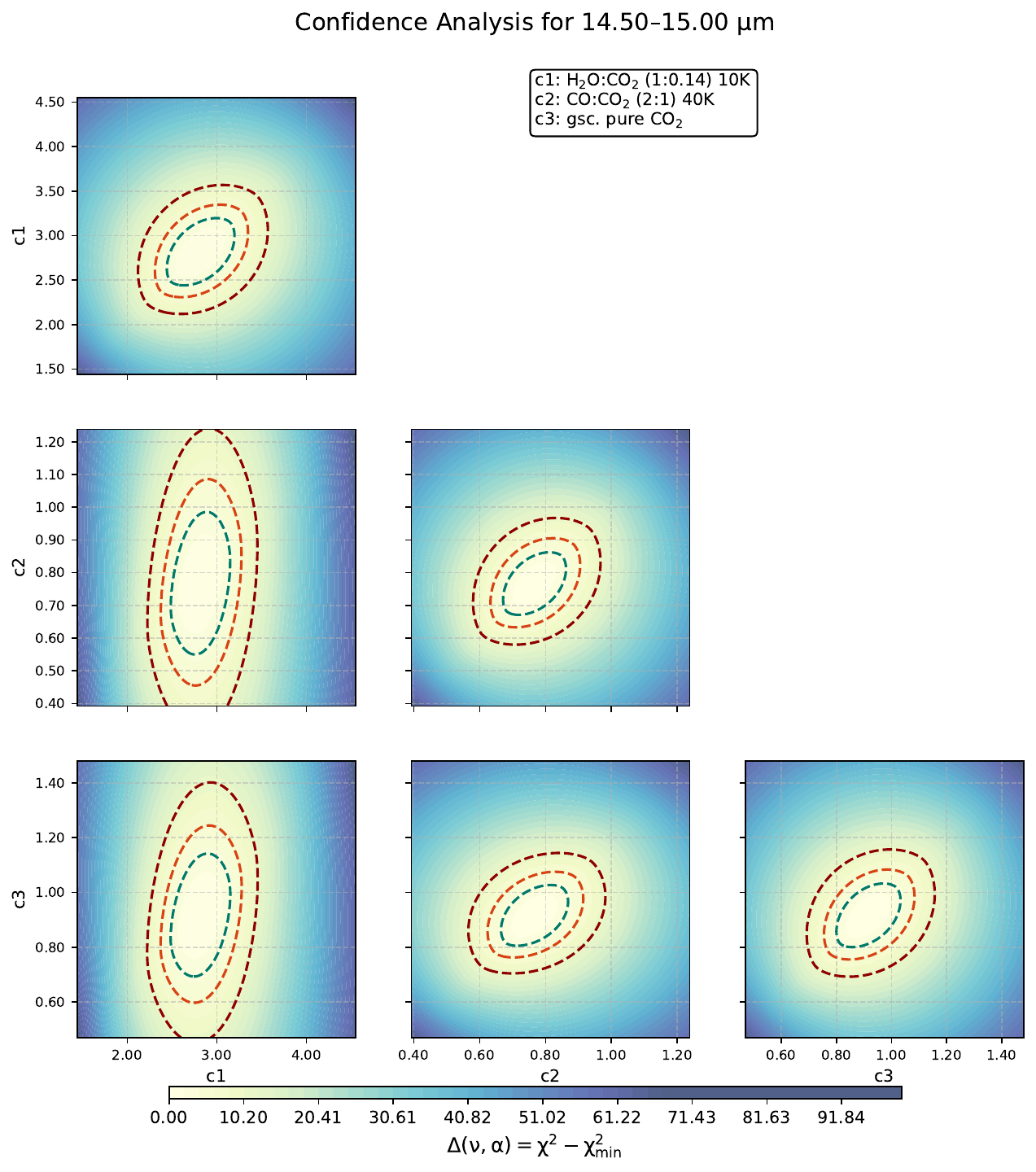}
    \caption{Confidence interval analysis of the CO$_{2}$ components based on $\chi^{2}$ in 14.50 - 15 $\mu$m region. The blue, red and brown dotted lines indicate the 1, 2 and 3 $\sigma$ levels. The coefficients of the remaining components are kept constant while performing this analysis.}
    \label{fig:corner_co2_1}
\end{figure*}
\begin{figure*}[h!]
    \includegraphics[width=\textwidth]{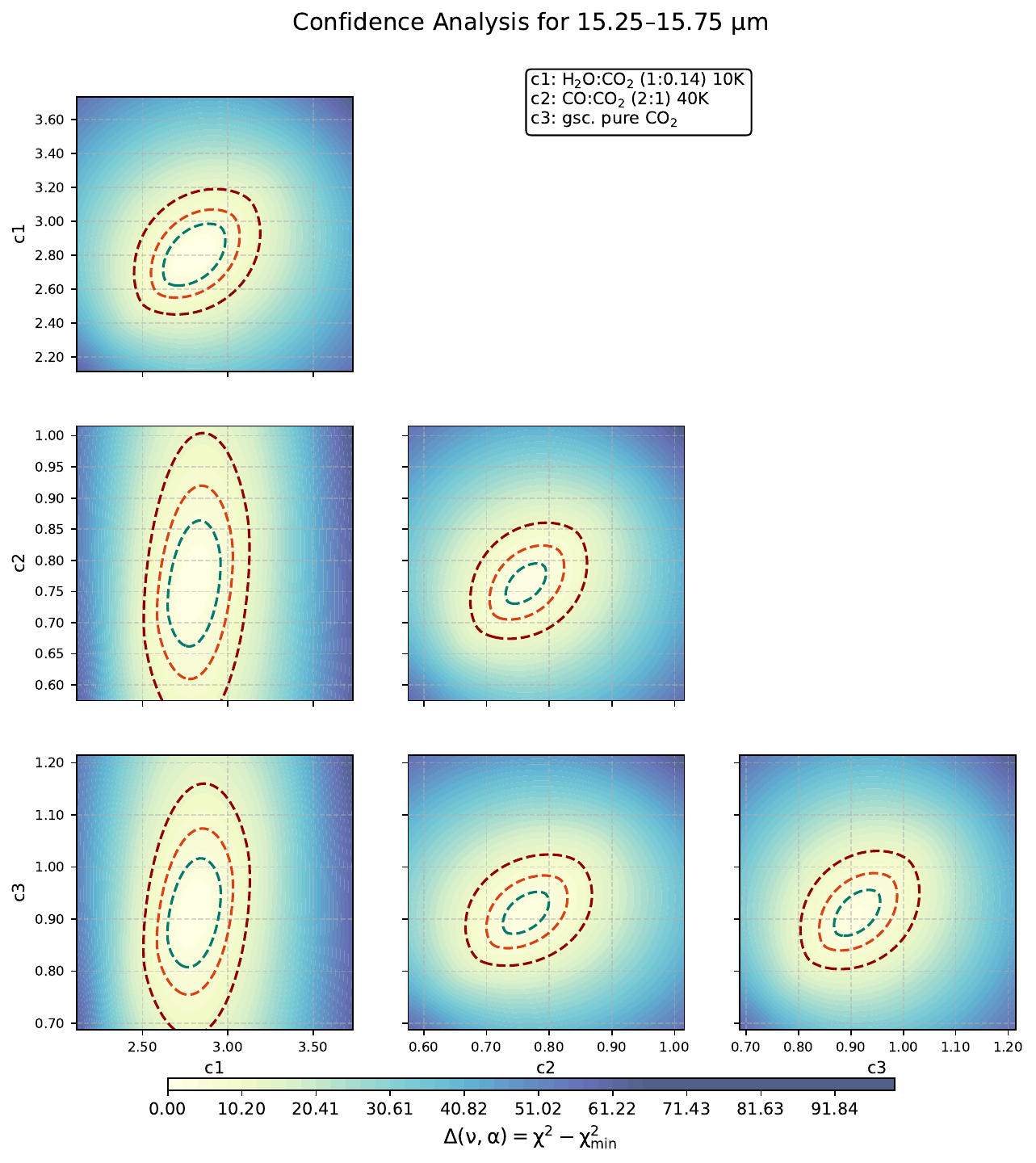}
    \caption{Confidence interval analysis of the CO$_{2}$ components based on $\chi^{2}$ in 15.25 - 15.75 $\mu$m region. The blue, red and brown dotted lines indicate the 1, 2 and 3 $\sigma$ levels. The coefficients of the remaining components are kept constant while performing this analysis.}
    \label{fig:corner_co2_2}
\end{figure*}
\begin{figure*}[h!]
    \includegraphics[width=\textwidth]{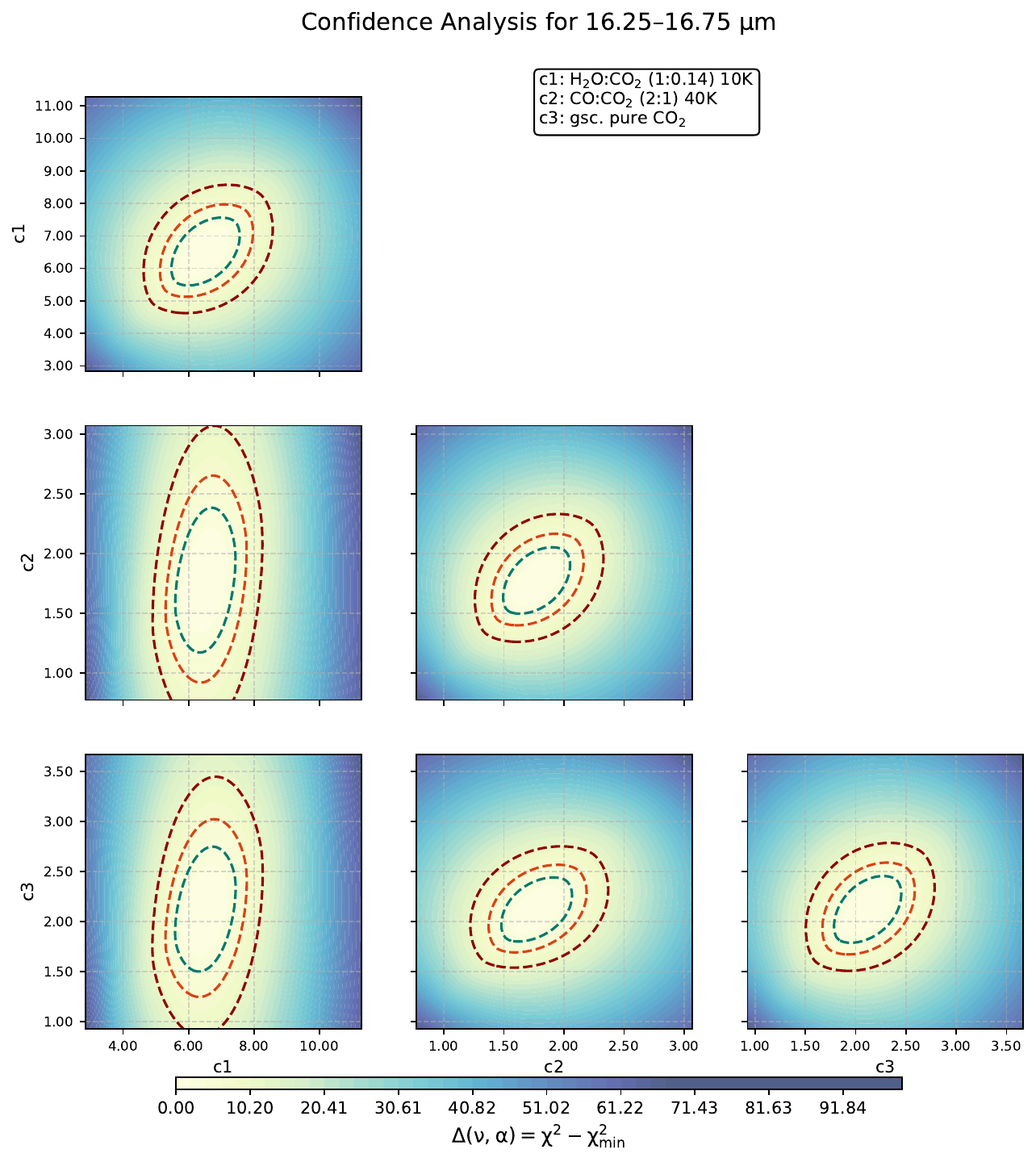}
    \caption{Confidence interval analysis of the CO$_{2}$ components based on $\chi^{2}$ in 16.25 - 16.75 $\mu$m region. The blue, red and brown dotted lines indicate the 1, 2 and 3 $\sigma$ levels. The coefficients of the remaining components are kept constant while performing this analysis.}
    \label{fig:corner_co2_3}
\end{figure*}
\begin{figure*}[h!]
    \includegraphics[width=\textwidth]{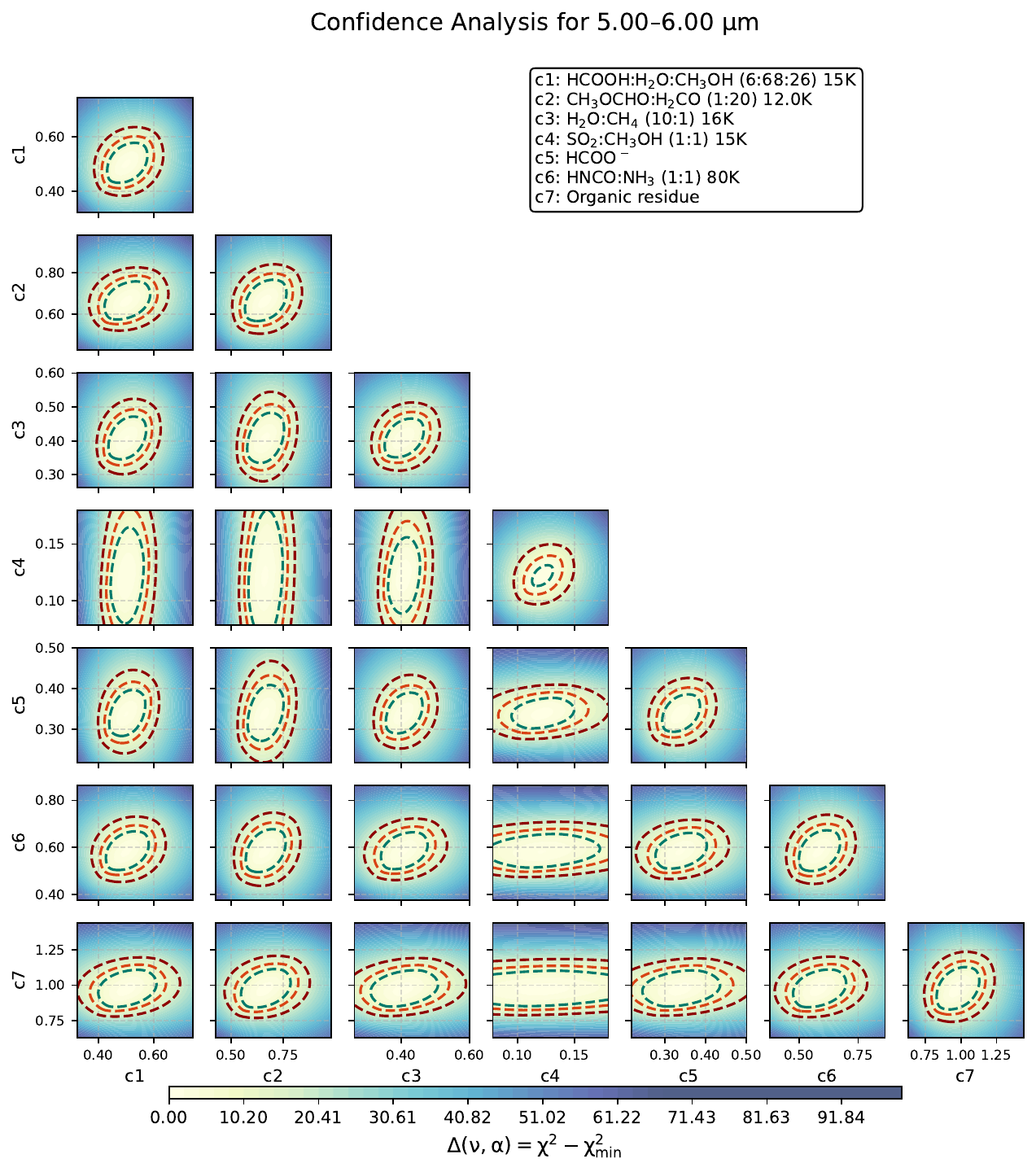}
    \caption{Confidence interval analysis of the components without H$_{2}$O and CO$_{2}$ bands based on $\chi^{2}$ contributing in 5-6 $\mu$m region. The blue, red and brown dotted lines indicate the 1, 2 and 3 $\sigma$ levels. The coefficients of the remaining components are kept constant while performing this analysis.}
    \label{fig:corner_minor_1}
\end{figure*}
\begin{figure*}[h!]
    \includegraphics[width=\textwidth]{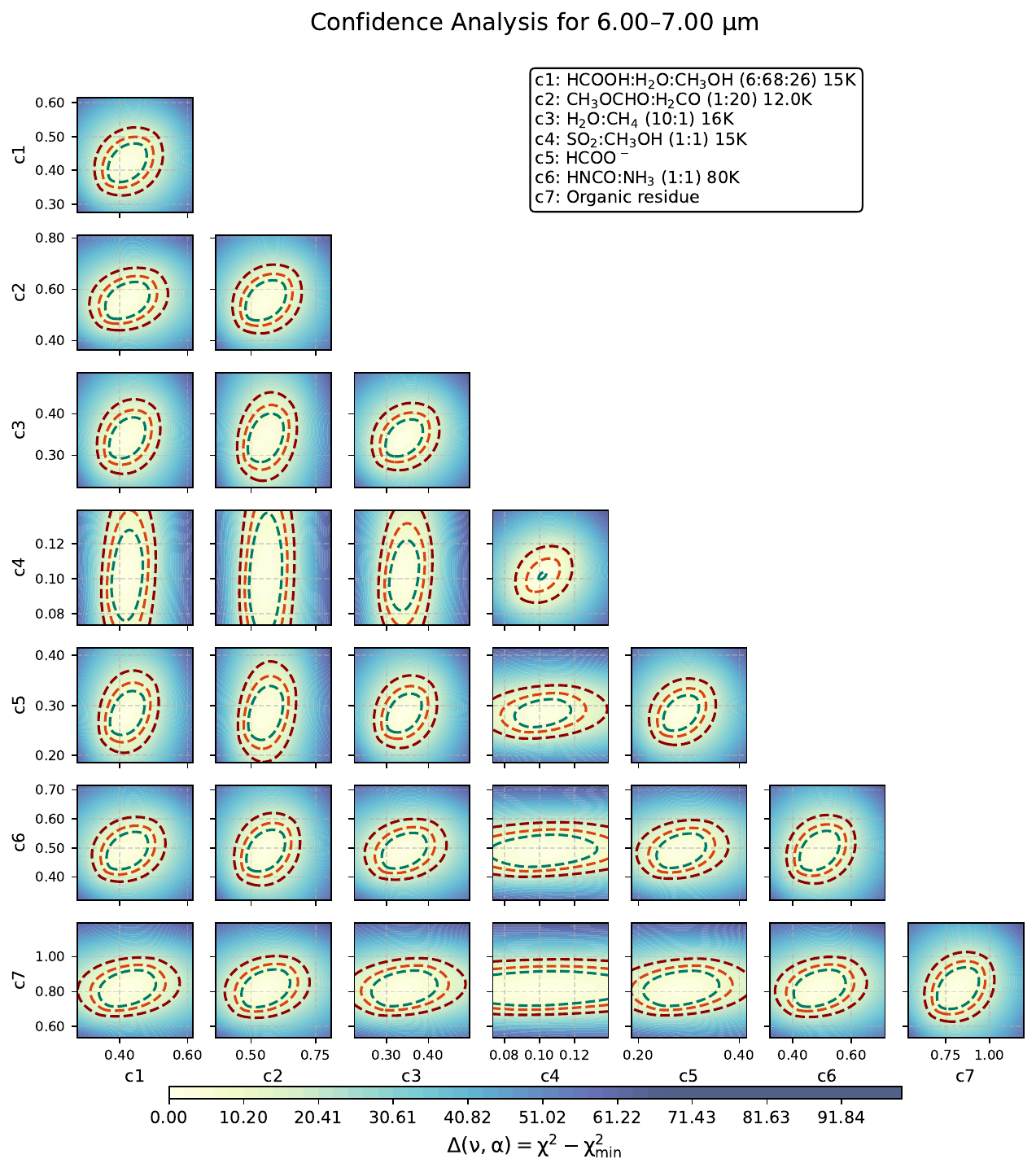}
    \caption{Confidence interval analysis of the components without H$_{2}$O and CO$_{2}$ bands based on $\chi^{2}$ that are contributing in 14.5-17.5  $\mu$m region. The blue, red and brown dotted lines indicate the 1, 2 and 3 $\sigma$ levels. The coefficients of the remaining components are kept constant while performing this analysis.}
    \label{fig:corner_minor_2}
\end{figure*}
\begin{figure*}[h!]
    \includegraphics[width=\textwidth]{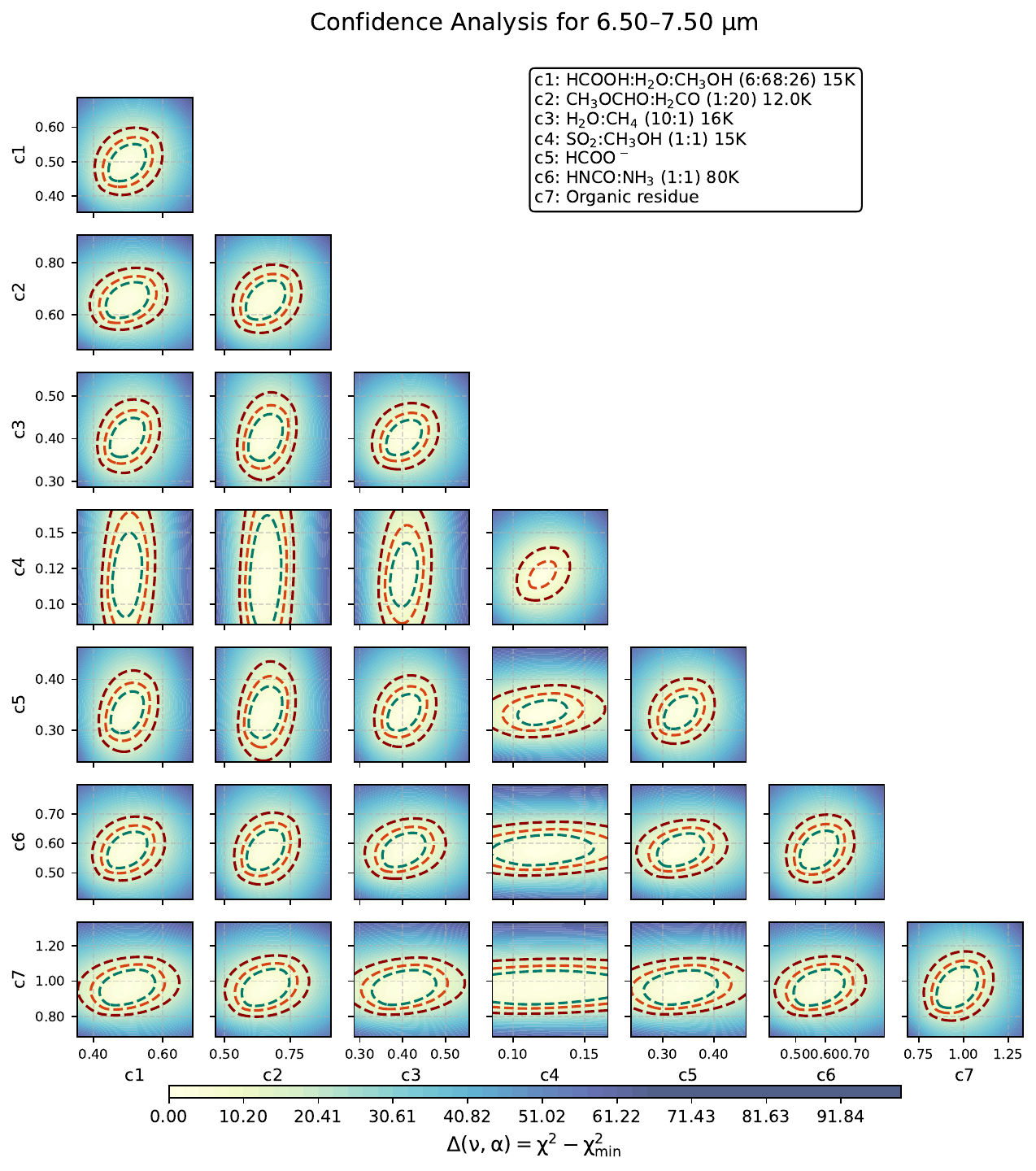}
    \caption{Confidence interval analysis of the components without H$_{2}$O and CO$_{2}$ bands based on $\chi^{2}$ in 5-11 $\mu$m region. The blue, red and brown dotted lines indicate the 1, 2 and 3 $\sigma$ levels. The coefficients of the remaining components are kept constant while performing this analysis.}
    \label{fig:corner_minor_3}
\end{figure*}
\clearpage
\bibliographystyle{aasjournal} 
\bibliography{bibliography} 
\end{document}